\documentclass{aa}  

\usepackage{graphicx}
\usepackage{txfonts}
\usepackage{lipsum}
\usepackage{subcaption}         
\usepackage{lscape}             
\usepackage{placeins}           

\newcommand{\uvec}[1]{\hat{\vec{#1}}}
\newcommand{\dinf}{{\rm d}}
\newcommand{\eulernumb}{{\rm e}}

\newcommand{\edited}[1]{#1}

\begin{document}

\title{Are the Parker and focused transport equations equivalent for galactic cosmic ray modulation?}


%

\author{J.P. van den Berg\inst{1,2}\fnmsep\thanks{Corresponding author: 24182869@mynwu.ac.za}
\and N.E. Engelbrecht\inst{1,2}
\and R.D. Strauss\inst{1,2}}

\institute{Centre for Space Research, North-West University, Potchefstroom, South Africa
\and National Institute for Theoretical and Computational Physics (NITheCS), South Africa}

 
\abstract
{The Parker transport equation (TPE) has been the equation of choice for the past 60 years in studies of galactic cosmic ray (GCR) modulation. Conversely, the focused TPE describes the same processes on a more fundamental level than the Parker TPE by modelling an anisotropic distribution rather than an isotropic one. It is usually assumed that the Parker TPE is valid for modelling GCRs, but the two TPEs have not been tested against each other in this context.}
{We conduct a first-of-its-kind comparison of these TPEs without particle drifts to test whether they produce the same results under identical diffusion conditions.} 
{A new model for protons during solar minimum conditions is developed to numerically solve the TPEs using stochastic differential equations. The TPEs are designed to be as consistent as possible for diffusion by normalising the pitch-angle-dependent diffusion coefficients used in the focused TPE to the isotropic diffusion coefficients used in the Parker TPE.}
{The Parker TPE overestimates the GCR intensity at Earth's orbit for low energies by $\sim 30\%$, and by $\sim 40\%$ over the poles. This stems from a small first-order anisotropy caused by particle fluxes over the poles. Particles gain easier access to the inner heliosphere by streaming in over the poles, where pitch-angle scattering is generally weaker, and the magnetic field is typically less wound. The focused TPE also yields nearly identical results for different pitch-angle dependencies of the diffusion coefficients.}
{The description of particle streaming and weak pitch-angle scattering as effective parallel diffusion in the Parker TPE makes it overly diffusive. This suggests that diffusion coefficients derived from fitting the Parker TPE to observations are likely underestimated. Furthermore, GCR spectral and anisotropy data alone cannot distinguish between scattering theories with similar mean free paths but different pitch-angle dependencies.}


\maketitle
\nolinenumbers


\section{Introduction}
\label{sec:Intro}

High-energy charged particles originating outside the heliosphere, called galactic cosmic rays (GCRs), diffuse and drift into the heliosphere against the outward-directed solar wind (SW). In general, cosmic rays (CRs) continuously lose energy as they are scattered by turbulence in the heliospheric magnetic field (HMF), a process known as solar modulation. The constant influx of GCRs into the heliosphere, together with long-term solar modulation, allows them to be treated as a quasi-stationary source. The many scatterings that GCRs experience over the large scales of the heliosphere, together with their entry from all directions, lead to a fairly isotropic and somewhat homogeneous distribution throughout the heliosphere. Since the seminal paper by \citet{Parker1965}, the so-called Parker transport equation \citep[TPE; see also][]{JokipiiParker1970, WebbGleeson1979, Moraal2013} has been widely used for more than 50 years to advance our understanding of CR transport in the isotropic limit \citep[see the reviews by][]{Quenby1984, Kota2013, EngelbrechtEA2022a}. Numerical modulation codes, which solve this equation to varying degrees of complexity, have yielded results that are sometimes extremely close to spacecraft observations \citep[see e.g.][]{Potgieter2013, RankinEA2022}.

It was soon realised, shortly after Parker's paper, that a more general TPE is needed for anisotropic distributions. For example, because solar energetic particles (SEPs) propagate preferentially away from the Sun and the HMF focuses them near the Sun, it is necessary to use the so-called focused transport equation \citep[FTE; see e.g.][]{Roelof1969, Skilling1971, Ruffolo1995}, which includes pitch-angle information. \edited{It} is widely accepted that the Parker TPE is inadequate for modelling SEPs, even when the focusing effect is included \citep[see e.g.][]{LitvinenkoNoble2013, EffenbergerLitvinenko2014, Malkov2017, Malkov2018, vandenBergEA2020}. This has led to the use of the so-called telegraph equation and other higher-order approximations of the FTE (see the aforementioned references and references therein). \edited{Ultimately, it became} clear that the FTE, with its pitch-angle information, is required to model the initial phase of an SEP event while the distribution is isotropising. The very nature of GCRs, however, suggests that their distribution is sufficiently isotropic for the Parker TPE to apply for typical modulation conditions. It has been widely and implicitly assumed that it does, with the only exceptions being anomalous \citep[][]{Isenberg1997, leRouxEA2007} and ultra-high-energy \citep{MaalalZhang2025} CRs, as well as CRs propagating very close to the Sun \citep{StraussEA2022, StraussEA2024}. However, the Parker TPE would be reliable for GCRs only if it produced intensities identical to those of the FTE under typical modulation conditions.

It follows from \edited{deriving the Parker TPE from the FTE, which is discussed in the next section,} that any process or situation that renders the distribution insufficiently isotropic makes the Parker TPE an approximation. This was recently illustrated by \citet{vandenBerg2023} through modelling of Jovian electrons in specific hypothetical scenarios. This is noteworthy, as Jovian electrons are a CR species similar to GCRs, in that Jupiter is a quasi-stationary source of electrons \citep{PyleSimpson1977, ChenetteEA1977, Moses1987} that are usually assumed to be sufficiently isotropic so that their transport is commonly described by the Parker TPE \cite[see e.g.][]{FerreiraEA2001, VogtEA2020, StraussEA2024, Engelbrecht2024} but can potentially deviate from isotropy due to jets \citep{SmithEA1976, McKibbenEA2007, DunzlaffEA2010} and more pronounced spatial gradients. \citet{vandenBerg2023} conclude that the Parker TPE is potentially invalid if deterministic pitch-angle changes (such as focusing or mirroring) are not negligible, the parallel \edited{mean free path (MFP)} is large, perpendicular diffusion is as important as parallel diffusion, there is interplay between pitch-angle scattering and perpendicular diffusion in the previous case, or the \edited{pitch-angle diffusion coefficient (PADC)} is asymmetric. This does not merely show that the Parker TPE may be invalid for Jovian electrons, but also identifies the physical mechanisms by which it can fail. Some of these limiting conditions are expected to be met in the broader heliosphere, particularly those pertaining to \edited{diffusion coefficients} (i.e. large parallel MFPs, sizeable \edited{ratios of the perpendicular to the parallel MFP}, and even asymmetric PADCs due to non-zero magnetic or cross helicities\footnote{Non-zero magnetic or cross helicities are observed in the inner heliosphere \citep[see][]{BreechEA2005}, although this is not considered in this work \cite[see][for how this leads to asymmetric PADCs]{Schlickeiser2002}.}).

To the best of our knowledge, the validity of the Parker TPE for studying GCR modulation has not been examined before. It is then the aim of this paper to highlight how the Parker and focused TPEs could differ, using typical, well-tested inputs for transport coefficients and heliospheric plasma parameters in a first-of-its-kind comparison of GCR modulation with both TPEs. The focus of this investigation is to compare the results of the Parker TPE with those of the FTE under \edited{identical} transport conditions, \edited{rather than} to reproduce specific GCR observations. The relative difference between solutions of the TPEs is considered and explained, starting by neglecting drifts and focusing only on diffusion. \edited{The theoretical background required for this work is summarised in Sect.~\ref{sec:Background}.} Sect.~\ref{sec:Model} describes the numerical model used in the study, with some of its mathematical details in Appendix~\ref{apndx:Maths} and its validation presented in Appendix~\ref{apndx:TestModel}. The results of the comparison are presented in Sect.~\ref{sec:Results}, with some mathematical details in Appendix~\ref{apndx:PADanisoDmm}. Sect.~\ref{sec:Discussion} discusses the results and their implications, while a summary of this investigation is given in Sect.~\ref{sec:Summary}.


\section{\edited{Background}}
\label{sec:Background}

\edited{A} hierarchy of TPEs exists: first, a Fokker-Planck-type equation describes the evolution of the distribution over the entire phase space and is the most complete model; then the \edited{FTE} for a gyrotropic distribution retains pitch-angle information; and finally, \edited{a so-called diffusion-advection equation, like} the Parker TPE, describes an isotropic distribution averaged over pitch angle. Assuming that particle gyration about the magnetic field is the fastest process, an average over gyrophase (the gyration angle about the magnetic field) can be performed to retain pitch-angle information. The pitch angle, $\alpha$, is the angle between the particle velocity vector and the magnetic field vector, and it describes the extent to which the particle moves along the field. Usually, the cosine of the pitch angle, $\mu = \cos \alpha$, is used, where a particle streaming along (only gyrating around) the magnetic field has $\alpha = 0^{\circ}$ or $\alpha = 180^{\circ} \Longrightarrow |\mu| = 1$ ($\alpha = 90^{\circ} \Longrightarrow \mu = 0$). If the momentum vectors of the particles are scattered sufficiently by turbulence, such that their pitch angles are uniformly distributed, an average over pitch cosine can be performed to neglect pitch-angle information.

The FTE is an evolution equation for the gyrotropic distribution function, $f (\vec{r}, p, \mu, t)$, of the particle number density per unit momentum volume at position $\vec{r}$ and time $t$, with particle momentum $p$ and pitch cosine measured in a non-relativistic flow frame. It is customary to denote quantities measured in the flow frame with a prime, but the prime will be omitted here for convenience, as transformations between frames will not be considered. Similarly, it is customary to denote the distribution function in a mixed frame with a superscript asterisk, but this will also be omitted. The equation reads \citep{Zhang2006, Zank2014, Wijsen2020, vandenBerg2023}
\begin{align}
\label{eq:FocusTPE}
& \frac{\partial f}{\partial t} + \vec{\nabla} \cdot \left( \vec{v}_{\rm gc} f \right) + \frac{1}{p^2} \frac{\partial}{\partial p} \left[ p^2 \frac{\dinf p}{\dinf t} f \right] + \frac{\partial}{\partial \mu} \left[ \frac{\dinf \mu}{\dinf t} f \right] \nonumber \\
& = \frac{\partial}{\partial \mu} \left[ D_{\mu\mu} \frac{\partial f}{\partial \mu} \right] + \vec{\nabla} \cdot \left( \tens{D}_{\perp} \cdot \vec{\nabla} f \right) ,
\end{align}
where $D_{\mu\mu}(\mu)$ is the \edited{PADC} and $\tens{D}_{\perp} = D_{\perp}(\mu) \, (\tens{I} - \uvec{b} \uvec{b})$ is the perpendicular diffusion tensor for axisymmetric turbulence, with $D_{\perp}(\mu)$ the pitch-angle-dependent cross-field diffusion coefficient (CFDC), $\uvec{b} = \vec{B} / B$ a unit vector in the direction of the background magnetic field, and $\tens{I}$ the $3 \times 3$ identity matrix. The terms in Eq.~\ref{eq:FocusTPE}, from left to right, describe temporal changes, the motion of the instantaneous guiding centre (GC), momentum and pitch-angle changes due to pseudo-forces arising from measuring momentum in the non-inertial frame of the plasma flow, and pitch-angle and spatial diffusion, respectively. The perpendicular diffusion being the same in both directions perpendicular to the magnetic field in the diffusion tensor here is only valid for axisymmetric turbulence. It is assumed that mixed spatial and pitch-angle diffusion coefficients (i.e. $D_{\mu\perp}$ and $D_{\perp\mu}$) are negligible. This is a simplification, but according to \citet{Schlickeiser2011}, these terms appear to be zero for incompressible, axisymmetric turbulence.

Furthermore, in Eq.~\ref{eq:FocusTPE},
\begin{subequations}
\begin{align}
\label{eq:drdt}
\vec{v}_{\rm gc} & = \mu v \, \uvec{b} + \vec{u} + \vec{v}_{\rm d}(\mu) , \\
\label{eq:dpdt}
\frac{\dinf p}{\dinf t}    & = p \left\lbrace \frac{1 - 3 \mu^2 }{2} \, \uvec{b} \uvec{b} : \vec{\nabla} \vec{u} - \frac{1 - \mu^2}{2} \, \vec{\nabla} \cdot \vec{u} \right. - \nonumber \\
 & \;\;\;\; \left. \frac{\mu}{v} \, \uvec{b} \cdot \left[ \frac{\partial \vec{u}}{\partial t} + \left( \vec{u} \cdot \vec{\nabla} \right) \vec{u} \right] \right\rbrace , \\
\label{eq:dmudt}
\frac{\dinf \mu}{\dinf t}  & = \left. \frac{1 - \mu^2}{2} \right\lbrace v \, \vec{\nabla} \cdot \uvec{b} + \mu \, \vec{\nabla} \cdot \vec{u} - 3 \mu \, \uvec{b} \uvec{b} : \vec{\nabla} \vec{u} \; - \nonumber \\
 & \;\;\;\; \left. \frac{2}{v} \, \uvec{b} \cdot \left[ \frac{\partial \vec{u}}{\partial t} + \left( \vec{u} \cdot \vec{\nabla} \right) \vec{u} \right] \right\rbrace ,
\end{align}
\end{subequations}
with $v$ the particle speed in the flow frame, $\vec{u}$ the flow velocity, $\vec{v}_{\rm d}(\mu)$ the gyrophase-averaged GC drift velocity in the flow frame, and $\vec{a} \vec{b} : \vec{c} \vec{d} = a_i b_j c_j d_i$ indicating a tensor contraction. The motion of the GC, described by the terms in Eq.~\ref{eq:drdt} from left to right, includes streaming of GCs along the magnetic field, advection by the plasma flow in which the magnetic field is embedded, and drifts of the GCs relative to the magnetic field. The drift velocity, $\vec{v}_{\rm d}(\mu)$, is mainly due to GC drift perpendicular to the background magnetic field, but there may also be some drift along the field \citep[see][for a discussion]{RossiOlbert1970, BurgerEA1985}. The momentum changes in Eq.~\ref{eq:dpdt} include the inverse Fermi and betatron effects \citep[first two terms; see][]{WebbGleeson1979} and the acceleration of the flow (last two terms). These pseudo-forces are also included in Eq.~\ref{eq:dmudt} for pitch-angle changes, with magnetic mirroring or focusing due to a gradient along the magnetic field contributing as well (the first term). Detailed interpretations and discussions of the various physical processes included in the FTE are provided in \citet{WebbGleeson1979}, \citet{Ruffolo1995}, \citet{leRouxEA2007}, \citet{Lampa2011}, and \citet{leRouxWebb2012}, with a summary to be found in \citet{vandenBerg2023}. Eqs~\ref{eq:dpdt} and \ref{eq:dmudt} could also include terms arising from an electric field along the magnetic field, which might not be removed by the transformation into the flow frame \citep{Zhang2006, leRouxWebb2012}. It is, however, usually assumed that the SW is infinitely conductive, so that there are no large-scale field-aligned electric fields. The dependence of all these terms on the pitch cosine implies that the pitch angle not only governs the efficiency of spatial transport, but also energy changes.

The Parker TPE can be derived from the FTE using a perturbation approach, assuming the distribution is nearly isotropic. Through an arduous mathematical process \citep[see][]{HasselmannWibberenz1970, Schlickeiser2002, LitvinenkoSchlickeiser2013, HeSchlickeiser2014, vandenBerg2023}, this derivation requires that pitch-angle scattering be efficient enough to keep the distribution nearly isotropic, implying that deterministic pitch-angle changes must be negligible (formally, by setting $\dinf \mu / \dinf t = 0$), that all other physical processes (i.e. perpendicular diffusion, advection, drifts, and momentum changes) must be slower than or unimportant compared to parallel diffusion, and that there is no interplay between pitch-angle scattering and perpendicular diffusion. This then yields the pitch-angle-independent quantities
\begin{subequations}
\label{eq:ParkerFocusRelation}
\begin{align}
\label{eq:ODI}
F_0 & = \frac{1}{2} \int_{-1}^1 \!\! f(\mu) \, \dinf \mu , \\
\label{eq:IsoDrift}
\vec{V}_{\rm d} & = \frac{1}{2} \int_{-1}^1 \!\! \vec{v}_{\rm d}(\mu) \, \dinf \mu , \\
\label{eq:AdiabaticEnergyChanges}
\frac{1}{2} \int_{-1}^1 \! \frac{\dinf p}{\dinf t} \, \dinf \mu & = - \frac{p}{3} \left( \vec{\nabla} \cdot \vec{u} \right) , \\
\label{eq:KappaParl}
\kappa_{\parallel} & = \frac{v^2}{8} \int_{-1}^1 \! \frac{(1 - \mu^2)^2}{D_{\mu \mu} (\mu)} \, \dinf \mu , \\
\label{eq:KappaPerp}
\kappa_{\perp} & = \frac{1}{2} \int_{-1}^1 \!\! D_{\perp} (\mu) \, \dinf \mu ,
\end{align}
\end{subequations}
to be used in the Parker TPE, with $\kappa_{\parallel}$ and $\kappa_{\perp}$ the isotropic \edited{DCs} parallel and perpendicular to the background magnetic field, respectively. If focusing or mirroring is included in this derivation \citep[by keeping only the first term in Eq.~\ref{eq:dmudt}; see][]{BeeckWibberenz1986, LitvinenkoSchlickeiser2013, HeSchlickeiser2014, WangQin2018, vandenBerg2023}, the definitions of the DCs are modified to include the so-called focusing length\footnote{The inclusion of an electric field parallel to the magnetic field would also cause an anisotropic distribution, with the electric field playing a role analogous to that of focusing.}, $L^{-1} = \vec{\nabla} \cdot \uvec{b}$, and additional advection terms appear due to the interplay between scattering and focusing (not shown here). For this, it is then required that $\lambda_{\parallel} \ll |L|$ to keep the distribution nearly isotropic, where $\lambda = 3 \kappa / v$ is the particle \edited{MFP}. Additional correction terms enter the TPE if the perturbation is not small \citep[see e.g.][]{HeSchlickeiser2014, WangQin2018}. However, these correction terms seem to vanish for a constant focusing length, are of fourth order in $\lambda_{\parallel} / |L|$ for a spatially varying focusing length and isotropic pitch-angle scattering in the weak focusing limit \citep{HeSchlickeiser2014}, and only seem to modify the parallel DC as long as pitch-angle scattering, perpendicular diffusion, and the focusing length are constant in space \citep{WangQin2018}.

The first requirement (that $\dinf \mu / \dinf t = 0$ or that $\lambda_{\parallel} \ll |L|$) holds for high-energy particles in the outer heliosphere, but not over the poles, where magnetic focusing may still be significant relative to large parallel MFPs. The second requirement (that parallel diffusion must be the fastest process) implies that either $\kappa_{\perp} \ll \kappa_{\parallel}$ or the distribution should not have large gradients perpendicular to the magnetic field (as the diffusive perpendicular transport is proportional to both the CFDC and the gradient perpendicular to the magnetic field). Although $\kappa_{\perp} / \kappa_{\parallel}$ is sensitive to the predictions of the employed scattering theories and the turbulence transport models used as input to those theories \citep[see e.g.][]{EngelbrechtEA2022a}, it is generally slightly larger over the poles than in the equatorial regions \citep[see e.g.][]{EngelbrechtBurger2013a, EngelbrechtBurger2013b, MolotoEA2018} and might approach a sizeable fraction of one in the outer heliosphere depending on how the effect of pick-up-ion-generated turbulence is incorporated into the parallel MFP \citep[compare][]{Engelbrecht2017, ZhaoEA2017}. At least galactic protons do not exhibit a large spatial variation between $0.4 - 2~{\rm GV}$ ($0.082 - 1.27~{\rm GeV}$), as their radial and latitudinal gradients are less than $\sim 4\%~{\rm au}^{-1}$ and $\sim 0.2\%~{\rm deg}^{-1}$, respectively \citep[see][and references therein]{RankinEA2022}. The last requirement (that there should be no interplay between pitch-angle scattering and perpendicular diffusion) implies that perpendicular diffusion must be pitch-angle independent. However, limited full-orbit simulations \citep[see][]{QinShalchi2014} and theoretical considerations \citep[see][]{StraussEA2016, Engelbrecht2019} indicate that this is not the case.

It can be concluded from these considerations that there may be subtleties in GCR transport beyond the assumption of isotropy, especially in the polar regions of the heliosphere. In contrast, the classical derivation of the Parker TPE, where a continuity equation is the starting point \citep[i.e.][]{Parker1965, JokipiiParker1970, WebbGleeson1979, Moraal2013}, seems to imply that the Parker TPE is valid for an isotropic distribution with position and time measured in a stationary observer’s frame, and momentum measured in a non-relativistic flow frame. The form of the diffusion tensor also seems not to be theoretically derived, but rather to be justified by the empirical observation that particles diffuse parallel and perpendicular to the magnetic field. It is also usually assumed that $\kappa_{\parallel} \gg \kappa_{\perp}$, based on some physical considerations \citep[see e.g.][]{Parker1965, Parker1967, JokipiiParker1970}, but also justified by theory, observations, and test-particle simulations \cite[see e.g.][]{Shalchi2020, EngelbrechtEA2022a, LangEA2024, ElsEA2024}. This classical derivation is less revealing about the restrictions on the Parker TPE and its application, as it does not necessarily reveal the underlying assumptions required to derive the Parker TPE from the FTE.

The Parker TPE is an evolution equation for the omnidirectional intensity, $F_0 (\vec{r}, p, t)$, of the particle number density per unit momentum volume, with momentum measured in the flow frame. It is given by \citep{JokipiiParker1970, WebbGleeson1979, Moraal2013}
\begin{align}
\label{eq:ParkerTPE}
& \frac{\partial F_0}{\partial t} + \vec{\nabla} \cdot \left[ (\vec{u} + \vec{V}_{\rm d}) \, F_0 \right] - \frac{1}{p^2} \frac{\partial}{\partial p} \left[ p^2 \frac{p}{3} (\vec{\nabla} \cdot \vec{u}) \, F_0 \right] \nonumber \\
& = \vec{\nabla} \cdot \left( \tens{\kappa} \cdot \vec{\nabla} F_0 \right) ,
\end{align}
where $\tens{\kappa} = \kappa_{\parallel} \, \uvec{b} \uvec{b} + \kappa_{\perp} (\tens{I} - \uvec{b} \uvec{b})$ is the isotropic diffusion tensor for axisymmetric turbulence. The terms in Eq.~\ref{eq:ParkerTPE}, from left to right, describe temporal changes, advection by plasma flow, drifts of the GC relative to the magnetic field, adiabatic energy changes, and spatial diffusion, respectively. Note that momentum diffusion can also be included \citep[see e.g.][]{Quenby1984, Schlickeiser2002}, that the drift velocity can be absorbed into the diffusion tensor if it is divergence-free \citep[see e.g.][]{JokipiiParker1970, MinnieEA2007}, and that the form of the TPE differs when working in a stationary observer's frame \citep[as is also the case for the FTE; see][]{WebbGleeson1979}. Although momentum diffusion may be important for certain applications, such as anomalous CRs \citep[see e.g.][]{StraussEA2010, StraussEA2011}, it is a second-order process in the flow frame and is usually neglected for GCR transport because diffusion, adiabatic energy changes, and drifts are more important for high-energy GCRs (hence the neglect of momentum diffusion in the FTE as well).

Some interesting insights can be gained by deriving the Parker TPE from the FTE, as discussed by \citet{vandenBerg2023}. Firstly, to lowest order, anisotropy arises from a local balance between pitch-angle scattering, focusing, and the momentum and spatial gradients of $F_0$, which act as source terms \citep[see also][]{HasselmannWibberenz1970}. This implies that the absence of focusing or field-aligned electric fields does not guarantee an isotropic CR distribution. Secondly, Eq.~\ref{eq:AdiabaticEnergyChanges} implies that particles undergo some energy changes that should average out for an isotropic distribution, but the adiabatic energy change in the Parker TPE would be only an approximation if any process causes an anisotropic distribution. Hence, the Parker TPE may under- or overestimate CR energy losses if the distribution is anisotropic. Although the exact forms of $\vec{v}_{\rm d}(\mu)$ and $\vec{V}_{\rm d}$ are not specified here, as this paper will focus on diffusion, a similar conclusion can be drawn from Eq.~\ref{eq:IsoDrift} about the drift terms. However, the effect of this on CR modulation is unclear \citep[][for example, conclude that the isotropic drift velocity could still be acceptable even if $95\%$ of the particles propagate in the same direction, with half of them having $|\mu| > 0.72$]{BurgerEA1985}. Thirdly, the parallel diffusion term of Eq.~\ref{eq:KappaParl} follows from the streaming term ($\mu v \, \uvec{b}$; hence its explicit inclusion in Eq.~\ref{eq:drdt}) and the pitch-angle diffusion term, indicating that parallel spatial diffusion is due to pitch-angle scattering disrupting the streaming of particles along the field \citep[see also][]{vandenBergEA2020}. Hence, it should be realised that a larger parallel DC, due to reduced pitch-angle scattering, actually implies more streaming, which is a non-diffusive process. The Parker TPE might therefore be only an approximation for CR transport in the outer heliosphere and over the poles, where the parallel MFP is generally expected to be very large \citep[see again][]{EngelbrechtBurger2013a, EngelbrechtBurger2013b, ZhaoEA2017, MolotoEA2018, EngelbrechtEA2022a}. Fourthly, including the interplay between pitch-angle scattering and other physical pitch-angle-dependent processes would introduce additional diffusion and advection terms in both momentum and configuration space \citep[see the discussion and summary in][]{vandenBerg2023}. One could argue that the definitions of the parallel and perpendicular MFPs should be revised to account for these additional spatial DCs. Moreover, pitch-angle scattering and perpendicular diffusion might be coupled and cannot be treated separately: this is not merely a case of parallel and perpendicular diffusion being dependent on one another \citep[see the theories of][]{MatthaeusEA2003, Qin2007, Shalchi2010}, but also of missing diffusion terms in the Parker TPE. Lastly, any process causing an anisotropic distribution would modify the definition of the MFPs, depending on the pitch-angle dependence of the PADC and CFDC (e.g. a distribution with more field-aligned particles would have a larger parallel MFP since $D_{\mu\mu} \propto 1 - \mu^2$ in general). Taken together, this and the previous realisation imply that any parallel (perpendicular) DC inferred from fitting the Parker TPE to observations is an `effective parallel (perpendicular) DC' that includes diffusion contributions from all physical processes, not just from pitch-angle scattering (cross-field diffusion).

Since the Parker TPE follows from the FTE, the limitations of the FTE should also be stated for transparency. Firstly, the inclusion of perpendicular transport (i.e. drifts and cross-field diffusion) in the FTE has been questioned, as summarised by \citet{vandenBergEA2020}, because all spatial transport across the magnetic field is absent from the original derivations \citep[see e.g.][]{Skilling1971, Isenberg1997, Zank2014}. However, \citet{leRouxEA2007} show that drift effects persist in the momentum changes, and alternative derivations \citep[see][]{Schlickeiser2002, Zhang2006, Wijsen2020} show that perpendicular transport can be retained if a transformation is made to the GC position before performing a gyrophase average \citep[see][for a possible explanation as to why this is]{vandenBerg2023}. It is ironic, then, that the inclusion of perpendicular transport in the Parker TPE has not been questioned to the same extent as its inclusion in the FTE. The FTE with perpendicular transport, and by implication the Parker TPE, is valid only when working with GCs, but the inclusion of perpendicular transport need not be questioned. However, caution should be exercised when the fields vary over distances smaller than the Larmor radius, as the instantaneous GC velocity is not relativistically consistent \citep[see the discussions by][]{IsenbergJokipii1979, BurgerEA1985}. Secondly, just as isotropy is the limiting factor for the Parker TPE, the FTE assumes a gyrotropic distribution. \citet{leRouxEA2014} show that a non-uniform magnetic field or flow, an accelerating flow, and an electric field could cause slow variations in the gyrophase (processes that are especially important at perpendicular shocks), which might lead to a non-gyrotropic distribution. Additionally, \citet{Wijsen2020} points out that neutral sheet drift could also cause a non-gyrotropic distribution. As with all models, some approximations are necessary to keep the problem tractable, and the gyrotropic assumption appears unavoidable to avoid solving a much more complicated gyrophase TPE. Lastly, the most technical assumption is that both the Parker and focused TPEs assume diffusive behaviour. This implies that the evolution of the distribution must be considered over a time interval that is short enough for the distribution not to change significantly, but not so short that the microphysical processes are quasi-deterministic \citep[in light of the gyrotropic assumption, this implies for the FTE that the pitch-angle decorrelation time should be longer than the cyclotron period; see e.g.][]{HasselmannWibberenz1970, Schlickeiser2002, Zank2014, vandenBerg2023}. The assumption of diffusive pitch-angle scattering in the FTE, however, is less restrictive than the assumption of diffusive spatial transport in the Parker TPE, since the FTE still includes particle streaming along the magnetic field \citep[see e.g.][for a review of non-diffusive transport and its applications]{EffenbergerEA2025}. Although the limitations of the FTE are real and should be heeded, they are on a different physical level from those of the Parker TPE, so that the FTE is the more complete TPE.


\section{Numerical method}
\label{sec:Model}

The flow velocity, HMF, and DCs must be specified to apply Eqs~\ref{eq:FocusTPE} or \ref{eq:ParkerTPE} to a particular scenario. Luckily, due to the quasi-stationarity of GCRs (i.e. intensities changing over long time intervals), the temporal evolution of the TPEs need not be considered. The DCs and heliospheric structure are chosen to represent those typically used successfully in GCR modulation studies during solar minimum periods. These are similar to those of \citet{Vos2011} and \citet{Raath2015}, with specific differences introduced to simplify the problem and reduce computational complexity. Many aspects of these models are also discussed in \citet{Potgieter2013} and utilised in numerous other works. Although these DCs are somewhat arbitrary, they have been shown to reproduce GCR observations at Earth without overcomplicating the matter. The parameters used in this study are the averages of those employed by \citet{Vos2011} and \citet{PotgieterEA2014} to replicate the proton observations made by PAMELA between 2006 and 2009 \citep[see also][]{Raath2015}.


\subsection{Heliospheric structure and diffusion coefficients}
\label{subsec:HMFSWDCs}

Following \citet{GleesonUrch1971}, the flux through the solar surface at a radial distance of $r_{\odot} = 0.005~{\rm au}$ is required to be zero. This implies a reflective inner boundary if the SW speed is zero at the solar surface and drifts are neglected. An outer boundary is placed at a radial distance of $R_{\rm o} = 90~{\rm au}$ to avoid complications from pitch-angle-dependent diffusive shock acceleration and a potentially non-gyrotropic distribution at the termination shock \citep[see e.g.][]{leRouxEA2007, leRouxEA2014}. Here, it is assumed that GCRs enter the modulation cavity isotropically (i.e. uniformly in latitude, longitude, and pitch cosine, with $\mu$ uniformly distributed between $[-1; 1]$). The spectrum of GCR protons from \citet{Vos2011} is specified here as a function of kinetic energy $K$ by
\begin{align}
\label{eq:VLIS}
j_{\rm o}(K) = \; & \#~{\rm m}^{-2}~{\rm s}^{-1}~{\rm sr}^{-1}~ {\rm MeV}^{-1} \\
& \left\lbrace \begin{array}{lll}
0.8 \, \eulernumb^{4.64 - 0.08 [\ln (K / K_{\rm r})]^2 - 2.91 \sqrt{K / K_{\rm r}}} & \mbox{if} & K < 1.4~{\rm GeV} \\
0.775 \, \eulernumb^{3.22 - 2.78 \ln (K / K_{\rm r}) - 1.5 (K_{\rm r} / K)} & \mbox{if} & K \ge 1.4~{\rm GeV}
 \end{array} \right. \nonumber ,
\end{align}
where the differential intensity is related to the omnidirectional intensity by $j_K = p^2 F_0/2$ \citep[see][for useful relations among different quantities]{Moraal2013} and $K_{\rm r} = 1~{\rm GeV}$. Note that this so-called very-local interstellar spectrum is usually specified at a radial distance of $\sim 120~{\rm au}$, implying that the results from the current model would yield too high intensities because modulation in the heliosheath \cite[see e.g.][]{StoneEA2013} and by the termination shock is absent here. The aim of the present study, however, is not to reproduce a specific set of observations but rather to compare the results obtained from the Parker and focused TPEs for a set of \edited{commonly used} transport coefficients.

\begin{figure*}[t!]
\includegraphics[trim=30mm 12mm 40mm 22mm, clip, scale=0.44]{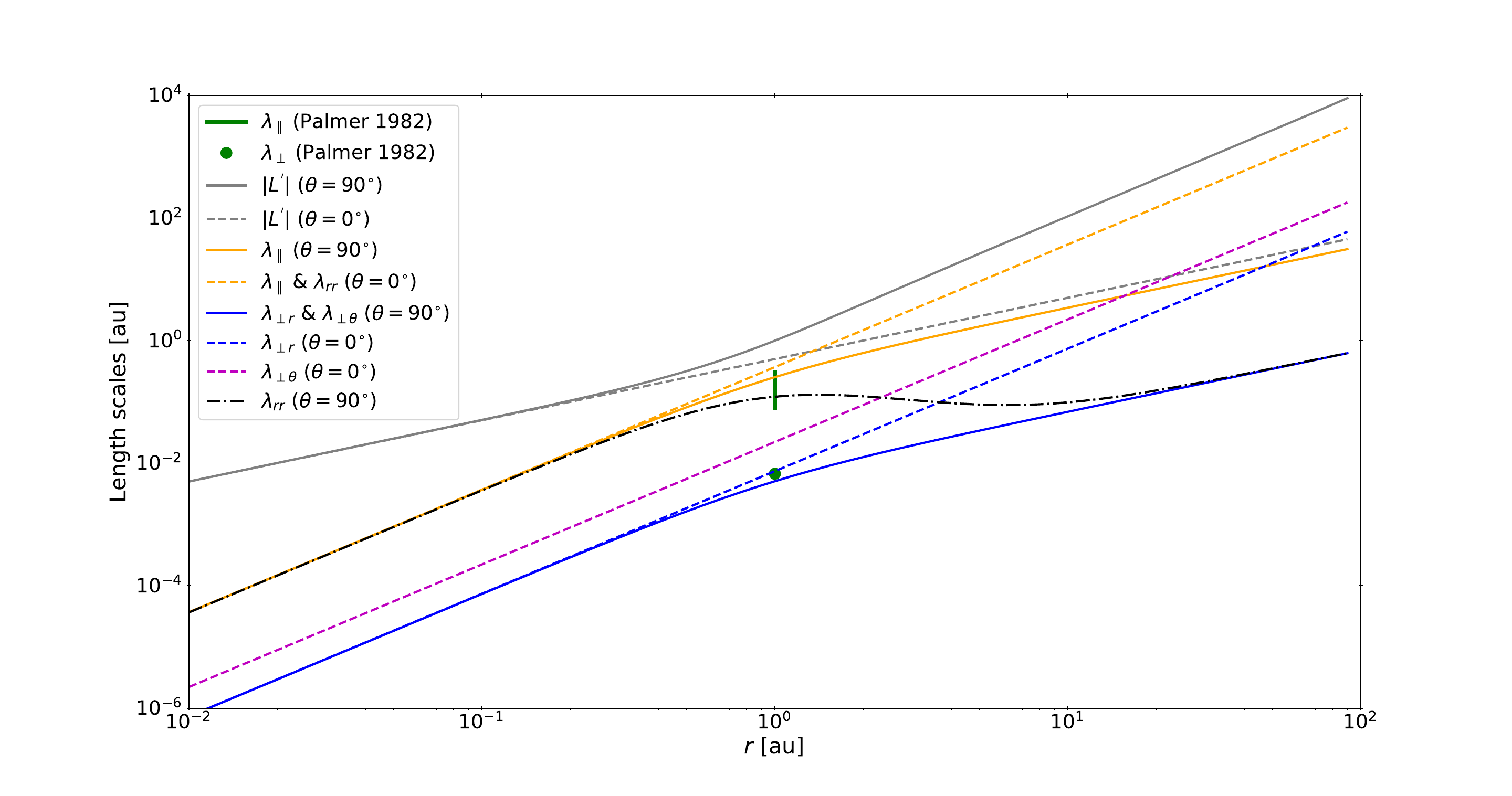}
\caption{\label{fig:LengthScales}Modified focusing length (Eq.~\ref{eq:FocusLength}; grey lines) and various MFPs for $1~{\rm GeV}$ protons as a function of radius in the equatorial plane (solid lines) and over the poles (dashed lines). The dashed orange line represents both $\lambda_{\parallel}$ and $\lambda_{rr}$ over the poles, but these two quantities differ in the equatorial plane (solid orange and dash-dotted black lines, respectively). Similarly, the solid blue line represents both $\lambda_{\perp r}$ and $\lambda_{\perp \theta}$ in the equatorial plane, but these two quantities differ over the poles (dashed blue and magenta lines, respectively). Lower-energy (higher-energy) particles will have smaller (larger) MFPs. For this particle energy, $L^{'} \approx L$ because the additional term in $L^{'}$ is of second order in $v_{\rm sw} / v$. \edited{The \citet{Palmer1982} consensus values at Earth are also indicated for reference.}}
\end{figure*}

Drifts are neglected (i.e. $\vec{v}_{\rm d}(\mu) = \vec{0}$ and $\vec{V}_{\rm d} = \vec{0}$) in this initial study for both simplicity and to systematically characterise the effects of different physical processes on any differences that may be observed in the results of the Parker and focused TPEs. The GC velocity being relativistically inconsistent and the potential departure from a gyrotropic distribution due to neutral sheet drift, as discussed in \edited{Sect.~\ref{sec:Background}}, are additional considerations for neglecting drifts here. The stationary solutions of Eqs~\ref{eq:FocusTPE} and \ref{eq:ParkerTPE} (i.e. with $\partial f / \partial t = 0$ and $\partial F_0 / \partial t = 0$, respectively) are calculated in a reference frame corotating with the Sun (the results in Sect.~\ref{sec:Results} are also presented in this reference frame). This specifies the flow velocity as \edited{\citep{Parker1958, Vos2011}}
\begin{equation}
\label{eq:FlowVelocity}
\vec{u} = v_{\rm sw}(r, \theta) \, (\uvec{r} - \tan \psi \, \uvec{\phi}) ,
\end{equation}
where $\tan \psi = \omega_{\odot} (r - r_{\odot}) \sin \theta / v_{\rm sw}$ specifies the spiral angle (the angle between the HMF line and the radial direction), with $\omega_{\odot} = 2 \pi / (25.4~{\rm days}) = 1.03 \times 10^{-2}~{\rm rad}~{\rm h}^{-1}$ the solar rotation rate, and $\uvec{r}$ and $\uvec{\phi}$ unit vectors in the radial and azimuthal direction, respectively. The \citet{Parker1958} HMF is valid in such a frame, given by
\begin{equation}
\label{eq:ParkerHMF}
\vec{B} = B_{\rm P} \left( \frac{r_{\oplus}}{r} \right)^2 (\uvec{r} - \tan \psi \, \uvec{\phi}) ,
\end{equation}
where $B_{\rm P} = B_{\oplus} / \sqrt{1 + \omega_{\odot}^2 (r_{\oplus} - r_{\odot})^2 / V_{\rm sw}^2}$ is normalised to the HMF magnitude at Earth \citep[$B_{\oplus} = 4.44~{\rm nT}$ at $r_{\oplus} = 1~{\rm au}$; see][for a summary of the Parker HMF]{OwensForsyth2013}. It is assumed that the SW is radially directed, accelerates near the Sun from $0~{\rm km}~{\rm s}^{-1}$ to a constant value of $V_{\rm sw} = 400~{\rm km}~{\rm s}^{-1}$ in the equatorial region \citep[in accordance with][]{SheeleyEA1997} and is a factor of two faster over the poles \citep[in accordance with][]{McComasEA2000}. This is parametrised by
\begin{equation}
\label{eq:SW}
v_{\rm sw} = V_{\rm sw} \, \frac{1 - \eulernumb^{\eta (1 - r/r_{\odot})}}{2} \left[ 3 + \tanh \left( \frac{|\theta - 90^{\circ}| - \alpha_{\odot} - \theta_{\rm s}}{\theta_{\rm t}} \right) \right] ,
\end{equation}
where $\eta = 1/15$ governs the radial transition rate, $\alpha_{\odot} = 13.5^{\circ}$ is the tilt angle between the magnetic and rotation axis of the Sun\footnote{Although no heliospheric current sheet is included in the model, $\alpha_{\odot}$ is usually between $\sim 5^{\circ}$ and $\sim 20^{\circ}$ during solar minimum conditions, such that $\alpha_{\odot} + \theta_{\rm s} \sim 30^{\circ}$ in most modulation studies \citep[see][]{Vos2011, Potgieter2013, PotgieterEA2014, Raath2015}.}, $\theta_{\rm s} = 15^{\circ}$, and $\theta_{\rm t} = 45^{\circ} / 2 \pi$ governs the latitudinal transition rate \citep[with the expressions of][having similar functional forms that differ slightly in numerical values]{Vos2011, Potgieter2013, PotgieterEA2014, Raath2015}.

The parallel DC is given as a function of rigidity $P$ by
\begin{equation}
\label{eq:KappaParlAdHoc}
\kappa_{\parallel} = \kappa_0 \, \frac{v B_0}{c B} \left( \frac{P}{P_{\rm r}} \right)^{\rho} \left[ \frac{(P / P_{\rm r})^{\gamma} + (P_{\rm k} / P_{\rm r})^{\gamma}}{1 + (P_{\rm k} / P_{\rm r})^{\gamma}} \right]^{\frac{\delta - \rho}{\gamma}} ,
\end{equation}
where $\kappa_0 = 1.3 \times 10^{23}~{\rm cm}^2~{\rm s}^{-1} = 2.1~{\rm au}^2~{\rm h}^{-1}$, $B_0 = 1~{\rm nT}$, $c$ is the speed of light in vacuum, $P_{\rm r} = 1~{\rm GV}$ ($= 0.43~{\rm GeV}$ for protons), $P_{\rm k} = 4~{\rm GV}$ ($= 3.17~{\rm GeV}$ for protons) is the rigidity where the break in the power law from low energies with index $\rho = 0.43$ to high energies with index $\delta = 1.95$ occurs, and $\gamma = 3$ governs the smoothness of the power law transition \citep{Vos2011, Potgieter2013, PotgieterEA2014, Raath2015}. The equivalent parallel MFP for a $1~{\rm GeV}$ proton is shown in Fig.~\ref{fig:LengthScales}\edited{, and its value at Earth is at the upper range of the \citet{Palmer1982} consensus values (vertical green line)}. The spatial dependence of the DC is determined by the magnetic field strength, causing the parallel MFP to increase $\propto r^2$ over the poles (dashed orange line) and for $r < 1~{\rm au}$, while increasing $\propto r$ in the equatorial plane for $r > 1~{\rm au}$ (solid orange line). \edited{This parallel DC has an energy dependence comparable to that of magnetostatic quasi-linear theory (QLT), which predicts $\lambda_{\parallel} \propto P^{1/3}$ at low energies, and $\lambda_{\parallel} \propto P^2$ at high energies. The spatial dependence of the QLT parallel MFP \citep[see e.g. the expression of][]{BurgerEA2008} is, however, governed by the assumed spatial behaviour of various turbulence quantities, and combining turbulence transport models with the QLT expression usually produces stronger radial variation than observed in Fig.~\ref{fig:LengthScales} \citep[see the summary in][]{EngelbrechtEA2022a}.}

The CFDC in the cone containing the HMF spiral is some fraction $\chi = 0.02$ of the parallel DC \citep[in accordance with][]{GiacaloneJokipii1999}, while the CFDC in the polar direction is enhanced by a factor of $\varsigma = 3$ over the poles \citep[in accordance with][]{BurgerEA2000, FerreiraEA2001} and is given the same latitudinal dependence as the SW for both simplicity and consistency \citep[in contrast to][which use slightly different latitudinal dependencies for the SW and perpendicular DC]{Vos2011, Potgieter2013, PotgieterEA2014, Raath2015}. This is parametrised by
\begin{subequations}
\label{eq:KappaPerpAdHoc}
\begin{align}
\kappa_{\perp r}      = \; & \chi \, \kappa_{\parallel} , \\
\kappa_{\perp \theta} = \; & \kappa_{\perp r} \left[ \frac{\varsigma + 1}{2} + \frac{\varsigma - 1}{2} \tanh \left( \frac{|\theta - 90^{\circ}| - \alpha_{\odot} - \theta_{\rm s}}{\theta_{\rm t}} \right) \right] ,
\end{align}
\end{subequations}
and relaxes the assumption of axisymmetric perpendicular diffusion\footnote{Theory has shown that non-axisymmetric perpendicular diffusion arising from non-axisymmetric turbulence leads to an increase of the perpendicular DC in one transverse direction and a corresponding decrease in the other transverse direction \cite[see e.g.][]{RuffoloEA2006, RuffoloEA2008, StraussEA2016}. This effect, however, is usually ignored in modulation studies using non-axisymmetric perpendicular DCs, as is done here as well.}. Fig.~\ref{fig:LengthScales} shows that the perpendicular MFPs of a $1~{\rm GeV}$ proton in the radial and polar directions are the same in the equatorial plane (solid blue line), but differ in magnitude over the poles (dashed blue and magenta lines). \edited{At Earth, the perpendicular MFP is slightly smaller than the \citet{Palmer1982} consensus value (green dot). Note that the perpendicular DCs share the same spatial and energy dependencies as the parallel DC. This is only somewhat reminiscent of the theoretical expression derived from the nonlinear guiding centre (NLGC) theory \citep[see e.g. the expression of][]{BurgerEA2008}, which has a weaker $\lambda_{\perp} \propto \lambda_{\parallel}^{1/3}$ dependence and would therefore exhibit a less pronounced spatial and energy dependence \citep[not taking into account the spatial dependence of the turbulence quantities; see the discussion of][]{EngelbrechtEA2022a}.}

Fig.~\ref{fig:LengthScales} also shows a modified focusing length,
\begin{equation}
\label{eq:FocusLength}
L^{'-1} = \vec{\nabla} \cdot \uvec{b} - \frac{2}{v^2} \, \uvec{b} \cdot \left[ \frac{\partial \vec{u}}{\partial t} + \left( \vec{u} \cdot \vec{\nabla} \right) \vec{u} \right] ,
\end{equation}
which has not been considered previously, but is more generally applicable in the heliosphere as only the $\mu \left( \vec{\nabla} \cdot \vec{u} - 3 \, \uvec{b} \uvec{b} : \vec{\nabla} \vec{u} \right)$ terms are neglected in Eq.~\ref{eq:dmudt} \edited{when deriving the Parker TPE from the FTE}. This form is considered for completeness and generally represents the focusing effect of the HMF for the particle energies considered here, as the additional term is of second order in $v_{\rm sw} / v$ (which is $\sim 0.67 - 1.33 \times 10^{-2}$ for the lowest energies considered). The diffusion condition ($\lambda_{\parallel} < |L^{'}|$) is met in the equatorial region (compare the solid orange and grey lines), but not over the poles (compare the dashed orange and grey lines). The ratio of $\lambda_{\parallel}$ to $|L^{'}|$ grows $\propto r$ from $\sim 7 \times 10^{-3}$ to $\sim 0.3$ at $r \approx 0.7~{\rm au}$ in the equatorial plane and then decreases $\propto r^{-1}$ to $\sim 3 \times 10^{-3}$ at the outer boundary. Over the poles, $\lambda_{\parallel} / |L^{'}|$ grows linearly with radial distance up to $\sim 70$ at the outer boundary. Although the focusing length is large in the outer heliosphere, the violation of the diffusion condition over the poles is mostly due to very large parallel MFPs (i.e. a combination of $|L^{'}| \propto r$ and $\lambda_{\parallel} \propto r^2$).

The FTE requires pitch-angle-dependent DCs, which, for this study, are normalised to the parallel and perpendicular DCs using Eqs~\ref{eq:KappaParl} and \ref{eq:KappaPerp}, respectively. This enables a direct, consistent comparison of results from the Parker and focused TPEs. Both isotropic and anisotropic pitch-angle scattering (not to be confused with isotropic or anisotropic distributions) will be investigated, given by
\begin{subequations}
\begin{align}
\label{eq:IsoDmm}
D_{\mu\mu}^{\rm iso}(\mu)   & = D_{\rm I} \, (1 - \mu^2) , \\
\label{eq:AnisoDmm}
D_{\mu\mu}^{\rm aniso}(\mu) & = D_{\rm A} \, (1 - \mu^2) \left( \frac{|\mu|}{1 + |\mu|} + \epsilon \right) ,
\end{align}
\end{subequations}
respectively, where $D_{\rm I} = v^2 / 6 \kappa_{\parallel}$,
\begin{align}
D_{\rm A} = \; & \frac{v^2}{4 \kappa_{\parallel} (1 + \epsilon)^4} \left[ \frac{1}{6} + 2 \epsilon + \frac{5}{2} \epsilon^2 + \frac{2}{3} \epsilon^3 \right. + \nonumber \\
 & \left. (1 + 2 \epsilon) \ln \left( \frac{1 + 2 \epsilon}{\epsilon} \right) \right]
\end{align}
for $\epsilon \neq 0$, and $\epsilon = 0.045$ parametrises dynamical effects. The anisotropic PADC of \citet{AguedaEA2008} is used here because its derivative is well behaved around $\mu = 0$ \citep[see][for the normalisation]{AguedaVainio2013}. Both $D_{\rm I}$ and $D_{\rm A}$ are calculated from Eq.~\ref{eq:KappaParlAdHoc} using Eq.~\ref{eq:KappaParl} to ensure parallel diffusion is consistent between the Parker and focused TPEs. \edited{Eq.~\ref{eq:AnisoDmm} is consistent with dynamical QLT and is often used in the FTE, whereas Eq.~\ref{eq:IsoDmm} would be expected only in strong turbulence or in magnetostatic QLT for particles resonating with a turbulence spectrum inversely proportional to the parallel wavenumber (i.e. $k_{\parallel}^{-1}$).} The CFDC is compactly written as
\begin{equation}
\label{eq:Dperp}
D_{\perp}(\mu) = h(\mu) \, \kappa_{\perp} ,
\end{equation}
where $h(\mu)$ governs the pitch-angle dependence and normalisation. Pitch-angle-independent perpendicular diffusion with $h(\mu) = 1$ will be used when Eq.~\ref{eq:IsoDmm} is employed, while $h(\mu) = 3 \mu^2$, representing nonlinear theories \citep[see][]{QinShalchi2014, Engelbrecht2019}, will be used when Eq.~\ref{eq:AnisoDmm} is employed. To ensure perpendicular diffusion is consistent between the Parker and focused TPEs, the numerical factor in $h(\mu)$ is calculated from Eq.~\ref{eq:KappaPerp} for Eq.~\ref{eq:KappaPerpAdHoc}. The choice of isotropic pitch-angle scattering with pitch-angle-independent perpendicular diffusion is the simplest option for the FTE, with no interplay between pitch-angle scattering and cross-field diffusion. The choice of anisotropic pitch-angle scattering with pitch-angle-dependent perpendicular diffusion represents a more realistic situation \citep[i.e. dynamical \edited{QLT} for the PADC and \edited{NLGC} theory for the CFDC; see e.g.][]{Schlickeiser2002, Engelbrecht2019}.


\subsection{Stochastic differential equations}
\label{subsec:SDEs}

As explained by \citet{StraussEffenberger2017}, the stationary solution for GCRs is easiest to calculate using time-backwards stochastic differential equations (SDEs). The general backwards Kolmogorov equation is of the form
\begin{equation}
\label{eq:BackwardKolmogorov}
- \, \frac{\partial \tilde{P}}{\partial \tau} = \vec{a} \cdot \vec{\nabla}_{\vec{q}} \tilde{P} + \frac{1}{2} \, \tens{C} : \vec{\nabla}_{\vec{q}} \vec{\nabla}_{\vec{q}} \tilde{P} - \L \tilde{P} ,
\end{equation}
where $\tau = T - t$ is a backwards time coordinate starting at some final time $t = T$, $\vec{q}$ is an $n$-dimensional vector with generalised  (i.e. phase-space) coordinates, $\vec{\nabla}_{\vec{q}}$ is the gradient operator in $n$-dimensional phase space, $\tilde{P} (\vec{q}, \tau \, | \, \vec{q}_{\rm f}, 0)$ is the conditional probability from a final state $\vec{q}_{\rm f}$ at time $t = T$ ($\tau = 0$), $\vec{a} (\vec{q}, \tau)$ is an $n$-dimensional vector describing drifts in phase space (not to be confused with the physical particle drift process), $\tens{C} (\vec{q}, \tau)$ is an $n \times n$ matrix describing diffusion in phase space (also not to be confused with the physical particle diffusion process), and $\L (\vec{q}, \tau)$ is the linear coefficient. The It\^{o}-type SDE equivalent to this equation is
\begin{equation}
\label{eq:ItoSDE}
\dinf \vec{Q}(\tau) = \vec{a} (\vec{Q}, \tau) \, \dinf\tau + \tens{B} (\vec{Q}, \tau) \cdot \dinf \vec{W}(\tau) ,
\end{equation}
where $\vec{Q}(\tau) = \vec{q}(\tau)$ are random variables corresponding to $\vec{q}$ (this relation implies that the variable $\vec{q}$, called a deterministic variable, is changing randomly, and it is this stochastic nature that is modelled by $\vec{Q}$), $\tens{C} = \tens{B} \cdot \tens{B}^{\rm T}$, and $\dinf \vec{W}(\tau)$ are $n$ independent Wiener processes (i.e. time-stationary stochastic L\'{e}vy processes where the time increments have a normal distribution with a mean of zero and a variance of $\dinf \tau$). In the modelling community, the temporal evolution of $\vec{Q}$ is called the `trajectory of a pseudo-particle'. However, `the temporal evolution of a phase-space density element' would be a more apt description, as the TPE describes the temporal evolution of a phase-space density element of an ensemble. A pseudo-particle trajectory represents only one possible realisation of how a pseudo-particle might evolve. An ensemble of solutions (\edited{each} independent of \edited{the others} if the stochastic drift and diffusion coefficients are independent of the distribution function) should therefore be used to \edited{statistically} compute quantities of interest, such as GCR differential intensities. The time-backwards approach explained here is mathematically equivalent to a time-forwards approach in the stationary limit. The time-backwards approach, however, is computationally more efficient because it ensures that all pseudo-particles contribute to the answer at the observer \citep{Gardiner1994, KloedenPlaten1995, KoppEA2012, StraussEffenberger2017}.

The trajectory of the pseudo-particle is calculated iteratively using the Euler-Maruyama scheme, i.e.
\begin{align}
\label{eq:EulerMaruyama}
\vec{Q} (\tau_k + \Delta \tau_k) \approx \; & \vec{Q} (\tau_k) + \vec{a} (\vec{Q}(\tau_k), \tau_k) \, \Delta \tau_k \; + \nonumber \\
 & \tens{B} (\vec{Q}(\tau_k), \tau_k) \cdot \vec{\Lambda} (\tau_k) \sqrt{\Delta \tau_k} ,
\end{align}
where $\Delta \tau_k$ is the time step of the $k^{\rm th}$ iteration, $\tau_k = \sum_{l=1}^k \Delta \tau_l$, and $\vec{\Lambda} (\tau_k)$ are $n$ independent normally-distributed pseudo-random numbers with zero mean and unit variance \citep{KloedenPlaten1995, StraussEffenberger2017}. Uniformly distributed pseudo-random numbers on $[0;1]$ are generated using the permuted congruential generator of \citet{ONeill2014}\footnote{\url{https://www.pcg-random.org/}} and transformed into normally distributed random numbers for the Wiener process using the Box-Muller transformation \citep[see][]{PressEA1992}. Following \citet{StraussEffenberger2017}, the linear term in Eq.~\ref{eq:BackwardKolmogorov} is handled by assigning a weight, $w$, to the pseudo-particle, which implies that
\begin{align}
\frac{\partial \tilde{P}}{\partial \tau} & \propto \L \tilde{P} \nonumber \\
\Longrightarrow \;\;\;\;\;\;\;\; \dinf [\ln w(\tau)] & = \L (\vec{q}, \tau) \, \dinf \tau \nonumber \\
\Longrightarrow \,\;\;\;\; w(\tau_k + \Delta \tau_k) & \approx w(\tau_k) \, \eulernumb^{{\scriptsize \L} (\vec{Q}(\tau_k), \tau_k) \, \Delta \tau_k} .
\end{align}
Note that the weight will not change if $\L = 0$. Eq.~\ref{eq:BackwardKolmogorov} can include sources, which can also be incorporated into the SDE formulation \citep[see e.g.][]{KoppEA2012, StraussEffenberger2017}, but this is not \edited{necessary} here. Although SDEs are unconditionally stable \citep[with bounded stochastic drift and diffusion coefficients, so that an initial error would not grow in the numerical scheme; see][]{KloedenPlaten1995, StraussEffenberger2017}, the time step should still be chosen sufficiently small to resolve physical structures or processes. To avoid time steps that are too large or too small, which would yield inaccurate results or long execution times, respectively, an adaptive time step can be used that changes at each iteration to ensure sufficient sampling of the length scale $l_i$ of $q_i$. This is written generally as \citep[see][]{StraussEffenberger2017}
\begin{equation}
\label{eq:VarTimeStep}
\Delta \tau_k (\vec{Q}(\tau_k), \tau_k) = \min \left\lbrace \frac{l_i (\vec{Q}(\tau_k), \tau_k)}{|a_i (\vec{Q}(\tau_k), \tau_k)|} ; \frac{l_i^2 (\vec{Q}(\tau_k), \tau_k)}{\sum_j \tens{B}_{ij}^2 (\vec{Q}(\tau_k), \tau_k)} \right\rbrace .
\end{equation}
This numerical scheme is implemented in C, and the pseudo-particles are traced independently using Open MPI \citep{GabrielEA2004}.

Most aspects of the SDE solution for the Parker TPE are covered in \citet{PeiEA2010}, \citet{KoppEA2012}, and \citet{StraussEffenberger2017}, while aspects of solving the FTE with SDEs are discussed by \citet{Wijsen2020} and \citet{vandenBerg2023}. For the current application, the phase-space coordinates with spherical spatial coordinates are $\vec{q} = [r, \theta, \phi, p, \mu]^{\rm T}$, and the random variables indicated by $\vec{Q} = [R, \Theta, \Phi, \Pi, M]^{\rm T}$ below. The exact forms of the SDEs in spherical spatial coordinates are given in Sects~\ref{subsubsec:ParkerSDEs} and \ref{subsubsec:FocusedSDEs}, with further mathematical details in Appendix~\ref{apndx:Maths}. The following discussion is generalised to the FTE, including $\mu$. Therefore, parts related to the pitch cosine do not apply to the Parker TPE. The reflective inner boundary condition implies that
\begin{equation}
\begin{array}{lclcl}
R \leftarrow 2 r_{\odot} - R & {\rm and} & M \leftarrow |M| & {\rm if} & R < r_{\odot} ,
\end{array}
\end{equation}
although this condition is only necessary numerically and in the Parker TPE, as magnetic mirroring should keep the pseudo-particles away from this boundary in the FTE. Coordinate re-normalisation is applied to restrict $\theta$, $\phi$, and $\mu$ to $[0;\pi]$, $[0;2\pi)$, and $[-1;1]$, respectively. This is done by
\begin{equation}
\left\lbrace \begin{array}{lclcl}
\Theta \leftarrow |\Theta|      & {\rm and} & \Phi \leftarrow \Phi + \pi & {\rm if} & \Theta < 0 \\
\Theta \leftarrow 2\pi - \Theta & {\rm and} & \Phi \leftarrow \Phi - \pi & {\rm if} & \Theta > \pi
\end{array} \right. ,
\end{equation}
followed by
\begin{equation}
\left\lbrace \begin{array}{lcl}
\Phi \leftarrow \Phi + 2\pi & {\rm if} & \Phi < 0 \\
\Phi \leftarrow \Phi - 2\pi & {\rm if} & \Phi \ge 2 \pi
\end{array} \right. ,
\end{equation}
together with
\begin{equation}
\begin{array}{lcl}
M \leftarrow {\rm sign}(M) \, 2 - M & {\rm if} & |M| > 1 ,
\end{array}
\end{equation}
where the re-normalisation of $\Theta$ and $M$ follows a reflecting boundary condition, while that of $\Phi$ follows a periodic boundary condition \citep{KoppEA2012, StraussEffenberger2017, vandenBerg2023}. The adaptive time step is
\begin{align}
\label{eq:AdaptiveTimeStep}
& \Delta \tau = \\
& \min \left\lbrace \frac{l_r}{|a_r|} ; \frac{l_{\theta}}{|a_{\theta}|} ; \frac{l_{\phi}}{|a_{\phi}|} ; \frac{l_p}{|a_p|} ; \frac{l_{\mu}}{|a_{\mu}|} ; \frac{l_r^2}{\tens{B}_{rr}^2 + \tens{B}_{r\phi}^2} ; \frac{l_{\theta}^2}{\tens{B}_{\theta \theta}^2} ; \frac{l_{\phi}^2}{\tens{B}_{\phi \phi}^2} ; \frac{l_{\mu}^2}{\tens{B}_{\mu \mu}^2} \right\rbrace , \nonumber
\end{align}
where the components of $\vec{a}$ and elements of $\tens{B}$ are discussed below for the Parker and focused TPEs separately. Following \citet{vandenBerg2023}, the length scales are mainly related to the MFPs and the focusing length by a small fraction $\ell = 2^{-6}$, i.e.
\begin{subequations}
\begin{align}
l_r        & = \ell \min \lbrace L_r ; \lambda_{rr} \rbrace \nonumber \\
           & = \ell \min \left\lbrace L \cos \psi ; 3 \, \frac{\kappa_{\parallel} \cos^2 \psi + \kappa_{\perp r} \sin^2 \psi}{v} \right\rbrace , \\
l_{\theta} & = \frac{\ell \lambda_{\theta \theta}}{r} = \frac{3 \ell \kappa_{\perp \theta}}{v r} , \\
l_{\phi}   & = \frac{\ell}{r \sin \theta} \, \min \lbrace L_{\phi} ; \lambda_{\phi \phi} \rbrace \nonumber \\
           & = \frac{\ell}{r \sin \theta} \, \min \left\lbrace L \sin \psi ; 3 \, \frac{\kappa_{\parallel} \sin^2 \psi + \kappa_{\perp r} \cos^2 \psi}{v} \right\rbrace , \\
l_p        & = \ell p , \\
l_{\mu}    & = \ell \sqrt{1 - \mu^2} .
\end{align}
\end{subequations}

In the time-backwards approach, all the pseudo-particles start at the observation point, say $(R(0), \Theta(0), \Phi(0)) = (r_{\rm f}, \theta_{\rm f}, \phi_{\rm f}) = (r_{\oplus}, \pi/2, 0)$ as an example for Earth, with some kinetic energy $K_{\rm f}$ (corresponding to momentum $\Pi(0) = p_{\rm f}$), pitch cosine $M(0) = \mu_{\rm f}$ (uniformly distributed on $[-1; 1]$), and weight $w(0) = w_{\rm f} = 1$. They are traced backwards in time until they reach the outer boundary at time $\tau_{\rm b}$, with weights $w(\tau_{\rm b}) = w_{\rm b}$ and energies $K_{\rm b} \ge K_{\rm f}$ (corresponding to momentum $\Pi(\tau_{\rm b}) = p_{\rm b}$; assuming there is no acceleration in the heliosphere, so the particles lose energy as they propagate from the outer boundary to the observer in the physical time-forward approach). The spectrum at the observation point at an energy $K_{\rm f}$ is then the weighted average over all pseudo-particles of Eq.~\ref{eq:VLIS} evaluated with their $K_{\rm b}$ \citep[see][]{StraussEffenberger2017}. Explicitly,
\begin{equation}
\label{eq:ODIfromSDE}
\hat{\jmath}_K(r_{\rm f}, \theta_{\rm f}, K_{\rm f}) = \frac{p_{\rm f}^2}{W} \sum_{l} \frac{w_{{\rm b}l} \, j_{\rm o} (K_{{\rm b}l})}{p_{{\rm b}l}^2} ,
\end{equation}
where the summation is over all pseudo-particles injected with $K_{\rm f}$ regardless of their pitch angles (this is consistent with the assumption of pitch-angle isotropy at the outer boundary), $W = \sum_{l} w_{{\rm b}l}$ is the sum of all the weights, and the ratio of momenta follows from the conversion between differential and omnidirectional intensities (note that the hat indicates that this is a discrete estimate of the continuous function it represents). The weighted standard deviation of the (assumed) independent and identically distributed random variables that comprise this weighted mean is used as an indication of its uncertainty, i.e.
\begin{equation}
\label{eq:Uncertainty}
\delta \hat{\jmath}_K \approx \frac{1}{W} \sqrt{\left[ \frac{p_{\rm f}^4}{W} \sum_{l} \frac{w_{{\rm b}l} \, j_{\rm o}^2 (K_{{\rm b}l})}{p_{{\rm b}l}^4} - \hat{\jmath}_K^2 \right] \sum_{l} w_{{\rm b}l}^2} .
\end{equation}
Note that even though the pseudo-particles are isotropically injected at the observer, each pseudo-particle will contribute a different weight, potentially resulting in an anisotropic distribution at the observer. Also note from Sect.~\ref{subsec:HMFSWDCs} and the rest of this section that none of the coefficients or calculations depends on $\phi$. Hence, if the solution is constructed only along a radial spoke as a function of $(r, \theta, K, \mu)$, the evolution of $\Phi$ can be neglected, and the number of SDEs can be reduced by one. The number of radial, latitudinal, kinetic energy, and pitch cosine bins is $90$, $22$, $30$, and $20$, respectively. When solving the Parker TPE, $150$ pseudo-particles are used per radial, latitudinal, and energy bin, whereas $300$ pseudo-particles are used per bin when solving the FTE. More particles should be used to reduce statistical noise and resolve the pitch-angle distribution (PAD). Initial estimates indicate that at least a factor of five more pseudo-particles are needed to calculate the anisotropy, and even more would be needed to resolve the PAD. However, it is computationally expensive to calculate the solution across the entire heliosphere using the FTE, where fast pitch-angle scattering must be resolved. To remedy this, a three-point average is applied to latitude, radius, and kinetic energy (in that order) to smooth the results. The testing and verification of this newly developed SDE model are discussed in Appendix~\ref{apndx:TestModel}.


\subsubsection{Stochastic Parker model}
\label{subsubsec:ParkerSDEs}

Eq.~\ref{eq:ParkerTPE} can be written compactly in the form of Eq.~\ref{eq:BackwardKolmogorov}, i.e.
\begin{equation}
\label{eq:BackwardsParker}
- \, \frac{\partial F_0}{\partial \tau} = \left( \vec{\nabla} \cdot \tens{\kappa} - \vec{u} \right) \cdot \vec{\nabla} F_0 + \frac{p \left( \vec{\nabla} \cdot \vec{u} \right)}{3} \, \frac{\partial F_0}{\partial p} + \frac{2 \, \tens{\kappa} : \vec{\nabla} \vec{\nabla} F_0}{2} ,
\end{equation}
with $\vec{V}_{\rm d}$ neglected since it will not be used here, but this becomes quite lengthy in spherical coordinates \citep[see][]{KoppEA2012}. For the Parker HMF, the isotropic diffusion tensor,
\begin{align}
\label{eq:KappaTensor}
\tens{\kappa} & = \left[ \begin{array}{ccc}
\tens{\kappa}_{rr}       & \tens{\kappa}_{r\theta}      & \tens{\kappa}_{r\phi} \\
\tens{\kappa}_{\theta r} & \tens{\kappa}_{\theta\theta} & \tens{\kappa}_{\theta\phi} \\
\tens{\kappa}_{\phi r}   & \tens{\kappa}_{\phi\theta}   & \tens{\kappa}_{\phi\phi}
\end{array} \right] \\
 & = \left[ \begin{array}{ccc}
\kappa_{\parallel} \cos^2 \psi + \kappa_{\perp r} \sin^2 \psi & 0                     & (\kappa_{\perp r} - \kappa_{\parallel}) \sin \psi \cos \psi \\
0                                                             & \kappa_{\perp \theta} & 0 \\
(\kappa_{\perp r} - \kappa_{\parallel}) \sin \psi \cos \psi   & 0                     & \kappa_{\parallel} \sin^2 \psi + \kappa_{\perp r} \cos^2 \psi
\end{array} \right] , \nonumber
\end{align}
is symmetric in spherical coordinates \citep{Vos2011, Potgieter2013, Raath2015}. With the heliosphere setup described in Sect.~\ref{subsec:HMFSWDCs}, Eq.~\ref{eq:BackwardsParker} in spherical coordinates simplifies to
\begin{align}
\label{eq:BackwardsParkerSpherical}
- \, \frac{\partial F_0}{\partial \tau} = \; & \frac{p}{3} \left( \frac{2 u_r}{r} + \frac{\partial u_r}{\partial r} \right) \frac{\partial F_0}{\partial p} + \frac{1}{r^2} \left( \tens{\kappa}_{\theta \theta} \cot \theta + \frac{\partial \tens{\kappa}_{\theta \theta}}{\partial \theta} \right) \frac{\partial F_0}{\partial \theta} \; + \nonumber \\
 & \left[ \frac{1}{r} \left( 2 \tens{\kappa}_{rr} + \csc \theta \, \frac{\partial \tens{\kappa}_{\phi r}}{\partial \phi} \right) + \frac{\partial \tens{\kappa}_{rr}}{\partial r} - u_r \right] \frac{\partial F_0}{\partial r} \; + \nonumber \\
 & \frac{\csc \theta}{r^2} \left[ \tens{\kappa}_{r\phi} + \csc \theta \, \frac{\partial \tens{\kappa}_{\phi \phi}}{\partial \phi} + r \left( \frac{\partial \tens{\kappa}_{r\phi}}{\partial r} - u_{\phi} \right) \right] \frac{\partial F_0}{\partial \phi} \; + \nonumber \\
 & \frac{1}{2} \left[ 2 \tens{\kappa}_{rr} \, \frac{\partial^2 F_0}{\partial r^2} + \frac{2 \tens{\kappa}_{\theta \theta}}{r^2} \, \frac{\partial^2 F_0}{\partial \theta^2} + \frac{2 \tens{\kappa}_{\phi \phi}}{r^2 \sin^2 \theta} \, \frac{\partial^2 F_0}{\partial \phi^2} \right. + \nonumber \\
 & \left. \frac{2 (\tens{\kappa}_{r\phi} + \tens{\kappa}_{\phi r})}{r \sin \theta} \, \frac{\partial^2 F_0}{\partial r \, \partial \phi} \right] .
\end{align}

The SDEs can be easily identified from the latter expression, i.e.
\begin{subequations}
\label{eq:StochasticDriftParker}
\begin{align}
a_r        & = \frac{1}{r} \left( 2 \tens{\kappa}_{rr} + \csc \theta \, \frac{\partial \tens{\kappa}_{\phi r}}{\partial \phi} \right) + \frac{\partial \tens{\kappa}_{rr}}{\partial r} - v_{\rm sw} , \\
a_{\theta} & = \frac{1}{r^2} \left( \tens{\kappa}_{\theta \theta} \cot \theta + \frac{\partial \tens{\kappa}_{\theta \theta}}{\partial \theta} \right) , \\
\label{eq:AphiParker}
a_{\phi}   & = \frac{\csc \theta}{r^2} \left[ \tens{\kappa}_{r\phi} + \csc \theta \, \frac{\partial \tens{\kappa}_{\phi \phi}}{\partial \phi} + r \left( v_{\rm sw} \tan \psi + \frac{\partial \tens{\kappa}_{r\phi}}{\partial r} \right) \right] , \\
a_p        & = \frac{p}{3} \left( \frac{2 v_{\rm sw}}{r} + \frac{\partial v_{\rm sw}}{\partial r} \right) ,
\end{align}
\end{subequations}
where Eq.~\ref{eq:FlowVelocity} was substituted, $2 v_{\rm sw} / r + \partial v_{\rm sw} / \partial r$ is the total divergence of an accelerating SW (Eq.~\ref{eq:divu}), and
\begin{align}
\tens{C} & = \left[ \begin{array}{ccc}
\tens{C}_{rr}     & 0                        & \tens{C}_{r \phi} \\
0                 & \tens{C}_{\theta \theta} & 0 \\
\tens{C}_{\phi r} & 0                        & \tens{C}_{\phi \phi} \\
\end{array} \right] \nonumber \\
 & = 2 \left[ \begin{array}{ccc}
\tens{\kappa}_{rr}                     & 0                                   & \tens{\kappa}_{r \phi} / r \sin \theta \\
0                                      & \tens{\kappa}_{\theta \theta} / r^2 & 0 \\
\tens{\kappa}_{r \phi} / r \sin \theta & 0                                   & \tens{\kappa}_{\phi \phi} / r^2 \sin^2 \theta \\
\end{array} \right]
\end{align}
is positive definite as long as $r \neq 0$ and $\theta \neq 0^{\circ}$ or $\theta \neq 180^{\circ}$. Note that only the spatial components of the diffusion tensors are given here because momentum diffusion is absent. Since the SW is specified only from the solar surface, $r = 0$ will not be reached if a boundary condition is imposed at $r_{\odot}$. By choosing $\tens{B}$ as upper triangular, it follows that \citep[][]{PeiEA2010, KoppEA2012}
\begin{align}
\tens{B} & = \left[ \begin{array}{ccc}
\tens{B}_{rr} & \tens{B}_{r \theta}      & \tens{B}_{r \phi} \\
0             & \tens{B}_{\theta \theta} & \tens{B}_{\theta \phi} \\
0             & 0                        & \tens{B}_{\phi \phi} \\
\end{array} \right] \nonumber \\
 & = \left[ \begin{array}{ccc}
\sqrt{\tens{C}_{rr} - \tens{C}_{r \phi}^2 / \tens{C}_{\phi \phi}} & 0                                         & \tens{C}_{r \phi} / \sqrt{\tens{C}_{\phi \phi}} \\
0                                                                 & \sqrt{\tens{C}_{\theta \theta}} & 0 \\
0                                                                 & 0                                          & \sqrt{\tens{C}_{\phi \phi}} \\
\end{array} \right] \nonumber \\
 & = \left[ \begin{array}{ccc}
\sqrt{2 (\tens{\kappa}_{rr} - \tens{\kappa}_{r \phi}^2 / \tens{\kappa}_{\phi \phi})} & 0                                          & \tens{\kappa}_{r \phi} \sqrt{2 / \tens{\kappa}_{\phi \phi}} \\
0                                                                                      & \sqrt{2 \tens{\kappa}_{\theta \theta}} / r & 0 \\
0                                                                                      & 0                                          & \sqrt{2 \tens{\kappa}_{\phi \phi}} / r \sin \theta \\
\end{array} \right] ,
\end{align}
where $\tens{\kappa}_{rr} \ge \tens{\kappa}_{r\phi}^2 / \tens{\kappa}_{\phi \phi}$ must hold (which it should as long as $\kappa_{\perp r} \le \kappa_{\parallel}$). Alternatively, an upper triangular matrix can be used, while an upper triangular matrix with $\tens{B}_{r\theta} = 0$ and $\tens{B}_{\theta r} \neq 0$ \citep[see][]{KoppEA2012} yields a similar matrix and the same condition for the Parker HMF used here. Another alternative given by \citet{PeiEA2010} yields a much more complex form. These alternatives might be helpful in keeping $\tens{B}$ real in the exceptional case when $\tens{\kappa}_{rr} < \tens{\kappa}_{r\phi}^2 / \tens{\kappa}_{\phi \phi}$ (it is theoretically not expected that $\kappa_{\perp r} > \kappa_{\parallel}$, and in the extreme case of isotropic turbulence, $\kappa_{\perp r} = \kappa_{\parallel}$, so that $\tens{\kappa}_{r\phi} = 0$), but do not resolve the coordinate singularity in spherical coordinates at the poles. To avoid division by zero, the problematic terms are set to zero directly over the poles (if the time step is small enough, however, physical processes should keep pseudo-particles away from the poles).


\subsubsection{Stochastic focused model}
\label{subsubsec:FocusedSDEs}

Eq.~\ref{eq:FocusTPE} in the form of Eq.~\ref{eq:BackwardKolmogorov} reads
\begin{align}
\label{eq:BackwardsFocus}
- \, \frac{\partial f}{\partial \tau} = \; & (\vec{\nabla} \cdot \tens{D}_{\perp} - \vec{v}_{\rm gc}) \cdot \vec{\nabla} f - \frac{\dinf p}{\dinf t} \, \frac{\partial f}{\partial p} + \left( \frac{\partial D_{\mu \mu}}{\partial \mu} - \frac{\dinf \mu}{\dinf t} \right) \frac{\partial f}{\partial \mu} \; + \nonumber \\
 & D_{\mu \mu} \, \frac{\partial^2 f}{\partial \mu^2} + \tens{D}_{\perp} : \vec{\nabla} \vec{\nabla} f \; - \nonumber \\
& \left( \vec{\nabla} \cdot \vec{v}_{\rm gc} + \frac{2}{p} \, \frac{\dinf p}{\dinf t} + \frac{\partial}{\partial p} \left[ \frac{\dinf p}{\dinf t} \right] + \frac{\partial}{\partial \mu} \left[ \frac{\dinf \mu}{\dinf t} \right] \right) f ,
\end{align}
with the perpendicular diffusion tensor in spherical coordinates \citep{StraussFichtner2015, vandenBerg2023},
\begin{align}
\label{eq:DperpTensor}
\tens{D}_{\perp} & = \left[ \begin{array}{ccc}
\tens{D}_{rr}^{\perp}       & \tens{D}_{r\theta}^{\perp}      & \tens{D}_{r\phi}^{\perp} \\
\tens{D}_{\theta r}^{\perp} & \tens{D}_{\theta\theta}^{\perp} & \tens{D}_{\theta\phi}^{\perp} \\
\tens{D}_{\phi r}^{\perp}   & \tens{D}_{\phi\theta}^{\perp}   & \tens{D}_{\phi\phi}^{\perp}
\end{array} \right] \\
 & = \left[ \begin{array}{ccc}
D_{\perp r}(\mu) \sin^2 \psi         & 0                     & D_{\perp r}(\mu) \sin \psi \cos \psi \\
0                                    & D_{\perp \theta}(\mu) & 0 \\
D_{\perp r}(\mu) \sin \psi \cos \psi & 0                     & D_{\perp r}(\mu) \cos^2 \psi
\end{array} \right] , \nonumber
\end{align}
being symmetric. Using the heliospheric setup of Sect.~\ref{subsec:HMFSWDCs} and $\dinf v / \dinf p = \dinf \left[ p c^2 / \sqrt{p^2 c^2 + E_0^2} \right] / \dinf p = v (1 - v^2 / c^2) / p$ for the simplification of the linear term, with $E_0$ the particle rest mass energy, Eq.~\ref{eq:BackwardsFocus} in spherical coordinates becomes
\begin{align}
\label{eq:BackwardsFocusSpherical}
- \, \frac{\partial f}{\partial \tau} = \; & - \left\lbrace\frac{\mu v}{c^2} \, \hat{b} \cdot \left[ \frac{\partial \vec{u}}{\partial t} + (\vec{u} \cdot \vec{\nabla}) \, \vec{u} \right] \right\rbrace f \; + \nonumber \\
 & \left( - \, \frac{\dinf p}{\dinf t} \right) \frac{\partial f}{\partial p} + \left( \frac{\partial D_{\mu \mu}}{\partial \mu} - \frac{\dinf \mu}{\dinf t} \right) \frac{\partial f}{\partial \mu} \; + \nonumber \\
 & \left[ \frac{1}{r} \left( 2 \tens{D}_{rr}^{\perp} + \csc \theta \, \frac{\partial \tens{D}_{\phi r}^{\perp}}{\partial \phi} \right) + \frac{\partial \tens{D}_{rr}^{\perp}}{\partial r} - \mu v b_r - u_r \right] \frac{\partial f}{\partial r} \; + \nonumber \\
 & \frac{1}{r^2} \left( \tens{D}_{\theta \theta}^{\perp} \cot \theta + \frac{\partial \tens{D}_{\theta \theta}^{\perp}}{\partial \theta} \right) \frac{\partial f}{\partial \theta} \; + \nonumber \\
 & \frac{\csc \theta}{r^2} \left[ \tens{D}_{r\phi}^{\perp} + \csc \theta \, \frac{\partial \tens{D}_{\phi \phi}^{\perp}}{\partial \phi} + r \left( \frac{\partial \tens{D}_{r\phi}^{\perp}}{\partial r} - \mu v b_{\phi} - u_{\phi} \right) \right] \frac{\partial f}{\partial \phi} \; + \nonumber \\
 & \frac{1}{2} \left[ 2 D_{\mu \mu} \, \frac{\partial^2 f}{\partial \mu^2} + 2 \tens{D}_{rr}^{\perp} \, \frac{\partial^2 f}{\partial r^2} + \frac{2 \tens{D}_{\theta \theta}^{\perp}}{r^2} \, \frac{\partial^2 f}{\partial \theta^2} \right. + \nonumber \\
 & \left. \frac{2 \tens{D}_{\phi \phi}^{\perp}}{r^2 \sin^2 \theta} \, \frac{\partial^2 f}{\partial \phi^2} + \frac{2 (\tens{D}_{r\phi}^{\perp} + \tens{D}_{\phi r}^{\perp})}{r \sin \theta} \, \frac{\partial^2 f}{\partial r \, \partial \phi} \right] ,
\end{align}
where $\vec{v}_{\rm d}(\mu)$ was neglected again and the linear term, $\dinf p / \dinf t$, and $\dinf \mu / \dinf t$ were not explicitly written in spherical coordinates due to their complicated forms (see Appendix~\ref{apndx:Maths} for these terms in spherical coordinates).

Identifying the SDEs from the latter equation yields
\begin{equation}
\L = - \frac{\mu v}{c^2} \, \hat{b} \cdot \left[ \frac{\partial \vec{u}}{\partial t} + \left( \vec{u} \cdot \vec{\nabla} \right) \vec{u} \right]
\end{equation}
for the linear coefficient,
\begin{subequations}
\label{eq:StochasticDriftFocus}
\begin{align}
a_r        = \; & \frac{1}{r} \left( 2 \tens{D}_{rr}^{\perp} + \csc \theta \, \frac{\partial \tens{D}_{\phi r}^{\perp}}{\partial \phi} \right) + \frac{\partial \tens{D}_{rr}^{\perp}}{\partial r} - \mu v \cos \psi - v_{\rm sw} , \\
a_{\theta} = \; & \frac{1}{r^2} \left( \tens{D}_{\theta \theta}^{\perp} \cot \theta + \frac{\partial \tens{D}_{\theta \theta}^{\perp}}{\partial \theta} \right) , \\
\label{eq:AphiFocus}
a_{\phi}   = \; & \frac{\csc \theta}{r^2} \left[ \tens{D}_{r\phi}^{\perp} + \csc \theta \, \frac{\partial \tens{D}_{\phi \phi}^{\perp}}{\partial \phi} \right. + \nonumber \\
 & \left. r \left( \mu v \sin \psi + v_{\rm sw} \tan \psi + \frac{\partial \tens{D}_{r\phi}^{\perp}}{\partial r}\right) \right] , \\
a_p        = \; & - \, \frac{\dinf p}{\dinf t} , \\
a_{\mu}    = \; & \frac{\partial D_{\mu \mu}}{\partial \mu} - \frac{\dinf \mu}{\dinf t}
\end{align}
\end{subequations}
for the stochastic drift coefficients, with $\vec{u}$ (Eq.~\ref{eq:FlowVelocity}) and $\vec{b}$ (Eq.~\ref{eq:ParkerUnitB}) substituted, and the perpendicular diffusion tensor,
\begin{align}
\tens{C}_{\perp} & = \left[ \begin{array}{ccc}
\tens{C}_{rr}^{\perp}     & 0                                & \tens{C}_{r \phi}^{\perp}    \\
0                         & \tens{C}_{\theta \theta}^{\perp} & 0             \\
\tens{C}_{\phi r}^{\perp} & 0                                & \tens{C}_{\phi \phi}^{\perp} \\
\end{array} \right] \nonumber \\
 & = 2 \left[ \begin{array}{ccc}
\tens{D}_{rr}^{\perp}                     & 0                                       & \tens{D}_{r \phi}^{\perp} / r \sin \theta        \\
0                                         & \tens{D}_{\theta \theta}^{\perp} / r^2 & 0                                         \\
\tens{D}_{r \phi}^{\perp} / r \sin \theta & 0                                       & \tens{D}_{\phi \phi}^{\perp} / r^2 \sin^2 \theta \\
\end{array} \right] ,
\end{align}
being positive definite as long as $r \neq 0$ and $\theta \neq 0^{\circ}$ or $\theta \neq 180^{\circ}$. For pitch-angle scattering, $\tens{B}_{\mu\mu} = \sqrt{2 D_{\mu\mu}}$, and
\begin{align}
\label{eq:DiffCoefUT}
\tens{B}_{\perp} & = \left[ \begin{array}{ccc}
\tens{B}_{rr}^{\perp} & \tens{B}_{r \theta}^{\perp}      & \tens{B}_{r \phi}^{\perp} \\
0                     & \tens{B}_{\theta \theta}^{\perp} & \tens{B}_{\theta \phi}^{\perp} \\
0                     & 0                                & \tens{B}_{\phi \phi}^{\perp} \\
\end{array} \right] \nonumber \\
 & = \left[ \begin{array}{ccc}
\sqrt{\tens{C}_{rr} - \tens{C}_{r \phi}^2 / \tens{C}_{\phi \phi}} & 0                               & \tens{C}_{r \phi} / \sqrt{\tens{C}_{\phi \phi}} \\
0                                                                 & \sqrt{\tens{C}_{\theta \theta}} & 0 \\
0                                                                 & 0                               & \sqrt{\tens{C}_{\phi \phi}} \\
\end{array} \right] \nonumber \\
 & = \left[ \begin{array}{ccc}
\sqrt{2 \left[ \tens{D}_{rr}^{\perp} - (\tens{D}_{r \phi}^{\perp})^2 / \tens{D}_{\phi \phi}^{\perp} \right]} & 0                                             & \tens{D}_{r \phi}^{\perp} \sqrt{2 / \tens{D}_{\phi \phi}^{\perp}} \\
0                                                                                                            & \sqrt{2 \tens{D}_{\theta \theta}^{\perp}} / r & 0 \\
0                                                                                                            & 0                                             & \sqrt{2 \tens{D}_{\phi \phi}^{\perp}} / r \sin \theta \\
\end{array} \right] \nonumber \\
 & = \left[ \begin{array}{ccc}
0 & 0                                        & \sqrt{2 D_{\perp r}(\mu)} \, \sin \psi \\
0 & \sqrt{2 D_{\perp \theta}(\mu)} \; r^{-1} & 0 \\
0 & 0                                        & \sqrt{2 D_{\perp r}(\mu)} \; r^{-1} \csc \theta \cos \psi \\
\end{array} \right]
\end{align}
for the spatial part of $\tens{B}$ chosen as upper triangular, where the condition $\tens{D}_{rr}^{\perp} \ge (\tens{D}_{r\phi}^{\perp})^2 / \tens{D}_{\phi \phi}^{\perp}$ is not necessary in the FTE for the Parker HMF since $(\tens{D}_{r\phi}^{\perp})^2 / \tens{D}_{\phi \phi}^{\perp} = D_{\perp r}^2(\mu) \sin^2 \psi \cos^2 \psi / D_{\perp r}(\mu) \cos^2 \psi = D_{\perp r}(\mu) \sin^2 \psi = \tens{D}_{rr}^{\perp}$ \citep{vandenBerg2023}. Note that the $v_{\rm sw} \tan \psi$ term in both Eqs~\ref{eq:AphiParker} and \ref{eq:AphiFocus} arises from solving the TPEs in the corotating frame (the only frame in which the Parker HMF form of Eq.~\ref{eq:ParkerHMF} is valid), and not from particles rigidity corotating with the Sun. However, this is inconsequential for GCRs in the steady state because the evolution of $\Phi$ is not considered in the SDEs, as described in the last paragraph of Sect.~\ref{subsec:SDEs} (it only needs to be taken into account if the solution is transformed into a stationary observer's frame).


\section{Results}
\label{sec:Results}

\begin{figure}[t!]
\includegraphics[trim=33mm 10mm 158mm 28mm, clip, scale=0.3]{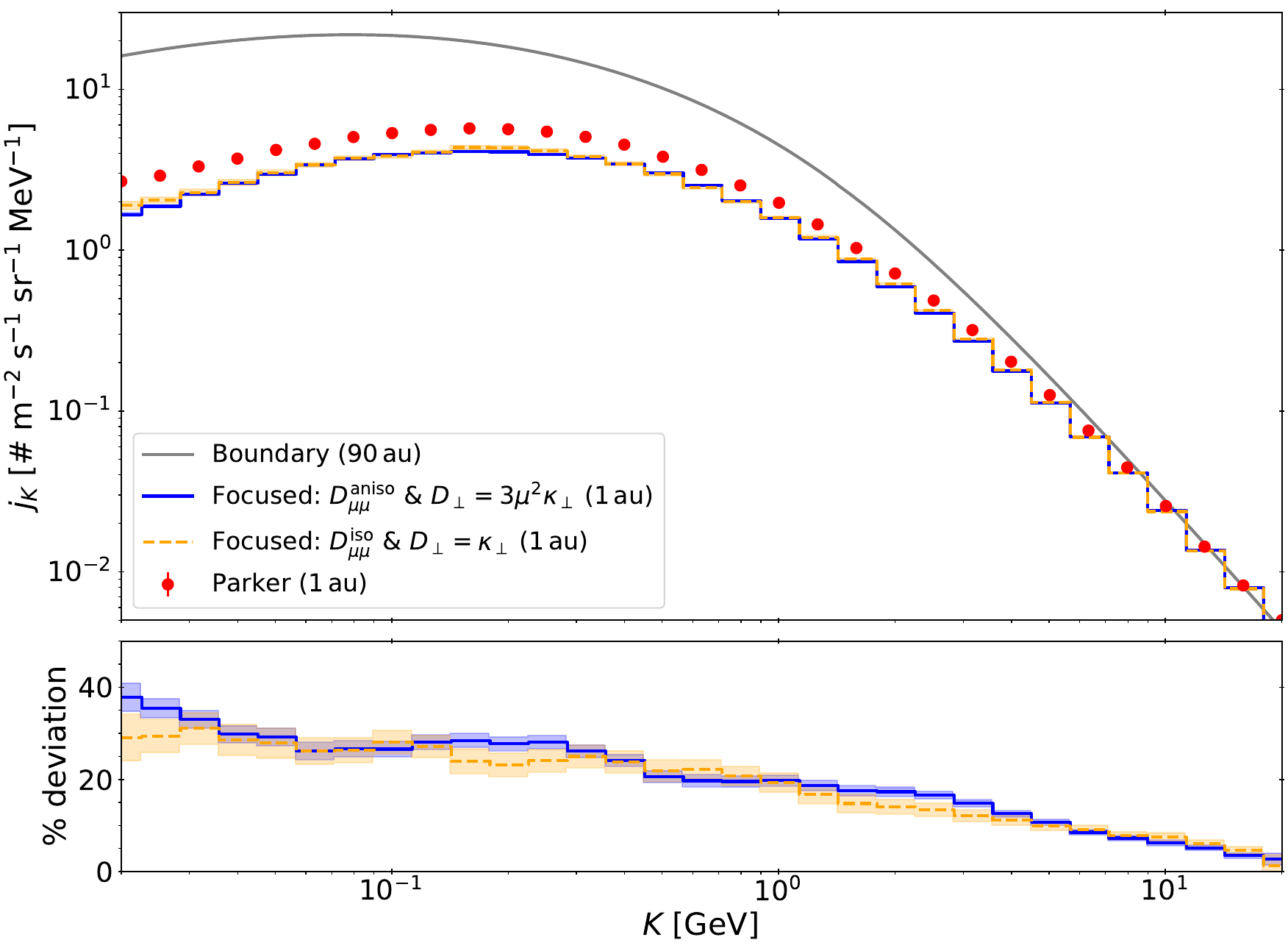}
\caption{\label{fig:EarthSpectra}Top: Proton spectrum at a radial distance of $1~{\rm au}$ in the equatorial plane as a function of kinetic energy resulting from the Parker (red dots) and focused (steps) TPEs. Results from the FTE for both isotropic pitch-angle scattering with pitch-angle-independent perpendicular diffusion (dashed orange) and anisotropic pitch-angle scattering with pitch-angle-dependent perpendicular diffusion (solid blue) are shown. The spectrum at the outer boundary is also shown for reference (solid grey line). Bottom: Percentage deviation ($100 \, |1 - j_K^{\rm Focus} / j_K^{\rm Parker}|$) between the results of the Parker and focused TPEs in the top panel. Shaded regions indicate uncertainties.}
\end{figure}

\begin{figure}[t!]
\includegraphics[trim=14mm 19mm 10mm 25mm, clip, scale=0.325]{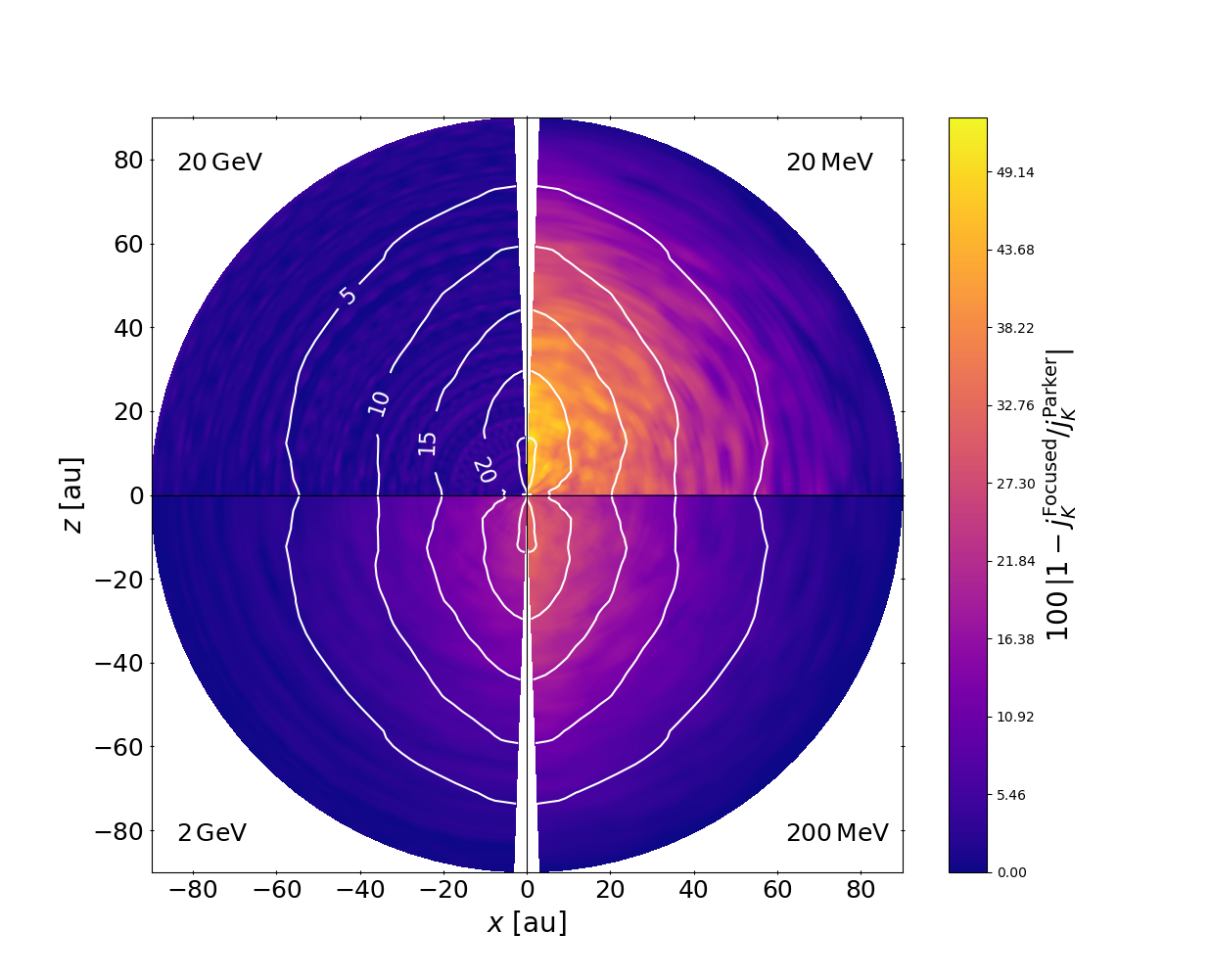}
\includegraphics[trim=14mm 10mm 10mm 25mm, clip, scale=0.325]{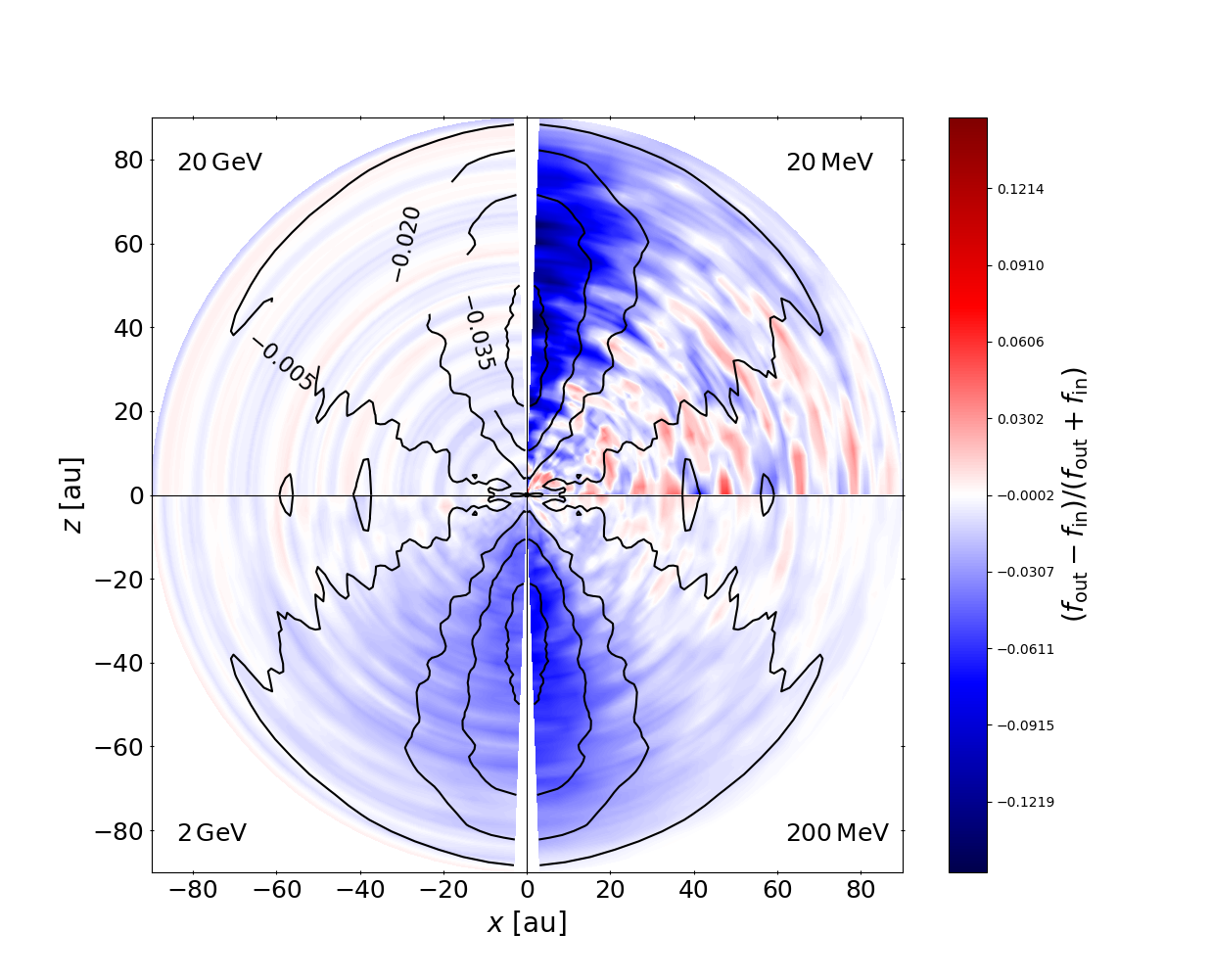}
\caption{\label{fig:DeviationAnisotropy}Top: Percentage deviation between the results of the Parker TPE and the FTE with anisotropic pitch-angle scattering and pitch-angle-dependent perpendicular diffusion throughout the heliosphere for four different energies (indicated in the quadrants) and averaged over all energies (contour lines). The equator and poles lie along the $x$- and $z$-axes, respectively. The innermost unlabelled contour line indicates $25 \%$. Bottom: Anisotropy proxy (Eq.~\ref{eq:AnisotropyProxy}) from the FTE with anisotropic pitch-angle scattering and pitch-angle-dependent perpendicular diffusion, presented in the same format as the top panel. The unlabelled contour line over the poles indicates a value of $-0.05$.}
\end{figure}

The spectrum of protons at a radial distance of $1~{\rm au}$ in the equatorial plane, as predicted by the Parker and focused TPEs, after smoothing the results by applying a three-point average over latitude, radius, and kinetic energy (in that order), is shown in the top panel of Fig.~\ref{fig:EarthSpectra}. Note that modulation in the heliosheath and at the termination shock is not included in the model. Therefore, these spectra cannot be directly compared with observations because the intensities are too high; however, the aim of this study is rather to directly compare the results computed using the two different TPEs than to compare them with a set of spacecraft observations. Unfortunately, the adiabatic limit (where $j_K \propto K$) shown in Appendix~\ref{apndx:TestModel} cannot be observed here, because the kinetic energies are too high. The choice of this limited energy range is made purely due to numerical constraints (a smaller $\ell$ is needed for the adaptive time step of Eq.~\ref{eq:AdaptiveTimeStep} at higher energies, and the execution time increases drastically at lower energies because efficient pitch-angle scattering must be resolved). The FTE predicts a similar spectrum whether isotropic pitch-angle scattering (Eq.~\ref{eq:IsoDmm}) with pitch-angle-independent perpendicular diffusion (Eq.~\ref{eq:Dperp} with $h = 1$) or anisotropic pitch-angle scattering (Eq.~\ref{eq:AnisoDmm}) with pitch-angle-dependent perpendicular diffusion (Eq.~\ref{eq:Dperp} with $h = 3 \mu^2$) is used. This is expected since \citet{vandenBerg2023} found that the interplay between pitch-angle scattering and perpendicular diffusion becomes noticeable only when $\kappa_{\perp} \sim \kappa_{\parallel}$, whereas $\kappa_{\perp r} / \kappa_{\parallel} = \chi = 0.02$ and $\kappa_{\perp \theta} / \kappa_{\parallel} = \chi \varsigma = 0.06$ at most over the poles in this work. Given the similarity of these results, unless stated otherwise, only anisotropic pitch-angle scattering with pitch-angle-dependent perpendicular diffusion will be discussed further, as this scenario is more realistic. Most importantly, the spectrum predicted by the Parker TPE is higher at low kinetic energies than that predicted by the FTE.

The bottom panel of Fig.~\ref{fig:EarthSpectra} shows the percentage deviation between the results from the Parker and focused TPEs, indicating that the difference between the two equations increases as energy decreases. Below $\sim 200~{\rm MeV}$, the Parker TPE overestimates the intensity by $\sim 30\%$. The apparent $\sim 5\%$ difference between the two FTE results at the lowest energies is an artefact of smoothing the results and does not reflect a physical difference (i.e. the unsmoothed results show no diverging trend within the uncertainties). To investigate the global behaviour of the Parker and focused TPEs, the top panel of Fig.~\ref{fig:DeviationAnisotropy} shows the percentage deviation between them across the heliosphere for four different energies. Note that the model results are independent of longitude because of the spherical modulation cavity assumed here, and that the southern and northern hemispheres are mirror images. At the highest energies, where the spectrum remains largely unmodulated, the difference between the Parker and focused TPEs is negligible. At the lowest energies, however, the difference increases towards the inner heliosphere and over the poles. Specifically, the percentage deviation over the poles reaches $\sim 40\%$ at a radial distance of $1~{\rm au}$. This pattern develops gradually, shifting from high to low energies, and is also clearly visible in the percentage deviation averaged across all kinetic energies (as indicated by the contour lines, which were additionally smoothed over radius and latitude).

\begin{figure}[t!]
\includegraphics[trim=33mm 70mm 148mm 28mm, clip, scale=0.295]{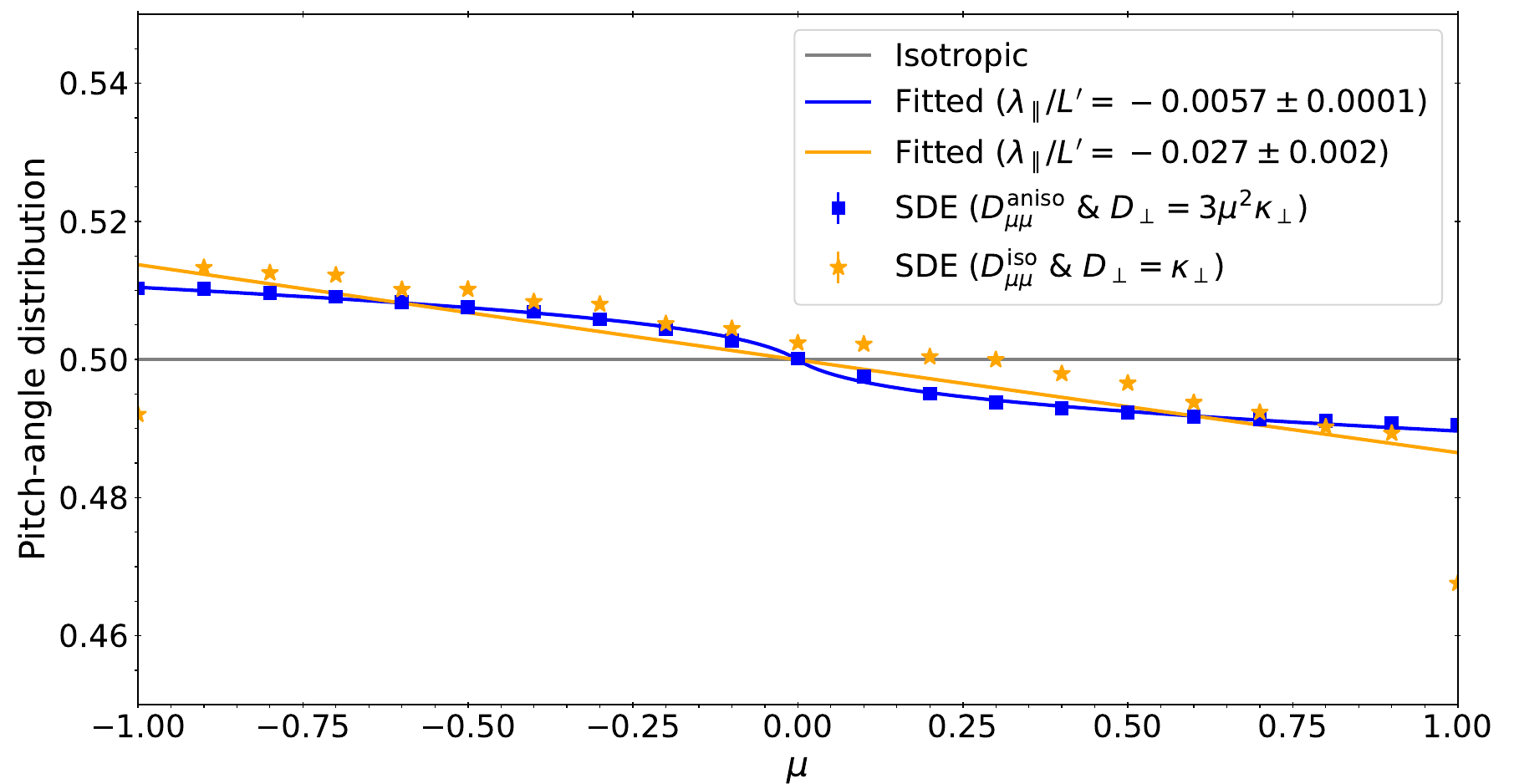}
\caption{\label{fig:PADs}Normalised PADs averaged over all latitudes, radial distances, and kinetic energies (in that order) resulting from the FTE for isotropic pitch-angle scattering with pitch-angle-independent perpendicular diffusion (orange stars) and anisotropic pitch-angle scattering with pitch-angle-dependent perpendicular diffusion (blue squares). The orange and blue lines indicate the analytical PADs of Eqs~\ref{eq:IsoPAD} and \ref{eq:AnisoPAD}, respectively, fitted to the simulations (with parameters indicated in the legend). Negative values of the pitch cosine signify particles moving towards the Sun.}
\end{figure}

To explain the difference between the Parker and focused TPEs, the initial quantities to examine are the first-,
\begin{equation}
\label{eq:A1}
A_1 = \frac{3}{2 F_0} \int_{-1}^1 \!\! \mu f(\mu) \, \dinf \mu ,
\end{equation}
and second-order,
\begin{equation}
\label{eq:A2}
A_2 = \frac{5}{4 F_0} \left[ 3 \int_{-1}^1 \!\! \mu^2 f(\mu) \, \dinf \mu - 2 F_0 \right] ,
\end{equation}
anisotropies \citep[see][for the origins and formal definitions]{BeeckWibberenz1986, Lampa2011}. The first-order anisotropy pertains to the (distribution-weighted) average pitch cosine and indicates a preferred direction of movement along the magnetic field \citep{vandenBergEA2020}. The second-order anisotropy becomes significant if the first-order anisotropy is nearly zero, but the distribution is not isotropic (i.e. if the distribution is an even function of the pitch cosine). Unfortunately, the statistical fluctuations inherent to SDEs are too large to discern any systematic behaviour in these anisotropy measurements. However, averaging the normalised PAD,
\begin{equation}
\tilde{f}(\mu) = \frac{f(\mu)}{\int_{-1}^1 f(\eta) \, \dinf \eta} = \frac{f(\mu)}{2 F_0} ,
\end{equation}
across regions of phase space reveals its general shape. The normalised PADs averaged over the entire phase space (and smoothed with a three-point average over pitch cosine) resulting from the FTE are shown in Fig.~\ref{fig:PADs}. The smaller values at $|\mu| = 1$, along with the overall noisiness of the PAD for isotropic pitch-angle scattering with pitch-angle-independent perpendicular diffusion, suggest that the time step might have been too large. It is clear that more particles are moving towards the Sun than away from it, as might be expected for GCRs diffusing into the heliosphere. Since the PADs are odd functions of the pitch cosine, it is expected that the second-order anisotropy will be smaller than the first-order anisotropy.

The shape of the PAD in Fig.~\ref{fig:PADs} varies depending on whether isotropic or anisotropic pitch-angle scattering occurs. These shapes should be explainable by the physical processes of pitch-angle transport. Balancing pitch-angle scattering and focusing in the steady-state solution, while neglecting all other transport processes,
\begin{equation}
\frac{\dinf}{\dinf \mu} \left[ \frac{(1 - \mu^2) v}{2 L^{'}} \, \tilde{f} \right] = \frac{\dinf}{\dinf \mu} \left[ D_{\mu\mu} \, \frac{\dinf \tilde{f}}{\dinf \mu} \right] ,
\end{equation}
and integrating twice from $-1$ to $\mu$, yields
\begin{equation}
\label{eq:StationaryPAD}
\tilde{f}(\mu) = \frac{\eulernumb^{G(\mu)}}{\int_{-1}^1 \eulernumb^{G(\eta)} \, \dinf \eta} ,
\end{equation}
where
\begin{equation}
\label{eq:DefineGmu}
G(\mu) = \frac{v}{2 L^{'}} \int_{-1}^{\mu} \frac{1 - \eta^2}{D_{\mu\mu}(\eta)} \, \dinf \eta ,
\end{equation}
the fact that $D_{\mu\mu} \propto 1 - \mu^2$ was used, and a normalisation condition (i.e. $\int_{-1}^1 \tilde{f}(\mu)  \, \dinf \mu = 1$) was applied to calculate the integration constant. Note that all physical processes affecting the PAD are parameterised here as $L^{'} D_{\mu\mu}$, and its effective value is therefore not the local value at the point where the PAD is being investigated \citep{BeeckWibberenz1986}. Alternatively, neglecting the temporal evolution and phase-space coordinates implies that this PAD represents a stationary solution of the distribution averaged over a part of phase space. For isotropic pitch-angle scattering (Eq.~\ref{eq:IsoDmm}), the PAD can be easily calculated, yielding
\begin{equation}
\label{eq:IsoPAD}
\tilde{f}_{\rm iso}(\mu) = \frac{\xi \, {\rm csch} \, \xi}{2} \, \eulernumb^{\xi \mu} ,
\end{equation}
where $\xi = \lambda_{\parallel} / L^{'}$, with $G(\mu) = \xi (1 + \mu)$ \citep[][defined $G(\mu)$ slightly differently but obtained the same PAD]{vandenBergEA2020}, from which the first-order anisotropy can be calculated to be
\begin{equation}
\label{eq:IsoA1}
A_1^{\rm iso} = 3 \left( \coth \xi - \frac{1}{\xi} \right) .
\end{equation}

\begin{figure*}[t!]
\includegraphics[trim=29mm 10mm 23mm 26mm, clip, scale=0.4375]{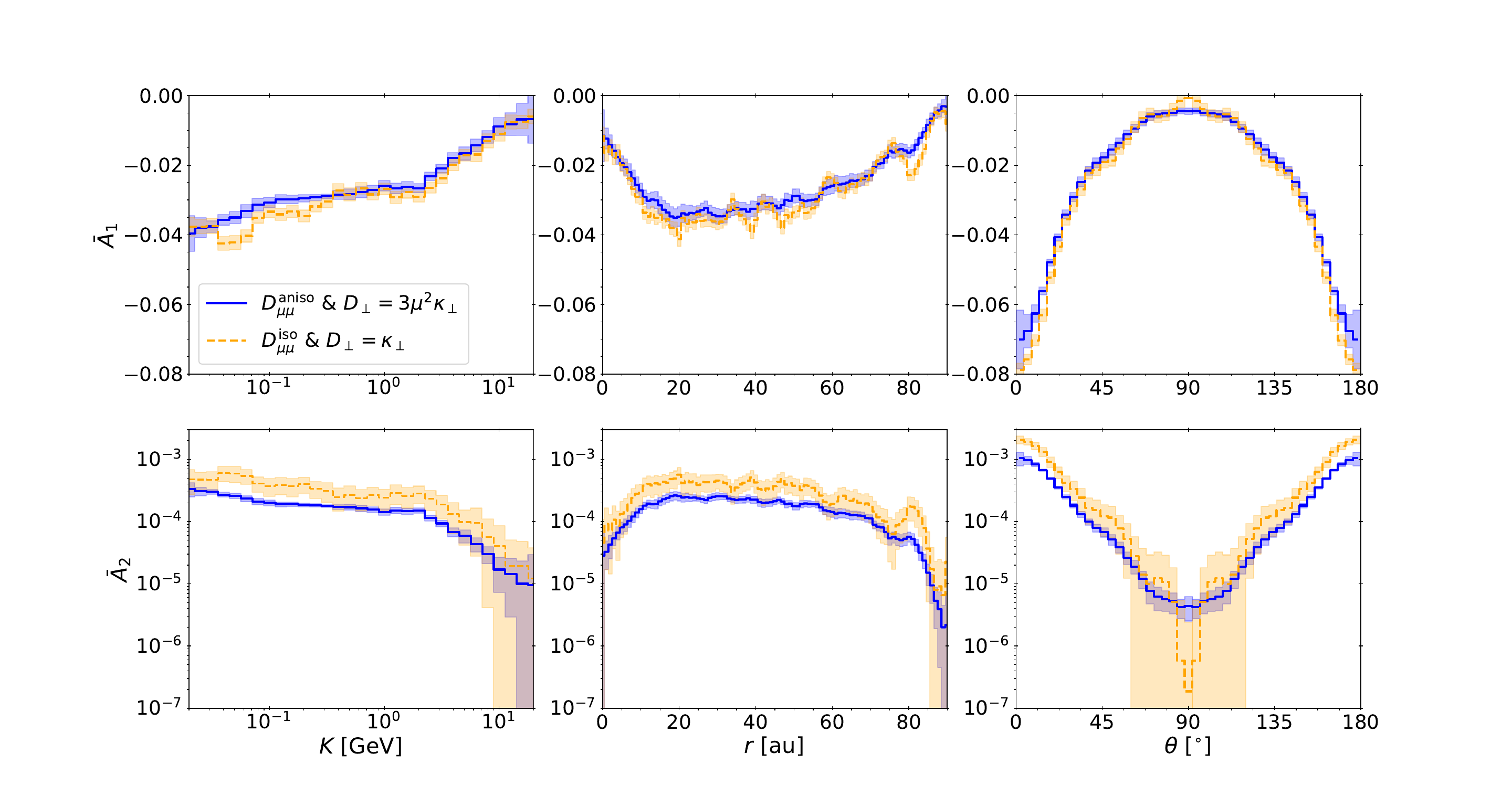}
\caption{\label{fig:FittedAnisotropy}First- (top) and second-order (bottom) anisotropies calculated from fitting Eqs~\ref{eq:IsoPAD} (dashed orange) or \ref{eq:AnisoPAD} (solid blue) to the simulated PADs averaged over all latitudes and radial distances (left), latitudes and kinetic energies (middle), or radial distances and kinetic energies (right). Eqs~\ref{eq:IsoA1} and \ref{eq:AnisoA1} are used to determine the first-order anisotropy, while the fitted distributions are integrated numerically as per Eq.~\ref{eq:A2} to calculate the second-order anisotropy (for simplicity, especially in the case of anisotropic pitch-angle scattering).}
\end{figure*}

Fitting this PAD to the isotropic pitch-angle scattering with pitch-angle-independent perpendicular diffusion results shown in Fig.~\ref{fig:PADs} (excluding the two points at $|\mu| = 1$; orange line) yields $\xi = - (2.7 \pm 0.2) \times 10^{-2}$. The PAD can be approximated by a straight line for such a small parameter, although the fit appears less accurate due to noise in the PAD. The first-order anisotropy can be calculated from the fitted $\xi$, with negative $\xi$- and $A_1$-values indicating that more particles are moving towards the Sun than away from it. Remember that the PAD includes signatures of all physical processes, meaning that the fitted value of $\xi$ is an `effective' value and does not represent the local value of $\lambda_{\parallel} / L^{'}$ at the point where the PAD is considered (i.e. compare this example to what might be expected from Fig.~\ref{fig:LengthScales}; also bear in mind that the PAD averaged over large regions of phase space is considered here). The PAD and first-order anisotropy can also be derived for anisotropic pitch-angle scattering (Eq.~\ref{eq:AnisoDmm}), but this process is more complex and is only detailed in Appendix~\ref{apndx:PADanisoDmm}. For anisotropic pitch-angle scattering with pitch-angle-dependent perpendicular diffusion, fitting the PAD (Eq.~\ref{eq:AnisoPAD}; blue line in Fig.~\ref{fig:PADs}) results in $\xi = - (5.7 \pm 0.1) \times 10^{-3}$. This smaller $\xi$-value, compared to that for isotropic pitch-angle scattering, despite both models sharing the same focusing length and parallel MFP, demonstrates that the fitted value is merely an effective estimate. The fact that the shapes of these PADs can be described by the analytical expectations derived from the PADC further confirms that the FTE model is functioning correctly.

By averaging the normalised PADs over only two of the three phase-space coordinates and fitting the analytical PADs, the anisotropies can be characterised as a function of a single coordinate. The average dependence of the first- and second-order anisotropies on kinetic energy, radial distance, and latitude is shown in Fig.~\ref{fig:FittedAnisotropy} (the fitted parameter was smoothed with a three-point average over kinetic energy or latitude and a five-point average over radial distance before calculating the anisotropies). The anisotropies derived from both isotropic and anisotropic pitch-angle scattering are similar, as expected given the comparable spectra of the two models. The second-order anisotropy from isotropic pitch-angle scattering is $\sim 2$ times larger, consistent with employing different pitch-angle dependencies for the CFDC. The second-order anisotropy has a smaller magnitude than the first-order anisotropy, as expected because of the odd functions of the PADs, and it is smaller when the magnitude of the first-order anisotropy is also smaller (such a small anisotropy is probably not measurable by particle detectors, but the second-order anisotropy is discussed here for its theoretical importance in explaining differences between the Parker and focused TPEs). The distributions are more anisotropic at low energies, between $\sim 20 - 40~{\rm au}$, and over the poles.

\begin{figure*}[t!]
\includegraphics[trim=14mm 19mm 25mm 26mm, clip, scale=0.335]{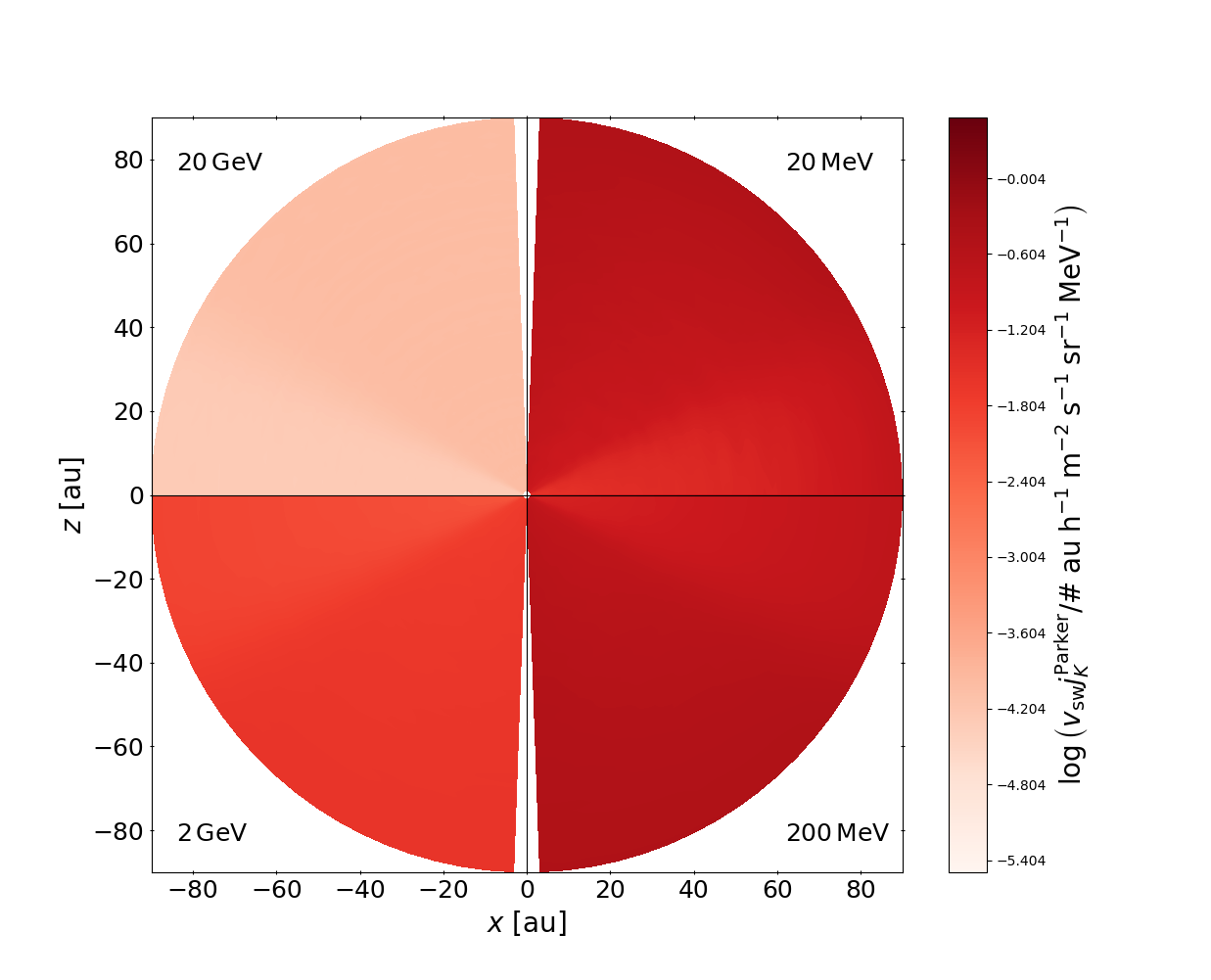}
\includegraphics[trim=23mm 19mm 25mm 26mm, clip, scale=0.335]{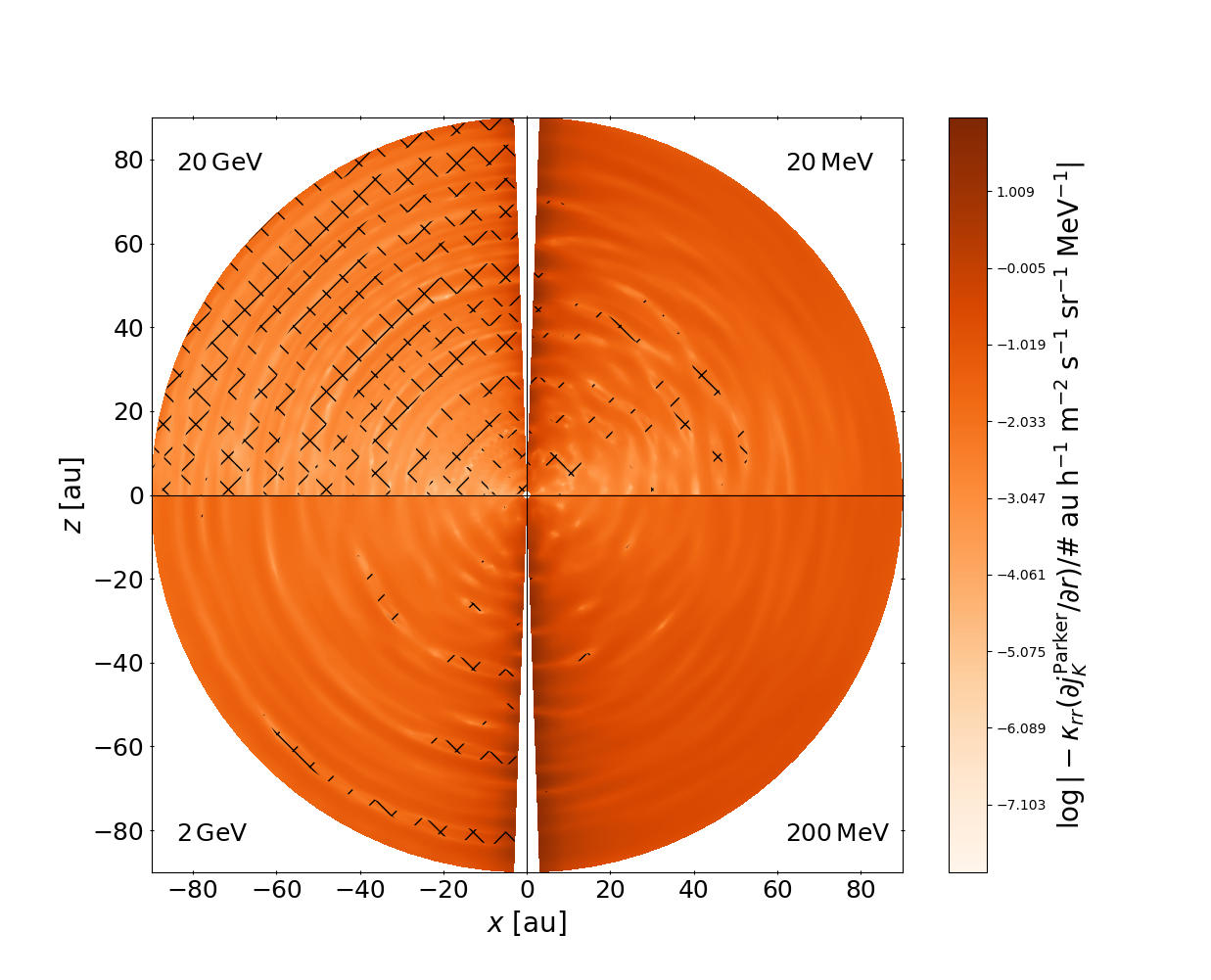}\\
\includegraphics[trim=14mm 8mm 25mm 16mm, clip, scale=0.335]{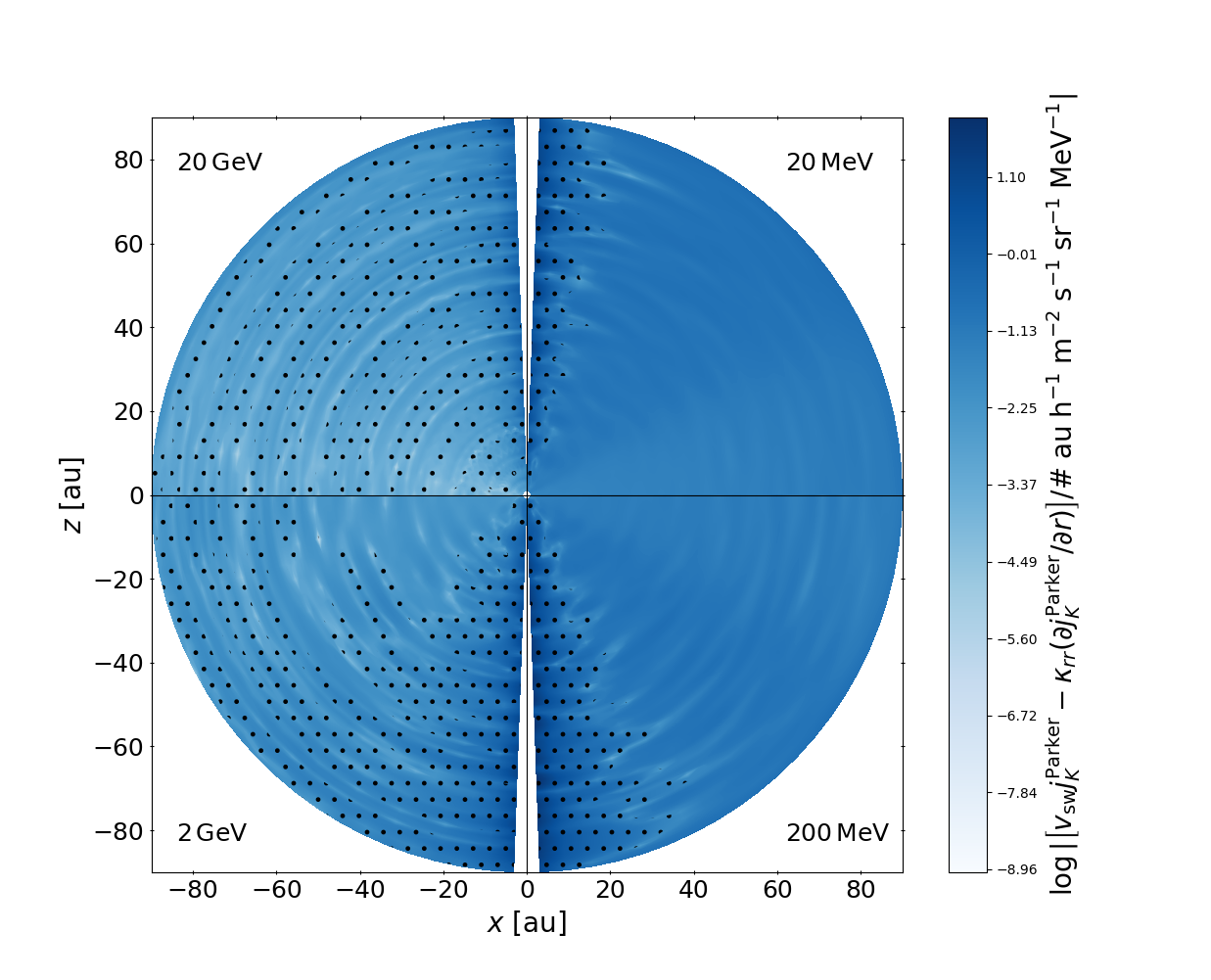}
\includegraphics[trim=23mm 8mm 25mm 16mm, clip, scale=0.335]{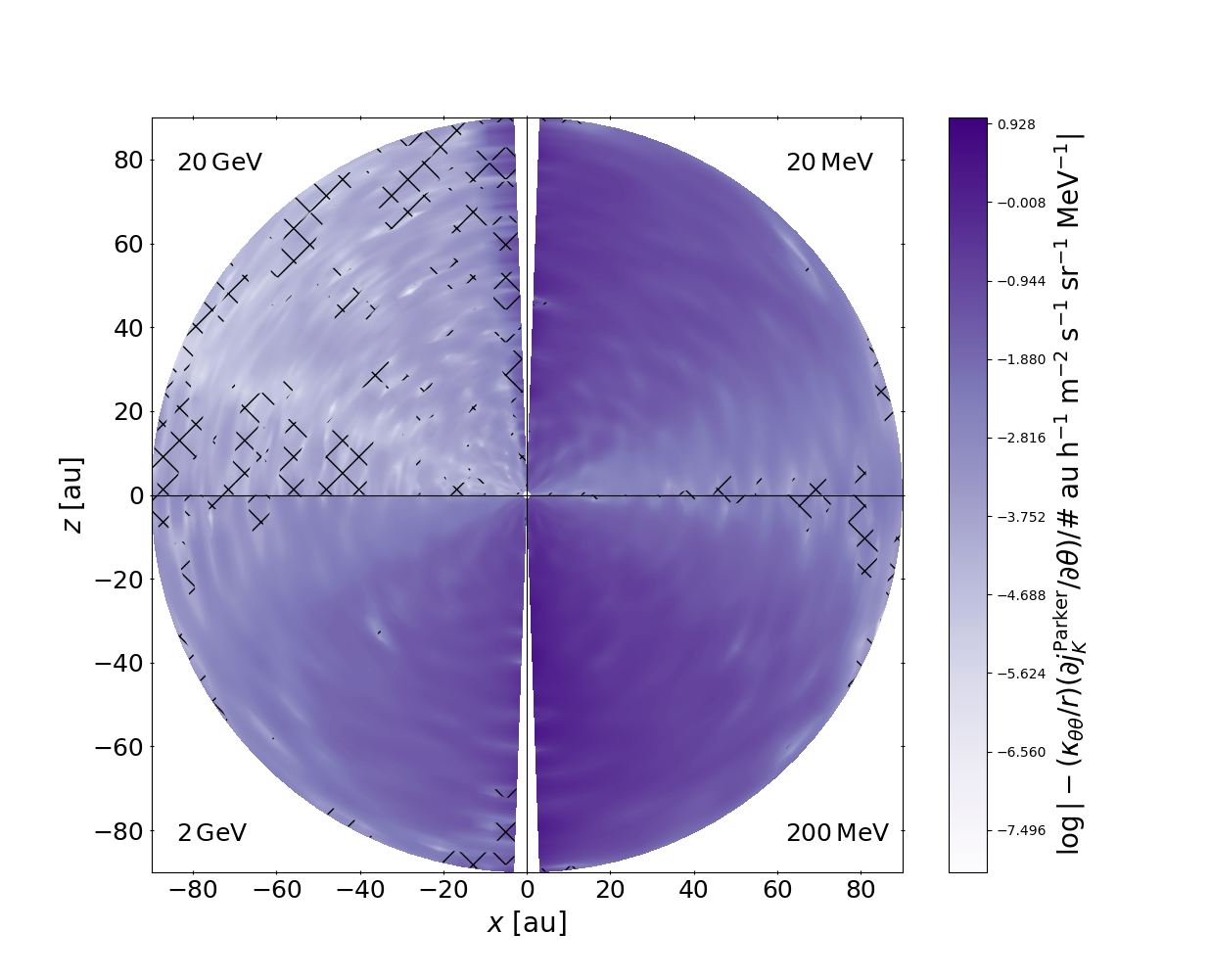}
\caption{\label{fig:ParkerFlux}Radial advective (top left), radial diffusive (top right), total radial (bottom left), and polar diffusive (bottom right) fluxes throughout the heliosphere resulting from the Parker TPE for four different energies (indicated in the quadrants). Hatches in the right panels indicate flux directed in the opposite direction than expected, and dots in the bottom left panel indicate inward flux.}
\end{figure*}

To assess how anisotropy varies across the heliosphere, the bottom panel of Fig.~\ref{fig:DeviationAnisotropy} shows the quantity
\begin{equation}
\label{eq:AnisotropyProxy}
{\cal R} = \frac{f_{\rm out} - f_{\rm in}}{f_{\rm out} + f_{\rm in}} = \frac{\int_0^1 f(\mu) \, \dinf \mu - \int_{-1}^0 f(\mu) \, \dinf \mu}{\int_{-1}^1 f(\mu) \, \dinf \mu} ,
\end{equation}
where $f_{\rm out} = \int_0^1 f(\mu) \, \dinf \mu$ and $f_{\rm in} = \int_{-1}^0 f(\mu) \, \dinf \mu$ represent the parts of the distribution of particles moving away and towards the Sun, respectively, as a proxy for anisotropy in the same format as the percentage deviation of the top panel (the original distribution function was first integrated, then $f_{\rm out}$, $f_{\rm in}$, and $f_{\rm out} + f_{\rm in}$ were smoothed separately over latitude, radius, and kinetic energy before calculating ${\cal R}$). This quantity was originally introduced by \citet{vandenBergEA2020} as the fraction of SEPs released at the Sun. Unfortunately, this quantity is not simply related to the first-order anisotropy \citep[as discussed by][]{vandenBergEA2020}\footnote{The anisotropy proxy would be one-third of the first-order anisotropy if the particles moved exactly along the magnetic field (i.e. if $\mu$ could only take values of $\pm 1$).}, but it displays distinct spatial and energy dependencies. This is because the partial integration over pitch cosine smooths the distribution somewhat, even though the distribution function remains too noisy to calculate the first-order anisotropy at each phase-space point. The streaky pattern of the anisotropy proxy in the bottom panel of Fig.~\ref{fig:DeviationAnisotropy} at high energies, compared to the patchy pattern at low energies, suggests that the time step may be too large at higher energies. Nonetheless, the proxy indicates that the distribution is nearly isotropic at the highest energies. At lower energies, however, particles tend to move preferentially towards the Sun over the poles and away from the Sun in the equatorial region. When the anisotropy proxy is averaged over energy (and smoothed over radius and latitude), the distribution appears nearly isotropic in the equatorial region, though the particles moving towards the Sun over the poles remain visible (see the contour lines).

\edited{The} anisotropy \edited{proxy} patterns resemble the trends observed in the first-order anisotropy: both $A_1$ and ${\cal R}$ are predominantly negative throughout the heliosphere, with the largest values near the poles between $\sim 20 - 60~{\rm au}$. Such an anisotropy pattern is expected because the diffusion condition ($\lambda_{\parallel} < |L^{'}|$) is met in the equatorial region (compare the solid orange and grey lines in Fig.~\ref{fig:LengthScales}), but not over the poles (compare the dashed orange and grey lines in Fig.~\ref{fig:LengthScales}), where focusing becomes more significant relative to pitch-angle scattering (due to very large MFPs). The anisotropy patterns also resemble those of the percentage deviation, though not exactly: the difference between the Parker and focused TPE results is greater closer to the Sun with a slight latitudinal dependence, while the anisotropy proxy shows a strong latitudinal dependence and peaks between $\sim 20 - 60~{\rm au}$ from the Sun. Naively, one might expect that the largest anisotropy, and thus the greatest difference between the Parker and focused TPEs, would occur in the outer heliosphere where the parallel MFP is largest, but this does not account for transport processes. This spatial mismatch could be interpreted as a sign of transport effects that lead to anisotropy causing a non-local difference between the Parker and focused TPEs at different locations. However, caution is needed, as this does not reflect the true anisotropy, and the differences between TPEs are driven by pitch-angle-weighted fluxes (see the following discussions). Still, the anisotropic distribution could influence particle energy losses and spatial transport, potentially leading the Parker TPE to underestimate the modulation. The presence of a non-negligible anisotropy may explain the disparity between the Parker and focused TPEs and warrants further investigation.

To explain these specific anisotropy patterns, consider the flux from the Parker TPE. A particular spatial component of the flux, $J_i$, has an associated anisotropy amplitude of $3 J_i / v F_0$ \citep[see e.g.][]{JokipiiParker1970, Quenby1984}, implying that a non-zero flux must be associated with a non-zero anisotropy, and vice versa. For the model setup considered here, the various spatial fluxes are
\begin{subequations}
\label{eq:ParkerFlux}
\begin{align}
\label{eq:ParkerRadialFlux}
J_r & = u_r F_0 - \tens{\kappa}_{rr} \, \frac{\partial F_0}{\partial r} , \\
\label{eq:ParkerPolarFlux}
J_{\theta} & = - \frac{\tens{\kappa}_{\theta \theta}}{r} \, \frac{\partial F_0}{\partial \theta} ,
\end{align}
\end{subequations}
where $u_r = v_{\rm sw}$ is the radial component of the flow velocity, $\tens{\kappa}_{rr} = \kappa_{\parallel} \cos^2 \psi + \kappa_{\perp r} \sin^2 \psi$ is the radial DC, and $\tens{\kappa}_{\theta \theta} = \kappa_{\perp \theta}$ is the polar DC (see Sect.~\ref{subsubsec:ParkerSDEs}; the longitudinal flux is not considered here, since a spherical heliosphere is assumed with no longitudinal dependencies). The radial flux results from an outward advective flux and an inward diffusive flux, whereas the polar flux results solely from a diffusive flux directed from the poles to the equator. Interestingly, these fluxes suggest that the advective anisotropy is $v_{\rm sw} / v$, which is small for most high-energy particles, while the diffusive anisotropies are the product of a MFP per unit radius and the spatial power-law index of the omni-directional intensity, i.e. $-(\lambda_{rr} / r) (\partial \ln F_0 / \partial \ln r)$ or $-(\lambda_{\theta \theta} / r) (\partial \ln F_0 / \partial \theta)$. Note that this flux anisotropy is not directly comparable to the usual first- or second-order anisotropy or the anisotropy proxy of Eq.~\ref{eq:AnisotropyProxy} because it is oriented along a specific spatial coordinate and is not measured relative to the HMF.

The various fluxes are shown for four different kinetic energies in Fig.~\ref{fig:ParkerFlux}. The latitudinal dependence of the SW is evident in the advective radial flux (top-left panel). The numerical derivative of $j_K$ is, unfortunately, noisy when calculating the diffusive flux (even after smoothing the derivative over latitude, radius, and kinetic energy). The diffusive flux directed \edited{opposite} to what is expected, especially at the highest energies, appears to \edited{arise} from numerical derivatives, as there is no physical mechanism in the model that could produce it (although unclear in the figure, the diffusive radial flux alternates between inward and outward). Therefore, the behaviour of both the radial and polar diffusive fluxes (right panels) is unclear at the highest energies because of the numerical noise. However, it is clear at lower energies that the radial diffusive flux (top-right panel) is greater over the poles than in the equatorial region. The polar diffusive flux (bottom-right panel) is also larger over the poles than in the equatorial region and displays the latitudinal dependence of the CFDC. The polar diffusive flux is generally smaller in magnitude than the radial diffusive flux because it reflects only slower perpendicular diffusion. Nevertheless, at low energies, the polar diffusive flux is significantly higher than at high energies over the poles, indicating that low-energy particles are transported efficiently from the poles to the equator. The total radial flux (bottom-left panel) exhibits a more complex structure: the inward (marked by dots) diffusive flux dominates the outward advective flux at high energies (note that the dots at the highest energies do not contradict the outward diffusive flux marked by the hatched areas, as the flux alternates direction due to numerical noise); however, at progressively lower energies, the outward advective flux begins to dominate the inward diffusive flux, except over the poles. This is caused by a combination of smaller MFPs for lower-energy particles and the radial diffusive flux being dominated by large parallel MFPs over the poles and small perpendicular MFPs in the equatorial region (see the black dash-dotted line in Fig.~\ref{fig:LengthScales}). SW advection (parallel diffusion) explains the outward-moving (inward-moving) particles observed in the anisotropy proxy of Fig.~\ref{fig:DeviationAnisotropy} at low energies in the equatorial region (over the poles).


\section{Discussion}
\label{sec:Discussion}

The previous section covered the results that could be realistically and reliably obtained from the SDE simulations. However, further discussion is needed to offer a more physical interpretation and summary of these results. Unfortunately, the fluxes in Eq.~\ref{eq:ParkerFlux} do not account for energy losses and can be considered as a `local spatial' flux. In reality, it is unlikely that a $20~{\rm MeV}$ particle diffused from the outer boundary into the inner heliosphere. Instead, a higher-energy particle would lose energy as it diffused against the outward-directed SW, experiencing different diffusion and energy-loss conditions along the way. To account for energy losses, \citet{Parker1967} calculates the flux from the radial component of the Parker TPE \citep[Eq.~\ref{eq:ParkerTPE1D}; see also the discussion in][]{JokipiiParker1970}. The radial part of the stationary solution of Eq.~\ref{eq:ParkerTPE} can be written as
\begin{equation}
\frac{1}{r^2} \frac{\partial}{\partial r} \left[ r^2 J_r \right] = \frac{1}{p^2} \frac{\partial}{\partial p} \left[ \frac{p^3}{3 r^2} \frac{\partial}{\partial r} \left( r^2 u_r \right) F_0 \right] ,
\end{equation}
and integration over radial distance yields
\begin{equation}
\label{eq:ParkerIntegratedRadialFlux}
J_r(r,p) = \frac{1}{3 r^2 p^2} \, \frac{\partial}{\partial p} \left[ p^3 \int_0^r \!\! r' \left( 2 u_r + r' \frac{\partial u_r}{\partial r'} \right) F_0 \, \dinf r' \right] ,
\end{equation}
keeping in mind that $u_r = v_{\rm sw}(r,\theta)$ and $F_0(r,\theta,p)$. Note, however, that this form of the radial flux is only valid for a spherically symmetric heliosphere, and has the implicit assumption of $\kappa_{\perp} = \kappa_{\parallel}$, $\vec{V}_{\rm d} = \vec{0}$, and no latitudinal dependencies in the DC or SW speed (see Appendix~\ref{apndx:TestModel}). Nonetheless, if this calculation is performed for each radial spoke at different latitudes (not shown), the latitudinal dependence of the SW becomes evident in the flux, due to the energy losses arising from its divergence. Furthermore, the flux is negative (directed towards the Sun) at high energies and positive (directed away from the Sun) at low energies. This is caused by energy losses being proportional to $\partial F_0 / \partial p$, with the shape of the spectrum yielding $\partial F_0 / \partial p > 0$ ($\partial F_0 / \partial p < 0$) at low (high) energies. Therefore, this approach does not account for the diffusive and advective processes but rather demonstrates that high-energy particles enter the heliosphere, lose energy, and are swept out by the SW at lower energies.

The fluxes from the FTE depend on pitch angle and could, in theory, be averaged over pitch cosine (indicated in what follows by angular brackets, i.e. $\langle \cdots \rangle_{\mu} = \int_{-1}^1 \cdots \, \dinf \mu / 2$) to be compared with Eq.~\ref{eq:ParkerFlux}. However, the distribution function is unfortunately too noisy to do so. Nonetheless, it remains insightful to consider the spatial fluxes of the FTE, namely
\begin{subequations}
\label{eq:FocusedFlux}
\begin{align}
\label{eq:FocusedRadialFlux}
J_r(\mu) & = (\mu v b_r + u_r) \, f(\mu) - \tens{D}_{rr}^{\perp}(\mu) \, \frac{\partial f(\mu)}{\partial r} , \\
\label{eq:FocusedPolarFlux}
J_{\theta}(\mu) & = - \frac{\tens{D}_{\theta \theta}^{\perp}(\mu)}{r} \, \frac{\partial f(\mu)}{\partial \theta} ,
\end{align}
\end{subequations}
where $b_r = \cos \psi$ is the radial component of the unit Parker HMF vector (see Eq.~\ref{eq:ParkerUnitB}), $\tens{D}_{rr}^{\perp}(\mu) = h(\mu) \, \kappa_{\perp r} \sin^2 \psi$ is the radial DC, and $\tens{D}_{\theta \theta}^{\perp}(\mu) = h(\mu) \, \kappa_{\perp \theta}$ is the polar DC (see Sect.~\ref{subsubsec:FocusedSDEs}; the longitudinal flux is again not considered, as the model is independent of longitude). Note that the diffusive flux is solely due to perpendicular diffusion, while the radial advective flux now incorporates the streaming of GCRs along the field. The distribution can, for simplicity, be expressed using Legendre polynomials \citep[see e.g.][]{BeeckWibberenz1986, Zank2014}. As an illustrative example, only the first three polynomials will be examined, i.e.
\begin{equation}
\label{eq:LegendreExpansion}
f(\mu) \approx \left( 1 + A_1 \mu + A_2 \frac{3 \mu^2 - 1}{2} \right) F_0 .
\end{equation}
Although the second-order term is expected to be insignificant in the current scenario, the first- and second-order terms serve as examples of how odd and even terms are anticipated to contribute in the following discussion. The pitch-angle averaged flux will then be
\begin{subequations}
\label{eq:AvgApproxFocusedFlux}
\begin{align}
\label{eq:MuAvgApproxRadialFlux}
\langle J_r(\mu) \rangle_{\mu} \approx \; & \left( \frac{v}{3} \, A_1 b_r + u_r - \frac{2}{5} \, \frac{\partial A_2}{\partial r} \, \tens{\kappa}_{rr}^{\perp} \right) F_0 \; - \nonumber \\
 & \tens{\kappa}_{rr}^{\perp} \left( 1 + \frac{2}{5} \, A_2 \right) \frac{\partial F_0}{\partial r} , \\
\label{eq:MuAvgApproxPolarFlux}
\langle J_{\theta}(\mu) \rangle_{\mu} \approx \; & - \frac{\tens{\kappa}_{\theta \theta}}{r} \left[ \frac{2}{5} \, \frac{\partial A_2}{\partial \theta} \, F_0 + \left( 1 + \frac{2}{5} \, A_2 \right) \frac{\partial F_0}{\partial \theta} \right] ,
\end{align}
\end{subequations}
where it was assumed that $h(\mu) = 3 \mu^2$ and $\tens{\kappa}_{rr}^{\perp} = \kappa_{\perp} \sin^2 \psi$ is the perpendicular part of the radial isotropic DC.

This has an interesting consequence if the distribution is perfectly isotropic (i.e. if $f(\mu) = F_0$ or $A_1 = A_2 = 0$): the SW advection and perpendicular diffusion will be the same as in the Parker TPE (see Eq.~\ref{eq:ParkerFlux}), but there will be no parallel diffusion. This is expected because the Parker TPE is derived by perturbing the isotropic distribution and solving for the perturbed anisotropic component, which then results in parallel diffusion (as discussed in \edited{Sect.~\ref{sec:Background}}). Therefore, a small anisotropy is necessary to describe parallel diffusion. Including the first-order anisotropy, the non-zero inward streaming flux ($A_1 v F_0 \cos \psi$ with $A_1 < 0$) accounts for pitch-angle scattering and other pitch-angle changes, but in the advective component of the flux rather than as parallel diffusion in the diffusive component. This follows from the fact that, to lowest order, anisotropy \edited{arises} from a local balance \edited{among} pitch-angle scattering, focusing, and the momentum and spatial gradients of $F_0$, \edited{which act} as source terms (as mentioned in \edited{Sect.~\ref{sec:Background}}). Therefore, this flux would be larger over the poles, due to the less wound \edited{HMF} and larger parallel MFPs, which enable particle streaming, and at low energies in the inner heliosphere, because the spatial gradients of $F_0$ are larger under these circumstances. Although the pitch-angle averaged fluxes from the FTE could thus differ quantitatively from those of the Parker TPE, they should be qualitatively similar, with only the interpretation of the fluxes differing. The qualitative difference between $A_1 v F_0 \cos \psi$ and $- \kappa_{\parallel} \cos^2 \psi (\partial F_0 / \partial r)$ likely explains part of the disparity between the Parker and focused TPEs, but quantifying this difference remains challenging because of the noise in the first-order anisotropy. Including the second-order anisotropy demonstrates that the perpendicular diffusive flux will be modified compared to what is expected from the Parker TPE, while additional advective fluxes might also arise due to the spatial gradient of the second-order anisotropy. These modifications, however, are expected to be negligible due to the smallness of the second-order anisotropy. The anisotropy would also modify the momentum change term of Eq.~\ref{eq:dpdt} to
\begin{align}
\label{eq:MuAvgApproxdpdt}
\left\langle \frac{\dinf p}{\dinf t} \right\rangle_{\mu} \approx \; & - p \left\lbrace \frac{A_2}{5} \, \uvec{b} \uvec{b} : \vec{\nabla} \vec{u} + \frac{1 - A_2 / 5}{3} \, \vec{\nabla} \cdot \vec{u} \right. + \nonumber \\
 & \left. \frac{A_1}{3 v} \, \uvec{b} \cdot \left[ \frac{\partial \vec{u}}{\partial t} + \left( \vec{u} \cdot \vec{\nabla} \right) \vec{u} \right] \right\rbrace .
\end{align}
Although the contribution of the second-order anisotropy will be minor for the GCRs considered here, the last term will result in slightly reduced energy losses (since $A_1 \uvec{b} \cdot [(\vec{u} \cdot \vec{\nabla}) \vec{u}] < 0$). However, it is apparent from Eq.~\ref{eq:buGradu} that this term is only significant near the Sun in the equatorial regions, and therefore, this correction can be expected to be negligible. These anisotropy-modified fluxes (note that the momentum flux is $(\dinf p / \dinf t) \, f(\mu)$ or $\langle \dinf p / \dinf t \rangle_{\mu} F_0$) highlight the additional terms needed in the Parker TPE if the interplay between pitch-angle scattering and various physical processes is taken into account (as mentioned in \edited{Sect.~\ref{sec:Background}}).

In conclusion, the non-zero fluxes, particle streaming, and increased importance of focusing over the poles imply an anisotropic distribution, as seen in the anisotropy proxy (bottom panels of Fig.~\ref{fig:DeviationAnisotropy}), which the Parker TPE does not model accurately, as indicated by the percentage deviation (top panels of Fig.~\ref{fig:DeviationAnisotropy}). Physically, particles tend to stream along the HMF over the poles, where they experience less pitch-angle scattering and more focusing, are slowly transported to the equator via perpendicular diffusion and advected outward by the SW in the equatorial regions after losing much of their energy. It appears more likely that GCRs diffuse from the poles towards the equator in the inner heliosphere than directly into the heliosphere in the equatorial region, because GCRs can more easily reach the inner heliosphere over the poles, owing to larger parallel MFPs compared to the smaller perpendicular MFPs in the equatorial regions (only the gradient of the small second-order anisotropy can oppose the polar diffusive flux from the poles to the equator, as per Eq.~\ref{eq:MuAvgApproxPolarFlux}). The streaming of GCRs over the poles and the focusing they undergo there are unavoidable consequences of the HMF geometry used here and the typical assumptions about DCs over the poles, which lead to large parallel MFPs. Although focusing might be expected to play a lesser role in the outer heliosphere compared to near the Sun, the large parallel MFPs over the poles imply that pitch-angle scattering is ineffective in isotropising the distribution if any other process causes anisotropy. One might wonder how this scenario over the poles differs from the so-called `scatter-free events' of SEPs (referring to events with very large parallel MFPs or field-aligned particles arriving first at an observer). Perhaps the two scenarios are similar, but we advise against using that term, as we believe there is no such thing as a `scatter-free event'. Turbulence is omnipresent in the heliosphere, and all charged particles are constantly subjected to small-scale pitch-angle scattering, even though the parallel MFP might be large \citep[see the discussions of][]{vandenBergEA2020, vandenBerg2023}. What remains true is that particles will stream more efficiently along the magnetic field if pitch-angle scattering is weak. The resulting anisotropy is very small compared to that observed during SEP events, yet the disproportionately large deviation between the Parker and focused TPEs is not due to the Parker TPE incorrectly modelling a small anisotropy at a single point but stems from the systematic accumulation of this incorrect modelling across the entire heliosphere. This also accounts for the spatial mismatch between the anisotropy proxy and the difference between the Parker and focused TPEs in Fig.~\ref{fig:DeviationAnisotropy}. It might be naively expected that streaming over the poles would be more efficient than parallel diffusion, implying that the FTE should produce a higher intensity than the Parker TPE. However, pitch-angle scattering disrupts streaming and causes parallel diffusion; if scattering is infrequent, the resulting parallel diffusion in the Parker TPE appears over-diffusive. If streaming with inefficient pitch-angle scattering is treated as effective parallel diffusion, the Parker TPE would overestimate the particle diffusion rate into the heliosphere, resulting in less modulation than in the FTE.

\edited{The question arises whether including the modifying terms that follow from deriving the Parker TPE from the FTE within the Parker TPE could reduce the difference between the results of the Parker and focused TPEs. Given that the omnidirectional intensity and first-order anisotropy resulting from the FTE were not statistically sensitive to the assumed pitch-angle dependencies of the DCs, the simplest case of isotropic pitch-angle scattering with pitch-angle-independent perpendicular diffusion can be considered in the current discussion. In this case, the modifying terms given by \citet{vandenBerg2023} simplify considerably, and the remaining terms arise from the interplay of pitch-angle scattering, focusing, and pitch-angle-dependent energy changes. The parallel DC is modified by focusing \citep[as expected from][]{BeeckWibberenz1986, HeSchlickeiser2014, WangQin2018}, while additional advection parallel to the magnetic field, momentum changes, mixed parallel and momentum diffusion, and momentum diffusion terms are added to the Parker TPE \citep[as expected from][]{WangQin2018}. Even though the modifying terms can be analytically calculated for these likely oversimplified DCs, they are still cumbersome to work with, and the direction and magnitude of these terms remain unclear. Without evaluating the terms, it can only be said that they will be polynomials of $\lambda_{\parallel} / L^{'}$. It may well be that a modified Parker TPE could come closer to the FTE result, but one would need to calculate many complicated transport coefficients for something that is still only an approximation of the FTE, in which case it seems easier to just solve the FTE. This new equation would most definitely not be the standard Parker equation used in numerous GCR modulation studies and would therefore yield different results. Solving this more complex TPE is beyond the scope of the current investigation, which aimed to determine whether the standard Parker TPE and FTE yield comparable results under typical GCR modulation conditions.}

Care has been taken in this work to make the Parker and focused TPEs as comparable as possible by normalising the pitch-angle-dependent DCs in the FTE to the isotropic DCs of the Parker TPE. However, the current model lacks two key components that require further investigation. Firstly, realistic scattering theories should be employed to calculate the DCs, using inputs from turbulence transport modelling. Given, however, the large MFPs in the polar regions and the outer heliosphere predicted by most scattering theories for realistic turbulence parameters \citep[see e.g. the comparison in][]{EngelbrechtEA2022a}, it is reasonable to expect the discrepancy between the Parker and focused TPEs to become even more pronounced in this case. Secondly, particle drifts were not included here, as a difference was already identified between the Parker and focused TPEs for diffusion alone. To realistically include drifts, the questions of pitch-angle-dependent neutral sheet drift and turbulent drift reduction\footnote{An initial theoretical and modelling investigation is discussed in \citet{vandenBergEA2021}; however, further work is necessary to verify and refine it.} should be addressed first. Nevertheless, it can be speculated that the difference between the Parker and focused TPEs would be greater for $qA > 0$, where $q$ is the particle charge and $A$ the polarity of the Sun's north pole. This is because positively charged GCRs drift in from the poles and out along the heliospheric current sheet when $qA > 0$, thereby adding to the flux patterns reported here. Although the drift pattern is reversed for $qA < 0$, it might not reduce the reported flux patterns, as the CRs still experience large MFPs in the outer heliosphere and over the poles. Additionally, including pitch-angle-dependent diffusive shock acceleration at the termination shock \citep[see e.g.][]{leRouxEA2007} or non-axisymmetric pitch-angle-dependent diffusion across the heliopause \citep[see e.g.][]{StraussFichtner2014, StraussEA2016} could produce other interesting differences between the solutions of the Parker and focused TPEs. The latter case could be more intricate if the very local interstellar spectrum is also anisotropic, as is evident from Voyager 1 observations of GCR proton anisotropies after crossing the heliopause \citep[see][]{RankinEA2019}. Finally, the differences between the Parker and focused TPE reported by \citet{vandenBerg2023} caused by the interplay between pitch-angle scattering and pitch-angle-dependent perpendicular diffusion when $\kappa_{\perp} \sim \kappa_{\parallel}$, might be evident in any astrophysical scenario where turbulence is isotropic. Unfortunately, the impact of modulation in the heliosheath, at the termination shock, due to drifts, or using more realistic MFPs on the difference between the Parker and focused TPEs cannot be quantified without further numerical modelling.


\section{Summary and conclusion}
\label{sec:Summary}

This work used transport coefficients and heliospheric model parameters commonly employed in GCR modulation studies that solve the Parker TPE for solar minimum conditions and accurately reproduce GCR intensity observations at Earth's orbit. In this model, the Parker TPE yields results that differ from those obtained by solving the FTE under identical conditions. These differences vary with particle energy and position in the heliosphere, reaching $\sim 30\%$ at lower kinetic energies when comparing intensities at Earth, and up to $\sim 40\%$ over the poles at the same energy and radial distance. Such a difference is comparable to the variation in observations between consecutive years caused by solar activity \citep[see e.g.][]{Vos2011, AdrianiEA2013, PotgieterEA2014, Raath2015, MartucciEA2018}, while this specific difference is likely influenced by the particular DCs and transport conditions employed and may not be a fixed value. Nonetheless, this deviation results from small first-order anisotropies that are especially significant at lower energies in the polar regions between $\sim 20 - 60~{\rm au}$. Limited observations of GCR anisotropies, typically at high energies of $> 1~{\rm TeV}$ \citep[see e.g.][]{AmenomoriEA2006, GuillianEA2007, BartoliEA2018, AbeysekaraEA2019, ChakrabortyEA2024, MaalalZhang2025}, suggest that the anisotropies are indeed small \citep[see e.g.][for lower energies $< 1~{\rm GeV}$]{AdrianiEA2015, AjelloEA2019}. Particle fluxes indicate that the anisotropy patterns are caused by inward particle streaming along the HMF in the polar regions, a natural and unavoidable consequence of the Parker HMF model used here. The less-wound magnetic field lines in the polar regions, combined with the larger parallel MFPs there, provide a natural highway for particles to stream into the inner heliosphere. The anisotropy pattern does not match the spatial pattern of the difference between the two TPEs. This is most likely not due to the Parker TPE incorrectly modelling the small anisotropy at a single point, but rather the cumulative effect of incorrect modelling across the entire heliosphere (particularly affecting low-energy particles, which spend more time being modulated). Additionally, the streaming of particles, coupled with weak pitch-angle scattering, cannot be accurately described as effective parallel diffusion \citep[see the discussion in][]{vandenBerg2023}, leading to the Parker TPE becoming over-diffusive \citep[this may partly explain why diffusive transport sometimes fails to account for certain observations; see][for a review]{EffenbergerEA2025}. Furthermore, the FTE produced nearly identical results across different pitch-angle dependencies of the DCs. This suggests that spectral and anisotropy data from GCR observations alone cannot distinguish between various scattering theories for the pitch-angle dependence of DCs that produce similar MFPs. This highlights the importance of testing scattering theories against full-orbit simulation results \citep[see e.g.][]{ElsEA2024, vandenBergEA2024} before their application in TPEs, and that the PAD should be studied both numerically and observationally.

Given the significant difference in GCR intensities at Earth calculated by solving the Parker TPE versus the FTE, it is questionable whether precision studies of GCR modulation, where precision means highly accurate fits of model results to detailed observations, can truly be considered meaningful. Certainly, there is no doubt that DCs can be chosen to precisely match the Parker TPE to observations. However, this work indicates that FTE results will no longer be comparable to observations if the same effective MFPs are used. Therefore, the spatial and energy dependencies, as well as the magnitudes of these effective DCs inferred from the Parker TPE, might not be comparable to even the best theoretical predictions for the DCs. Conversely, fitting observations with the FTE may yield MFPs larger than those inferred from fitting the Parker TPE. This should be taken into account when discussing and applying the \citet{Palmer1982} consensus values \citep[see also][]{BieberEA1994, EngelbrechtEA2022b}. Given these findings, it becomes even less clear when lower-dimensional approximate solutions to the Parker TPE, such as the commonly used force-field approximation \citep[see e.g.][]{GleesonAxford1968, Quenby1984, Moraal2013}, can provide meaningful insights when used to study and draw conclusions about the physics behind GCR modulation, considering their well-defined limitations as solutions to the Parker TPE \citep[see e.g.][]{CaballeroLopezMoraal2004, EngelbrechtDiFelice2020}. Nonetheless, for qualitative studies of GCR modulation, the Parker TPE, which describes the broader physical processes governing particle transport, is expected to produce reliable qualitative results. This, however, is subject to a caveat, especially when the Parker TPE is applied in exotic scenarios such as different astrospheres, where transport conditions can vary significantly from those in the heliosphere. This is especially true for the astrospheres of slowly rotating stars, where the less-wound magnetic field results in a greater contribution of parallel diffusion to the radial DC \citep[see e.g.][]{MesquitaEA2021, EngelbrechtEA2024}, thereby exacerbating the discrepancies discussed here. However, it should be noted that the larger uncertainties inherent in such endeavours \citep[see e.g.][]{HerbstEA2022} may outweigh those caused by using the Parker TPE.

The FTE describes transport processes in greater detail than the Parker TPE. Some of the crudest approximations in the Parker TPE \edited{are} assuming that $\dinf \mu / \dinf t = 0$ and neglecting the interplay between pitch-angle scattering and various pitch-angle dependent processes. Of course, one could try to improve this by including as many modifying terms as possible in the Parker TPE to account for the interplay among processes or by using a higher-order approximation of the FTE, such as a telegraph equation. However, such a TPE would still remain only a more refined approximation of the FTE. These additional DCs also imply that any parallel (perpendicular) DC inferred from fitting the Parker TPE to observations is therefore an `effective parallel (or perpendicular) DC' that includes diffusion contributions from all physical processes, not just from pitch-angle scattering (cross-field diffusion). For precision modelling in the heliosphere, eliminating the uncertainties implicit in the differences between the Parker and focused TPEs requires solving the full FTE. Although solving the FTE is computationally more challenging than the Parker TPE, the hope is that the CR modelling community will embrace the opportunity to deepen their understanding of CR modulation through particle transport modelling at a pitch-angle level. Previous conclusions about the global modulation of GCRs in the heliosphere might need to be revisited. Particularly, the effects of the various processes discussed in the last paragraph of Sect.~\ref{sec:Discussion}, which have not been included in the current study, should be explored, and their influence on the differences between the Parker and focused TPE results should be quantified.


\begin{acknowledgements}

JPvdB acknowledges partial support from the South African National Space Agency during the initial theoretical work. JPvdB also thanks J. Light for discussions on \edited{finite-difference} methods. This research was supported by the International Space Science Institute (ISSI) in Bern, through ISSI International Team project No. 24-608. \edited{The authors thank the anonymous reviewer for their comments.} The C code used the permuted congruential generator \citep{ONeill2014} and Open MPI \citep{GabrielEA2004}. Figures prepared with Matplotlib \citep{Hunter2007} and certain calculations done with NumPy \citep{HarrisEA2020} and SciPy \citep{VirtanenEA2020}. \edited{Claude (Sonnet 4.6; free plan) was used to organise parts of the manuscript and to identify where explanations are required. Grammarly (pro plan) was used to improve the language of the manuscript.}

\end{acknowledgements}

%

\bibliographystyle{aa}
\bibliography{References}

\begin{thebibliography}{121}
\expandafter\ifx\csname natexlab\endcsname\relax\def\natexlab#1{#1}\fi

\bibitem[{Abeysekara {et~al.}(2019)Abeysekara, Alfaro, Alvarez, Arceo,
  {Arteaga-Vel{\'a}zquez}, {Avila Rojas}, {Belmont-Moreno}, {BenZvi}, Brisbois,
  Capistr{\'a}n, Carramiana, Casanova, Cotti, Cotzomi, {D{\'\i}az-V{\'e}lez},
  {De Le{\'o}n}, {De la Fuente}, Dichiara, {DuVernois}, Espinoza, Fiorino,
  Fleischhack, Fraija, {Galv{\'a}n-G{\'a}mez}, {Garc{\'\i}a-Gonz{\'a}lez},
  Gonz{\'a}lez, Goodman, {Hampel-Arias}, Harding, Hernandez, Hona,
  {Hueyotl-Zahuantitla}, Iriarte, {Jardin-Blicq}, Joshi, Lara, {Le{\'o}n
  Vargas}, {Luis-Raya}, Malone, Marinelli, {Mart{\'\i}nez-Castro}, Martinez,
  Matthews, {Miranda-Romagnoli}, Moreno, Mostaf{\'a}, Nellen, Newbold, Nisa,
  {Noriega-Papaqui}, {P{\'e}rez-P{\'e}rez}, Pretz, Ren, Rho, Rivi{\`e}re,
  {Rosa-Gonz{\'a}lez}, Rosenberg, Salazar, {Salesa Greus}, Sandoval, Schneider,
  Schoorlemmer, Sinnis, Smith, Surajbali, Taboada, Tollefson, Torres,
  Villaseor, Weisgarber, Wood, Zepeda, Zhou, {\'A}lvarez, {HAWC Collaboration},
  Aartsen, Ackermann, Adams, Aguilar, Ahlers, Ahrens, Altmann, Andeen,
  Anderson, Ansseau, Anton, Arg{\"u}elles, Auffenberg, Axani, Backes,
  Bagherpour, Bai, Barbano, Barron, Barwick, Baum, Bay, Beatty, {Becker Tjus},
  Becker, {BenZvi}, Berley, Bernardini, Besson, Binder, Bindig, Blaufuss, Blot,
  Bohm, B{\"o}rner, Bos, B{\"o}ser, Botner, Bourbeau, Bourbeau, Bradascio,
  Braun, Bretz, Bron, {Brostean-Kaiser}, Burgman, Busse, Carver, Cheung,
  Chirkin, Clark, Classen, Collin, Conrad, Coppin, Correa, Cowen, Cross, Dave,
  Day, {de Andr{\'e}}, {De Clercq}, {DeLaunay}, Dembinski, Deoskar, {De
  Ridder}, Desiati, {de Vries}, {de Wasseige}, {de With}, {DeYoung},
  {D{\'\i}az-V{\'e}lez}, Dujmovic, Dunkman, Dvorak, Eberhardt, Ehrhardt,
  Eichmann, Eller, Evenson, Fahey, Fazely, Felde, Filimonov, Finley,
  Franckowiak, Friedman, Fritz, Gaisser, Gallagher, Ganster, Garrappa,
  Gerhardt, Ghorbani, Giang, Glauch, Gl{\"u}senkamp, Goldschmidt, Gonzalez,
  Grant, Griffith, Haack, Hallgren, Halve, Halzen, Hanson, Hebecker, Heereman,
  Helbing, Hellauer, Hickford, Hignight, Hill, Hoffman, Hoffmann, Hoinka,
  {Hokanson-Fasig}, Hoshina, Huang, Huber, Hultqvist, H{\"u}nnefeld, Hussain,
  In, \& Iovine}]{AbeysekaraEA2019}
Abeysekara, A.~U., Alfaro, R., Alvarez, C., {et~al.} 2019, \apj, 871, 96

\bibitem[{Adriani {et~al.}(2013)Adriani, Barbarino, Bazilevskaya, Bellotti,
  Boezio, Bogomolov, Bongi, Bonvicini, Borisov, Bottai, Bruno, Cafagna,
  Campana, Carbone, Carlson, Casolino, Castellini, {De Pascale}, {De Santis},
  {De Simone}, {Di Felice}, Formato, Galper, Grishantseva, Karelin, Koldashov,
  Koldobskiy, Krutkov, Kvashnin, Leonov, Malakhov, Marcelli, Mayorov, Menn,
  Mikhailov, Mocchiutti, Monaco, Mori, Nikonov, Osteria, Palma, Papini, Pearce,
  Picozza, Pizzolotto, Ricci, Ricciarini, Rossetto, Sarkar, Simon, Sparvoli,
  Spillantini, Stozhkov, Vacchi, Vannuccini, Vasilyev, Voronov, Yurkin, Wu,
  Zampa, Zampa, Zverev, Potgieter, \& Vos}]{AdrianiEA2013}
Adriani, O., Barbarino, G.~C., Bazilevskaya, G.~A., {et~al.} 2013, \apj, 765,
  91

\bibitem[{Adriani {et~al.}(2015)Adriani, Barbarino, Bazilevskaya, Bellotti,
  Boezio, Bogomolov, Bongi, Bonvicini, Bottai, Bruno, Cafagna, Campana,
  Carlson, Casolino, Castellini, {De Donato}, {De Santis}, {De Simone}, {Di
  Felice}, Formato, Galper, Giaccari, Karelin, Koldashov, Koldobskiy, Krutkov,
  Kvashnin, Leonov, Malakhov, Marcelli, Martucci, Mayorov, Menn, Merg{\'e},
  Mikhailov, Mocchiutti, Monaco, Mori, Munini, Osteria, Palma, Panico, Papini,
  Pearce, Picozza, Ricci, Ricciarini, Sarkar, Scotti, Simon, Sparvoli,
  Spillantini, Stozhkov, Vacchi, Vannuccini, Vasilyev, Voronov, Yurkin, Zampa,
  \& Zampa}]{AdrianiEA2015}
Adriani, O., Barbarino, G.~C., Bazilevskaya, G.~A., {et~al.} 2015, \apj, 811,
  21

\bibitem[{Agueda \& Vainio(2013)}]{AguedaVainio2013}
Agueda, N. \& Vainio, R. 2013, J. Space Weather Space Clim., 3, A10

\bibitem[{Agueda {et~al.}(2008)Agueda, Vainio, Lario, \&
  Sanahuja}]{AguedaEA2008}
Agueda, N., Vainio, R., Lario, D., \& Sanahuja, B. 2008, \apj, 675, 1601

\bibitem[{Ajello {et~al.}(2019)Ajello, Baldini, Barbiellini, Bastieri, Bechtol,
  Bellazzini, Bissaldi, Blandford, Bonino, Bottacini, Brandt, Bruel, Buson,
  Cameron, Caputo, Cavazzuti, Chen, Chiaro, Ciprini, {Cohen-Tanugi}, Costantin,
  Cuoco, Cutini, {D'Ammando}, {de la Torre Luque}, {de Palma}, Desai, Digel,
  {Di Lalla}, {Di Venere}, Dom{\'\i}nguez, Fegan, Fukazawa, Funk, Fusco,
  Gargano, Gasparrini, Giglietto, Giordano, Giroletti, Green, Grenier, Guiriec,
  Hayashi, Hays, Hewitt, Horan, J{\'o}hannesson, Kuss, Latronico, Li, Liodakis,
  Longo, Loparco, Lubrano, Maldera, Manfreda, {Mart{\'\i}-Devesa}, Mazziotta,
  Meehan, Mereu, Meyer, Michelson, Mirabal, Mitthumsiri, Mizuno, Morselli,
  Negro, Nuss, Omodei, Orienti, Orlando, Paliya, Paneque, Persic,
  {Pesce-Rollins}, Piron, Porter, Principe, Rain{\`o}, Rando, Razzano,
  Razzaque, Reimer, Reimer, Serini, Sgr{\`o}, Siskind, Spandre, Spinelli,
  Suson, Tajima, Thayer, Torres, Troja, Vandenbroucke, Yassine, Zimmer, \&
  {Fermi-LAT Collaboration}}]{AjelloEA2019}
Ajello, M., Baldini, L., Barbiellini, G., {et~al.} 2019, \apj, 883, 33

\bibitem[{Amenomori {et~al.}(2006)Amenomori, Ayabe, Bi, Chen, Cui,
  Danzengluobu, Ding, Ding, Feng, Feng, Feng, Gao, Geng, Guo, He, He, Hibino,
  Hotta, Hu, Hu, Huang, Huang, Jia, Kajino, Kasahara, Katayose, Kato, Kawata,
  Labaciren, Le, Li, Li, Lou, Lu, Lu, Meng, Mizutani, Mu, Munakata, Nagai,
  Nanjo, Nishizawa, Ohnishi, Ohta, Onuma, Ouchi, Ozawa, Ren, Saito, Saito,
  Sakata, Sako, Sasaki, Shibata, Shiomi, Shirai, Sugimoto, Takita, Tan,
  Tateyama, Torii, Tsuchiya, Udo, Wang, Wang, Wang, Wang, Wu, Xue, Yamamoto,
  Yan, Yang, Yasue, Ye, Yu, Yuan, Yuda, Zhang, Zhang, Zhang, Zhang, Zhang,
  Zhang, Zhaxisangzhu, Zhou, \& {Tibet AS{\ensuremath{\gamma}}
  Collaboration}}]{AmenomoriEA2006}
Amenomori, M., Ayabe, S., Bi, X.~J., {et~al.} 2006, Sci., 314, 439

\bibitem[{Bartoli {et~al.}(2018)Bartoli, Bernardini, Bi, Cao, Catalanotti,
  Chen, Chen, Cui, Dai, {D'Amone}, Danzengluobu, {De Mitri}, {D'Ettorre
  Piazzoli}, {Di Girolamo}, {Di Sciascio}, Feng, Feng, Gao, Gou, Guo, He, Hu,
  Hu, Iacovacci, Iuppa, Jia, Labaciren, Li, Liu, Liu, Liu, Lu, Ma, Ma,
  Mancarella, Mari, Marsella, Mastroianni, Montini, Ning, Perrone, Pistilli,
  Ruffolo, Salvini, Santonico, Shen, Sheng, Shi, Surdo, Tan, Vallania,
  Vernetto, Vigorito, Wang, Wu, Wu, Xue, Yang, Yang, Yao, Yuan, Zha, Zhang,
  Zhang, Zhang, Zhang, Zhao, Zhaxiciren, Zhaxisangzhu, Zhou, Zhu, Zhu, \&
  {ARGO-YBJ Collaboration}}]{BartoliEA2018}
Bartoli, B., Bernardini, P., Bi, X.~J., {et~al.} 2018, \apj, 861, 93

\bibitem[{Beeck \& Wibberenz(1986)}]{BeeckWibberenz1986}
Beeck, J. \& Wibberenz, G. 1986, \apj, 311, 437

\bibitem[{Bieber {et~al.}(1994)Bieber, Matthaeus, Smith, Wanner, Kallenrode, \&
  Wibberenz}]{BieberEA1994}
Bieber, J., Matthaeus, W.~H., Smith, C.~W., {et~al.} 1994, \apj, 420, 294

\bibitem[{Breech {et~al.}(2005)Breech, Matthaeus, Minnie, Oughton, Parhi,
  Bieber, \& Bavassano}]{BreechEA2005}
Breech, B., Matthaeus, W.~H., Minnie, J., {et~al.} 2005, \grl, 32, L06103

\bibitem[{Burger {et~al.}(2008)Burger, Kr{\"u}ger, Hitge, \&
  Engelbrecht}]{BurgerEA2008}
Burger, R.~A., Kr{\"u}ger, T.~P.~J., Hitge, M., \& Engelbrecht, N.~E. 2008,
  \apj, 674, 511

\bibitem[{Burger {et~al.}(1985)Burger, Moraal, \& Webb}]{BurgerEA1985}
Burger, R.~A., Moraal, H., \& Webb, G.~M. 1985, \apss, 116, 107

\bibitem[{Burger {et~al.}(2000)Burger, Potgieter, \& Heber}]{BurgerEA2000}
Burger, R.~A., Potgieter, M.~S., \& Heber, B. 2000, \jgr, 105, 27447

\bibitem[{{Caballero-Lopez} \& Moraal(2004)}]{CaballeroLopezMoraal2004}
{Caballero-Lopez}, R.~A. \& Moraal, H. 2004, \jgr (Space Phys.), 109, A01101

\bibitem[{Chakraborty {et~al.}(2024)Chakraborty, Ahmad, Chandra, Dugad,
  Goswami, Gupta, Hariharan, Hayashi, Jagadeesan, Jain, Jain, Kawakami, Koi,
  Kojima, Mahapatra, Mohanty, Moharana, Muraki, Nakamura, Nayak, Nonaka,
  Oshima, Pant, Pattanaik, Paul, Pradhan, Rameez, Ramesh, Saha, Sahoo, Scaria,
  Shibata, Tabata, Takamaru, Tanaka, Varsi, Yamazaki, \&
  Zuberi}]{ChakrabortyEA2024}
Chakraborty, M., Ahmad, S., Chandra, A., {et~al.} 2024, \apj, 961, 87

\bibitem[{Chenette {et~al.}(1977)Chenette, Conlon, Pyle, \&
  Simpson}]{ChenetteEA1977}
Chenette, D.~L., Conlon, T.~F., Pyle, K.~R., \& Simpson, J.~A. 1977, \apjl,
  215, L95

\bibitem[{Dunzlaff {et~al.}(2010)Dunzlaff, Kopp, \& Heber}]{DunzlaffEA2010}
Dunzlaff, P., Kopp, A., \& Heber, B. 2010, \jgr (Space Phys.), 115, A10106

\bibitem[{Effenberger \& Litvinenko(2014)}]{EffenbergerLitvinenko2014}
Effenberger, F. \& Litvinenko, Y.~E. 2014, \apj, 783, 15

\bibitem[{Effenberger {et~al.}(2025)Effenberger, Walter, Fichtner, Aerdker,
  Grauer, Laitinen, {le Roux}, Litvinenko, L{\"u}bke, Perri, Reichherzer,
  Shalchi, {van den Berg}, \& Zimbardo}]{EffenbergerEA2025}
Effenberger, F., Walter, D., Fichtner, H., {et~al.} 2025, \ssr, 221, 75

\bibitem[{Els {et~al.}(2024)Els, Engelbrecht, Lang, \& Strauss}]{ElsEA2024}
Els, P.~L., Engelbrecht, N.~E., Lang, J.~T., \& Strauss, R.~D. 2024, \apj, 975,
  134

\bibitem[{Engelbrecht(2017)}]{Engelbrecht2017}
Engelbrecht, N.~E. 2017, \apjl, 849, L15

\bibitem[{Engelbrecht(2019)}]{Engelbrecht2019}
Engelbrecht, N.~E. 2019, \apj, 880, 60

\bibitem[{Engelbrecht(2024)}]{Engelbrecht2024}
Engelbrecht, N.~E. 2024, \apj, 975, 227

\bibitem[{Engelbrecht \& Burger(2013{\natexlab{a}})}]{EngelbrechtBurger2013a}
Engelbrecht, N.~E. \& Burger, R.~A. 2013{\natexlab{a}}, \apj, 772, 46

\bibitem[{Engelbrecht \& Burger(2013{\natexlab{b}})}]{EngelbrechtBurger2013b}
Engelbrecht, N.~E. \& Burger, R.~A. 2013{\natexlab{b}}, \apj, 779, 158

\bibitem[{Engelbrecht \& {Di Felice}(2020)}]{EngelbrechtDiFelice2020}
Engelbrecht, N.~E. \& {Di Felice}, V. 2020, \prd, 102, 103007

\bibitem[{Engelbrecht {et~al.}(2022{\natexlab{a}})Engelbrecht, Effenberger,
  Florinski, Potgieter, Ruffolo, Chhiber, Usmanov, Rankin, \&
  Els}]{EngelbrechtEA2022a}
Engelbrecht, N.~E., Effenberger, F., Florinski, V., {et~al.}
  2022{\natexlab{a}}, \ssr, 218, 33

\bibitem[{Engelbrecht {et~al.}(2024)Engelbrecht, Herbst, Strauss, Scherer,
  Light, \& Moloto}]{EngelbrechtEA2024}
Engelbrecht, N.~E., Herbst, K., Strauss, R.~D.~T., {et~al.} 2024, \apj, 964, 89

\bibitem[{Engelbrecht {et~al.}(2022{\natexlab{b}})Engelbrecht, Vogt, Herbst,
  Strauss, \& Burger}]{EngelbrechtEA2022b}
Engelbrecht, N.~E., Vogt, A., Herbst, K., Strauss, R.~D.~T., \& Burger, R.~A.
  2022{\natexlab{b}}, \apj, 929, 8

\bibitem[{Ferreira {et~al.}(2001)Ferreira, Potgieter, Burger, Heber, \&
  Fichtner}]{FerreiraEA2001}
Ferreira, S.~E.~S., Potgieter, M.~S., Burger, R.~A., Heber, B., \& Fichtner, H.
  2001, \jgr, 106, 24979

\bibitem[{Gabriel {et~al.}(2004)Gabriel, Fagg, Bosilca, Angskun, Dongarra,
  Squyres, Sahay, Kambadur, Barrett, Lumsdaine, Castain, Daniel, Graham, \&
  Woodall}]{GabrielEA2004}
Gabriel, E., Fagg, G.~E., Bosilca, G., {et~al.} 2004, in Proc. 11th European
  PVM/MPI Users' Group Meeting, Budapest, Hungary, 97

\bibitem[{Gardiner(1994)}]{Gardiner1994}
Gardiner, C.~W. 1994, {Handbook of stochastic methods for physics, chemistry
  and the natural sciences} (Springer-Verlag)

\bibitem[{Giacalone \& Jokipii(1999)}]{GiacaloneJokipii1999}
Giacalone, J. \& Jokipii, J.~R. 1999, \apj, 520, 204

\bibitem[{Gleeson \& Axford(1968)}]{GleesonAxford1968}
Gleeson, L.~J. \& Axford, W.~I. 1968, \apj, 154, 1011

\bibitem[{Gleeson \& Urch(1971)}]{GleesonUrch1971}
Gleeson, L.~J. \& Urch, I.~H. 1971, \apss, 11, 288

\bibitem[{Gradshteyn \& Ryzhik(2007)}]{GradshteynRyzhik2007}
Gradshteyn, I.~S. \& Ryzhik, I.~M. 2007, {Table of integrals, series, and
  products}, 7th edn., ed. A.~Jeffrey \& D.~Zwillinger (Academic Press)

\bibitem[{Guillian {et~al.}(2007)Guillian, Hosaka, Ishihara, Kameda, Koshio,
  Minamino, Mitsuda, Miura, Moriyama, Nakahata, Namba, Obayashi, Ogawa,
  Shiozawa, Suzuki, Takeda, Takeuchi, Yamada, Higuchi, Ishitsuka, Kajita,
  Kaneyuki, Mitsuka, Nakayama, Nishino, Okada, Okumura, Saji, Takenaga, Desai,
  Kearns, Stone, Sulak, Wang, Goldhaber, Casper, Gajewski, Griskevich, Kropp,
  Liu, Mine, Smy, Sobel, Vagins, Ganezer, Hill, Keig, Scholberg, Walter,
  Ellsworth, Tasaka, Kibayashi, Learned, Matsuno, Messier, Hayato, Ichikawa,
  Ishida, Ishii, Iwashita, Kobayashi, Nakadaira, Nakamura, Nitta, Oyama,
  Totsuka, Suzuki, Hasegawa, Kato, Maesaka, Nakaya, Nishikawa, Sato, Yamamoto,
  Yokoyama, Haines, Dazeley, Hatakeyama, Svoboda, Blaufuss, Goodman, Sullivan,
  Turcan, Habig, Fukuda, Itow, Sakuda, Yoshida, Kim, Yoo, Okazawa, Ishizuka,
  Jung, Kato, Kobayashi, Malek, Mauger, {McGrew}, Sharkey, Yanagisawa, Gando,
  Hasegawa, Inoue, Shirai, Suzuki, Nishijima, Ishino, Watanabe, Koshiba,
  Kielczewska, Berns, Gran, Shiraishi, Stachyra, Washburn, Wilkes, \&
  Munakata}]{GuillianEA2007}
Guillian, G., Hosaka, J., Ishihara, K., {et~al.} 2007, \prd, 75, 062003

\bibitem[{Harris {et~al.}(2020)Harris, Millman, {van der Walt}, Gommers,
  Virtanen, Cournapeau, Wieser, Taylor, Berg, Smith, Kern, Picus, Hoyer, {van
  Kerkwijk}, Brett, Haldane, {del R{\'\i}o}, Wiebe, Peterson,
  {G{\'e}rard-Marchant}, Sheppard, Reddy, Weckesser, Abbasi, Gohlke, \&
  Oliphant}]{HarrisEA2020}
Harris, C.~R., Millman, K.~J., {van der Walt}, S.~J., {et~al.} 2020, \nat, 585,
  357

\bibitem[{Hasselmann \& Wibberenz(1970)}]{HasselmannWibberenz1970}
Hasselmann, K. \& Wibberenz, G. 1970, \apj, 162, 1049

\bibitem[{He \& Schlickeiser(2014)}]{HeSchlickeiser2014}
He, H.-Q. \& Schlickeiser, R. 2014, \apj, 792, 85

\bibitem[{Herbst {et~al.}(2022)Herbst, Baalmann, Bykov, Engelbrecht, Ferreira,
  Izmodenov, Korolkov, Levenfish, Linsky, Meyer, Scherer, \&
  Strauss}]{HerbstEA2022}
Herbst, K., Baalmann, L.~R., Bykov, A., {et~al.} 2022, \ssr, 218, 29

\bibitem[{Hunter(2007)}]{Hunter2007}
Hunter, J.~D. 2007, Computing Sci. Eng., 9, 90

\bibitem[{Isenberg(1997)}]{Isenberg1997}
Isenberg, P.~A. 1997, \jgr, 102, 4719

\bibitem[{Isenberg \& Jokipii(1979)}]{IsenbergJokipii1979}
Isenberg, P.~A. \& Jokipii, J.~R. 1979, \apj, 234, 746

\bibitem[{Jokipii \& Parker(1970)}]{JokipiiParker1970}
Jokipii, J.~R. \& Parker, E.~N. 1970, \apj, 160, 735

\bibitem[{Kloeden \& Platen(1995)}]{KloedenPlaten1995}
Kloeden, P.~E. \& Platen, E. 1995, {Numerical solution of stochastic
  differential equations}, 2nd edn. (Springer-Verlag)

\bibitem[{Kopp {et~al.}(2012)Kopp, B{\"u}sching, Strauss, \&
  Potgieter}]{KoppEA2012}
Kopp, A., B{\"u}sching, I., Strauss, R.~D., \& Potgieter, M.~S. 2012, Comp.
  Phys. Comm., 183, 530

\bibitem[{K{\'o}ta(2013)}]{Kota2013}
K{\'o}ta, J. 2013, \ssr, 176, 391

\bibitem[{Lampa(2011)}]{Lampa2011}
Lampa, F. 2011, PhD thesis, University of Osnabrück, Germany

\bibitem[{Lang {et~al.}(2024)Lang, Strauss, Engelbrecht, {van den Berg},
  Dresing, Ruffolo, \& Bandyopadhyay}]{LangEA2024}
Lang, J.~T., Strauss, R.~D., Engelbrecht, N.~E., {et~al.} 2024, \apj, 971, 105

\bibitem[{le~{Roux} \& Webb(2012)}]{leRouxWebb2012}
le~{Roux}, J.~A. \& Webb, G.~M. 2012, \apj, 746, 104

\bibitem[{le~{Roux} {et~al.}(2007)le~{Roux}, Webb, Florinski, \&
  Zank}]{leRouxEA2007}
le~{Roux}, J.~A., Webb, G.~M., Florinski, V., \& Zank, G.~P. 2007, \apj, 662,
  350

\bibitem[{le~{Roux} {et~al.}(2014)le~{Roux}, Webb, \& Ye}]{leRouxEA2014}
le~{Roux}, J.~A., Webb, G.~M., \& Ye, J. 2014, in Astro. Society Pacific Conf.
  Series, Vol. 484, Outstanding problems in heliophysics: From coronal heating
  to the edge of the heliosphere, ed. Q.~Hu \& G.~P. Zank, 110

\bibitem[{Litvinenko \& Noble(2013)}]{LitvinenkoNoble2013}
Litvinenko, Y.~E. \& Noble, P.~L. 2013, \apj, 765, 31

\bibitem[{Litvinenko \& Schlickeiser(2013)}]{LitvinenkoSchlickeiser2013}
Litvinenko, Y.~E. \& Schlickeiser, R. 2013, \aap, 554, A59

\bibitem[{Maalal \& Zhang(2025)}]{MaalalZhang2025}
Maalal, N.~D. \& Zhang, M. 2025, \apj, 992, 46

\bibitem[{Malkov(2017)}]{Malkov2017}
Malkov, M.~A. 2017, \prd, 95, 023007

\bibitem[{Malkov(2018)}]{Malkov2018}
Malkov, M.~A. 2018, Nuc. Part. Phys. Proc., 297-299, 152

\bibitem[{Martucci {et~al.}(2018)Martucci, Munini, Boezio, {Di Felice},
  Adriani, Barbarino, Bazilevskaya, Bellotti, Bongi, Bonvicini, Bottai, Bruno,
  Cafagna, Campana, Carlson, Casolino, Castellini, {De Santis}, Galper,
  Karelin, Koldashov, Koldobskiy, Krutkov, Kvashnin, Leonov, Malakhov,
  Marcelli, Marcelli, Mayorov, Menn, Merg{\`e}, Mikhailov, Mocchiutti, Monaco,
  Mori, Osteria, Panico, Papini, Pearce, Picozza, Ricci, Ricciarini, Simon,
  Sparvoli, Spillantini, Stozhkov, Vacchi, Vannuccini, Vasilyev, Voronov,
  Yurkin, Zampa, Zampa, Potgieter, \& Raath}]{MartucciEA2018}
Martucci, M., Munini, R., Boezio, M., {et~al.} 2018, \apjl, 854, L2

\bibitem[{Matthaeus {et~al.}(2003)Matthaeus, Qin, Bieber, \&
  Zank}]{MatthaeusEA2003}
Matthaeus, W.~H., Qin, G., Bieber, J.~W., \& Zank, G.~P. 2003, \apjl, 590, L53

\bibitem[{{McComas} {et~al.}(2000){McComas}, Barraclough, Funsten, Gosling,
  {Santiago-Mu{\~n}oz}, Skoug, Goldstein, Neugebauer, Riley, \&
  Balogh}]{McComasEA2000}
{McComas}, D.~J., Barraclough, B.~L., Funsten, H.~O., {et~al.} 2000, \jgr, 105,
  10419

\bibitem[{{McKibben} {et~al.}(2007){McKibben}, Zhang, Heber, Kunow, \&
  Sanderson}]{McKibbenEA2007}
{McKibben}, R.~B., Zhang, M., Heber, B., Kunow, H., \& Sanderson, T.~R. 2007,
  \planss, 55, 21

\bibitem[{Mesquita {et~al.}(2021)Mesquita, {Rodgers-Lee}, \&
  Vidotto}]{MesquitaEA2021}
Mesquita, A.~L., {Rodgers-Lee}, D., \& Vidotto, A.~A. 2021, \mnras, 505, 1817

\bibitem[{Minnie {et~al.}(2007)Minnie, Bieber, Matthaeus, \&
  Burger}]{MinnieEA2007}
Minnie, J., Bieber, J.~W., Matthaeus, W.~H., \& Burger, R.~A. 2007, \apj, 670,
  1149

\bibitem[{Moloto {et~al.}(2018)Moloto, Engelbrecht, \& Burger}]{MolotoEA2018}
Moloto, K.~D., Engelbrecht, N.~E., \& Burger, R.~A. 2018, \apj, 859, 107

\bibitem[{Moraal(2013)}]{Moraal2013}
Moraal, H. 2013, \ssr, 176, 299

\bibitem[{Moses(1987)}]{Moses1987}
Moses, D. 1987, \apj, 313, 471

\bibitem[{O'Neill(2014)}]{ONeill2014}
O'Neill, M.~E. 2014, {PCG: A family of simple fast space-efficient
  statistically good algorithms for random number generation}, Tech. Rep.
  HMC-CS-2014-0905, Harvey Mudd College, Claremont, CA

\bibitem[{Owens \& Forsyth(2013)}]{OwensForsyth2013}
Owens, M.~J. \& Forsyth, R.~J. 2013, Living Rev. Solar Phys., 10, 5

\bibitem[{Palmer(1982)}]{Palmer1982}
Palmer, I.~D. 1982, Rev. Geophys. Space Phys., 20, 335

\bibitem[{Parker(1958)}]{Parker1958}
Parker, E.~N. 1958, \apj, 128, 664

\bibitem[{Parker(1965)}]{Parker1965}
Parker, E.~N. 1965, \planss, 13, 9

\bibitem[{Parker(1967)}]{Parker1967}
Parker, E.~N. 1967, \planss, 15, 1723

\bibitem[{Pei {et~al.}(2010)Pei, Bieber, Burger, \& Clem}]{PeiEA2010}
Pei, C., Bieber, J.~W., Burger, R.~A., \& Clem, J. 2010, \jgr (Space Phys.),
  115, A12107

\bibitem[{Potgieter(2013)}]{Potgieter2013}
Potgieter, M.~S. 2013, Living Rev. Solar Phys., 10, 3

\bibitem[{Potgieter {et~al.}(2014)Potgieter, Vos, Boezio, {De Simone}, {Di
  Felice}, \& Formato}]{PotgieterEA2014}
Potgieter, M.~S., Vos, E.~E., Boezio, M., {et~al.} 2014, \solphys, 289, 391

\bibitem[{Press {et~al.}(1992)Press, Teukolsky, Vetterling, \&
  Flannery}]{PressEA1992}
Press, W.~H., Teukolsky, S.~A., Vetterling, W.~T., \& Flannery, B.~P. 1992,
  {Numerical recipes in C: The art of scientific computing} (Cambridge
  University Press)

\bibitem[{Pyle \& Simpson(1977)}]{PyleSimpson1977}
Pyle, K.~R. \& Simpson, J.~A. 1977, \apjl, 215, L89

\bibitem[{Qin(2007)}]{Qin2007}
Qin, G. 2007, \apj, 656, 217

\bibitem[{Qin \& Shalchi(2014)}]{QinShalchi2014}
Qin, G. \& Shalchi, A. 2014, Applied Phys. Res., 6, 1

\bibitem[{Quenby(1984)}]{Quenby1984}
Quenby, J.~J. 1984, \ssr, 37, 201

\bibitem[{Raath(2015)}]{Raath2015}
Raath, J.~L. 2015, {A comparative study of cosmic ray modulation models}

\bibitem[{Rankin {et~al.}(2022)Rankin, Bindi, Bykov, Cummings, {Della Torre},
  Florinski, Heber, Potgieter, Stone, \& Zhang}]{RankinEA2022}
Rankin, J.~S., Bindi, V., Bykov, A.~M., {et~al.} 2022, \ssr, 218, 42

\bibitem[{Rankin {et~al.}(2019)Rankin, Stone, Cummings, {McComas}, Lal, \&
  Heikkila}]{RankinEA2019}
Rankin, J.~S., Stone, E.~C., Cummings, A.~C., {et~al.} 2019, \apj, 873, 46

\bibitem[{Roelof(1969)}]{Roelof1969}
Roelof, E.~C. 1969, in Lec. High-Energy Astrophys., ed. H.~{\"O}gelman \& J.~R.
  Wayland, 111

\bibitem[{Rossi \& Olbert(1970)}]{RossiOlbert1970}
Rossi, B. \& Olbert, S. 1970, {Introduction to the physics of space.}
  (McGraw-Hill)

\bibitem[{Ruffolo(1995)}]{Ruffolo1995}
Ruffolo, D. 1995, \apj, 442, 861

\bibitem[{Ruffolo {et~al.}(2006)Ruffolo, Chuychai, \&
  Matthaeus}]{RuffoloEA2006}
Ruffolo, D., Chuychai, P., \& Matthaeus, W.~H. 2006, \apj, 644, 971

\bibitem[{Ruffolo {et~al.}(2008)Ruffolo, Chuychai, Wongpan, Minnie, Bieber, \&
  Matthaeus}]{RuffoloEA2008}
Ruffolo, D., Chuychai, P., Wongpan, P., {et~al.} 2008, \apj, 686, 1231

\bibitem[{Schlickeiser(2002)}]{Schlickeiser2002}
Schlickeiser, R. 2002, {Cosmic ray astrophysics} (Springer-Verlag)

\bibitem[{Schlickeiser(2011)}]{Schlickeiser2011}
Schlickeiser, R. 2011, \apj, 732, 96

\bibitem[{Shalchi(2010)}]{Shalchi2010}
Shalchi, A. 2010, \apjl, 720, L127

\bibitem[{Shalchi(2020)}]{Shalchi2020}
Shalchi, A. 2020, \ssr, 216, 23

\bibitem[{Sheeley {et~al.}(1997)Sheeley, Wang, Hawley, Brueckner, Dere, Howard,
  Koomen, Korendyke, Michels, Paswaters, Socker, {St. Cyr}, Wang, Lamy,
  Llebaria, Schwenn, Simnett, Plunkett, \& Biesecker}]{SheeleyEA1997}
Sheeley, N.~R., Wang, Y.-M., Hawley, S.~H., {et~al.} 1997, \apj, 484, 472

\bibitem[{Skilling(1971)}]{Skilling1971}
Skilling, J. 1971, \apj, 170, 265

\bibitem[{Smith {et~al.}(1976)Smith, Tsurutani, Chenette, Conlon, \&
  Simpson}]{SmithEA1976}
Smith, E.~J., Tsurutani, B.~T., Chenette, D.~L., Conlon, T.~F., \& Simpson,
  J.~A. 1976, \jgr, 81, 65

\bibitem[{Steenkamp(1995)}]{Steenkamp1995}
Steenkamp, R. 1995, PhD thesis, Potchefstroom University for Christian Higher
  Education

\bibitem[{Stone {et~al.}(2013)Stone, Cummings, McDonald, Heikkila, Lal, \&
  Webber}]{StoneEA2013}
Stone, E.~C., Cummings, A.~C., McDonald, F.~B., {et~al.} 2013, Sci., 341, 150

\bibitem[{Strauss {et~al.}(2024)Strauss, Dresing, Engelbrecht, Mitchell,
  K{\"u}hl, Jensen, Fleth, {S{\'a}nchez-Cano}, Posner, Rankin, Lee, {van den
  Berg}, Ferreira, \& Heber}]{StraussEA2024}
Strauss, R.~D., Dresing, N., Engelbrecht, N.~E., {et~al.} 2024, \apj, 961, 57

\bibitem[{Strauss \& Fichtner(2014)}]{StraussFichtner2014}
Strauss, R.~D. \& Fichtner, H. 2014, \aap, 572, L3

\bibitem[{Strauss \& Fichtner(2015)}]{StraussFichtner2015}
Strauss, R.~D. \& Fichtner, H. 2015, \apj, 801, 29

\bibitem[{Strauss {et~al.}(2016)Strauss, {le Roux}, Engelbrecht, Ruffolo, \&
  Dunzlaff}]{StraussEA2016}
Strauss, R.~D., {le Roux}, J.~A., Engelbrecht, N.~E., Ruffolo, D., \& Dunzlaff,
  P. 2016, \apj, 825, 43

\bibitem[{Strauss {et~al.}(2011)Strauss, Potgieter, \&
  Ferreira}]{StraussEA2011}
Strauss, R.~D., Potgieter, M.~S., \& Ferreira, S.~E.~S. 2011, Adv. Space Res.,
  48, 65

\bibitem[{Strauss {et~al.}(2010)Strauss, Potgieter, Ferreira, \&
  Hill}]{StraussEA2010}
Strauss, R.~D., Potgieter, M.~S., Ferreira, S.~E.~S., \& Hill, M.~E. 2010,
  \aap, 522, A35

\bibitem[{Strauss {et~al.}(2022)Strauss, {van den Berg}, \&
  Rankin}]{StraussEA2022}
Strauss, R.~D., {van den Berg}, J.~P., \& Rankin, J.~S. 2022, \apj, 928, 22

\bibitem[{Strauss \& Effenberger(2017)}]{StraussEffenberger2017}
Strauss, R.~D.~T. \& Effenberger, F. 2017, \ssr, 212, 151

\bibitem[{van {den Berg}(2023)}]{vandenBerg2023}
van {den Berg}, J.~P. 2023, PhD thesis, North-West University, South Africa

\bibitem[{{van den Berg} {et~al.}(2024){van den Berg}, Els, \&
  Engelbrecht}]{vandenBergEA2024}
{van den Berg}, J.~P., Els, P.~L., \& Engelbrecht, N.~E. 2024, \apj, 977, 174

\bibitem[{van {den Berg} {et~al.}(2021)van {den Berg}, Engelbrecht, Wijsen, \&
  Strauss}]{vandenBergEA2021}
van {den Berg}, J.~P., Engelbrecht, N.~E., Wijsen, N., \& Strauss, R.~D. 2021,
  \apj, 922, 200

\bibitem[{van {den Berg} {et~al.}(2020)van {den Berg}, Strauss, \&
  Effenberger}]{vandenBergEA2020}
van {den Berg}, J.~P., Strauss, R.~D.~T., \& Effenberger, F. 2020, \ssr, 216,
  146

\bibitem[{Virtanen {et~al.}(2020)Virtanen, Gommers, Oliphant, Haberland, Reddy,
  Cournapeau, Burovski, Peterson, Weckesser, Bright, {van der Walt}, Brett,
  Wilson, Millman, Mayorov, Nelson, Jones, Kern, Larson, Carey, Polat, Feng,
  Moore, {VanderPlas}, Laxalde, Perktold, Cimrman, Henriksen, Quintero, Harris,
  Archibald, Ribeiro, Pedregosa, {van Mulbregt}, \& {SciPy 1. 0
  Contributors}}]{VirtanenEA2020}
Virtanen, P., Gommers, R., Oliphant, T.~E., {et~al.} 2020, Nat. Meth., 17, 261

\bibitem[{Vogt {et~al.}(2020)Vogt, Engelbrecht, Strauss, Heber, Kopp, \&
  Herbst}]{VogtEA2020}
Vogt, A., Engelbrecht, N.~E., Strauss, R.~D., {et~al.} 2020, \aap, 642, A170

\bibitem[{Vos(2011)}]{Vos2011}
Vos, E.~E. 2011, {Cosmic ray modulation processes in the heliosphere}

\bibitem[{Wang \& Qin(2018)}]{WangQin2018}
Wang, J.~F. \& Qin, G. 2018, \apj, 868, 139

\bibitem[{Webb \& Gleeson(1979)}]{WebbGleeson1979}
Webb, G.~M. \& Gleeson, L.~J. 1979, \apss, 60, 335

\bibitem[{Wijsen(2020)}]{Wijsen2020}
Wijsen, N. 2020, PhD thesis, KU Leuven, Belgium \& Universitat de Barcelona,
  Spain

\bibitem[{Wolfram$|$Alpha(2026)}]{WolframAlpha2026}
Wolfram$|$Alpha. 2026, (accessed 25 March 2026)
  \url{https://www.wolframalpha.com/}

\bibitem[{Zank(2014)}]{Zank2014}
Zank, G.~P. 2014, {Transport processes in space physics and astrophysics}, Vol.
  877 (Springer)

\bibitem[{Zhang(2006)}]{Zhang2006}
Zhang, M. 2006, \jgr (Space Phys.), 111, A04208

\bibitem[{Zhao {et~al.}(2017)Zhao, Adhikari, Zank, Hu, \& Feng}]{ZhaoEA2017}
Zhao, L.-L., Adhikari, L., Zank, G.~P., Hu, Q., \& Feng, X.~S. 2017, \apj, 849,
  88

\end{thebibliography}
 
\begin{appendix}


\section{Mathematical details of the model}
\label{apndx:Maths}

In Sect.~\ref{subsec:SDEs}, Eqs~\ref{eq:BackwardsParkerSpherical} and \ref{eq:BackwardsFocusSpherical} still require various derivatives, which are given here for completeness. The magnitude of the HMF is \edited{\citep{Vos2011, Raath2015}}
\begin{equation}
\label{eq:ParkerMag}
B = B_{\rm P} \left( \frac{r_{\oplus}}{r} \right)^2 \sec \psi ,
\end{equation}
and the unit vector along the magnetic field is \citep{Vos2011, Raath2015, StraussFichtner2015}
\begin{equation}
\label{eq:ParkerUnitB}
\uvec{b} = \cos \psi \, \uvec{r} - \sin \psi \, \uvec{\phi} ,
\end{equation}
where
\begin{subequations}
\label{eq:ParkerCosSin}
\begin{align}
\cos \psi & = \frac{1}{\sqrt{1 + \tan^2 \psi}}         = \frac{1}{\sqrt{1 + \omega_{\odot}^2 (r - r_{\odot})^2 \sin^2 \theta / v_{\rm sw}^2}} , \\
\sin \psi & = \frac{\tan \psi}{\sqrt{1 + \tan^2 \psi}} = \frac{\omega_{\odot} (r - r_{\odot}) \sin \theta / v_{\rm sw}}{\sqrt{1 + \omega_{\odot}^2 (r - r_{\odot})^2 \sin^2 \theta / v_{\rm sw}^2}} ,
\end{align}
\end{subequations}
and it should be kept in mind that $v_{\rm sw}(r, \theta)$.

Using
\begin{subequations}
\begin{align}
\frac{\partial}{\partial r} \left[ \cos \psi \right]      & = - \left( \frac{1}{r - r_{\odot}} - \frac{1}{v_{\rm sw}} \, \frac{\partial v_{\rm sw}}{\partial r} \right) \sin^2 \psi \cos \psi , \\
\frac{\partial}{\partial \theta} \left[ \cos \psi \right] & = - \left( \cot \theta - \frac{1}{v_{\rm sw}} \, \frac{\partial v_{\rm sw}}{\partial \theta} \right) \sin^2 \psi \cos \psi , \\
\frac{\partial}{\partial r} \left[ \sin \psi \right]      & = \left( \frac{1}{r - r_{\odot}} - \frac{1}{v_{\rm sw}} \, \frac{\partial v_{\rm sw}}{\partial r} \right) \sin \psi \cos^2 \psi , \\
\frac{\partial}{\partial \theta} \left[ \sin \psi \right] & = \left( \cot \theta - \frac{1}{v_{\rm sw}} \, \frac{\partial v_{\rm sw}}{\partial \theta} \right) \sin \psi \cos^2 \psi ,
\end{align}
\end{subequations}
it can be calculated that
\begin{subequations}
\begin{align}
\frac{\partial b_r}{\partial r}           & = \left( \frac{1}{v_{\rm sw}} \, \frac{\partial v_{\rm sw}}{\partial r} - \frac{1}{r - r_{\odot}}\right) \sin^2 \psi \cos \psi , \\
\frac{\partial b_r}{\partial \theta}      & = \left( \frac{1}{v_{\rm sw}} \, \frac{\partial v_{\rm sw}}{\partial \theta} - \cot \theta \right) \sin^2 \psi \cos \psi , \\
\frac{\partial b_{\phi}}{\partial r}      & = \left( \frac{1}{v_{\rm sw}} \, \frac{\partial v_{\rm sw}}{\partial r} - \frac{1}{r - r_{\odot}} \right) \sin \psi \cos^2 \psi , \\
\frac{\partial b_{\phi}}{\partial \theta} & = \left( \frac{1}{v_{\rm sw}} \, \frac{\partial v_{\rm sw}}{\partial \theta} - \cot \theta \right) \sin \psi \cos^2 \psi ,
\end{align}
\end{subequations}
and
\begin{subequations}
\begin{align}
\frac{\partial B}{\partial r}      & = B \left[ \left( \frac{1}{r - r_{\odot}} - \frac{1}{v_{\rm sw}} \, \frac{\partial v_{\rm sw}}{\partial r} \right) \sin^2 \psi - \frac{2}{r} \right] , \\
\frac{\partial B}{\partial \theta} & = B \left( \cot \theta - \frac{1}{v_{\rm sw}} \, \frac{\partial v_{\rm sw}}{\partial \theta} \right) \sin^2 \psi .
\end{align}
\end{subequations}
The latter derivatives are required for the DC derivatives (not shown here).

The various terms entering Eqs~\ref{eq:dpdt} and \ref{eq:dmudt} are \citep[see][for the radial parts of these quantities]{leRouxEA2007}
\begin{subequations}
\begin{align}
\vec{\nabla} \cdot \uvec{b} = \; & \frac{1}{r^2} \, \frac{\partial}{\partial r} \left[ r^2 b_r \right] + \frac{1}{r \sin \theta} \, \frac{\partial b_{\phi}}{\partial \phi} \nonumber \\
= \; & \left[ \frac{2}{r} - \left( \frac{1}{r - r_{\odot}} - \frac{1}{v_{\rm sw}} \, \frac{\partial v_{\rm sw}}{\partial r} \right) \sin^2 \psi \right] \cos \psi , \\
\label{eq:divu}
\vec{\nabla} \cdot \vec{u} = \; & \frac{1}{r^2} \, \frac{\partial}{\partial r} \left[ r^2 u_r \right] + \frac{1}{r \sin \theta} \, \frac{\partial u_{\phi}}{\partial \phi} \nonumber \\
= \; & \frac{2 v_{\rm sw}}{r} + \frac{\partial v_{\rm sw}}{\partial r} , \\
\uvec{b} \uvec{b} : \vec{\nabla} \vec{u} = \; & b_r \left( b_r \, \frac{\partial u_r}{\partial r} + \frac{b_{\phi}}{r \sin \theta} \, \frac{\partial u_r}{\partial \phi} - \frac{b_{\phi} u_{\phi}}{r} \right) + \nonumber \\
 & b_{\phi} \left( b_r \, \frac{\partial u_{\phi}}{\partial r} + \frac{b_{\phi}}{r \sin \theta} \, \frac{\partial u_{\phi}}{\partial \phi} + \frac{b_{\phi} u_r}{r} \right) \nonumber \\
= \; & v_{\rm sw} \left[ \left( \frac{1}{v_{\rm sw}} \, \frac{\partial v_{\rm sw}}{\partial r} - \frac{\tan^2 \psi}{r} \right) \cos^2 \psi \right. + \nonumber \\
 & \left. \left( \frac{1}{r} + \frac{1}{r - r_{\odot}} \right) \sin^2 \psi \right] , \\
\uvec{b} \cdot \frac{\partial \vec{u}}{\partial t} = \; & b_r \frac{\partial u_r}{\partial t} + b_{\phi} \frac{\partial u_{\phi}}{\partial t} \nonumber \\
= \; & 0 , \\
\label{eq:buGradu}
\hat{b} \cdot \left[ (\vec{u} \cdot \vec{\nabla}) \vec{u} \right] = \; & b_r \left( u_r \, \frac{\partial u_r}{\partial r} + \frac{u_{\phi}}{r \sin \theta} \, \frac{\partial u_r}{\partial \phi} - \frac{u_{\phi}^2}{r} \right) + \nonumber \\
 & b_{\phi} \left( u_r \, \frac{\partial u_{\phi}}{\partial r} + \frac{u_{\phi}}{r \sin \theta} \, \frac{\partial u_{\phi}}{\partial \phi} + \frac{u_r u_{\phi}}{r} \right) \nonumber \\
= \; & v_{\rm sw}^2 \left[ \left( \frac{1}{v_{\rm sw}} \, \frac{\partial v_{\rm sw}}{\partial r} - \frac{\tan^2 \psi}{r} \right) \cos \psi \right. + \nonumber \\
 & \left. \left( \frac{1}{r} + \frac{1}{r - r_{\odot}} \right) \sin \psi \tan \psi \right] .
\end{align}
\end{subequations}

The derivatives of the elements in the isotropic diffusion tensor (Eq.~\ref{eq:KappaTensor}) are
\begin{subequations}
\begin{align}
\frac{\partial \tens{\kappa}_{rr}}{\partial r} = \; & \frac{\partial \kappa_{\parallel}}{\partial r} \, \cos^2 \psi + \frac{\partial \kappa_{\perp r}}{\partial r} \, \sin^2 \psi \; + \nonumber \\
 & 2 \tens{\kappa}_{r\phi} \left( \frac{1}{r - r_{\odot}} - \frac{1}{v_{\rm sw}} \, \frac{\partial v_{\rm sw}}{\partial r} \right) \sin \psi \cos \psi , \\
\frac{\partial \tens{\kappa}_{r\phi}}{\partial r} = \; & \left( \frac{\partial \kappa_{\perp r}}{\partial r} - \frac{\partial \kappa_{\parallel}}{\partial r} \right) \sin \psi \cos \psi \; + \nonumber \\
 & \tens{\kappa}_{r\phi} \left( \frac{1}{r - r_{\odot}} - \frac{1}{v_{\rm sw}} \, \frac{\partial v_{\rm sw}}{\partial r} \right) (\cos^2 \psi - \sin^2 \psi) , \\
\frac{\partial \tens{\kappa}_{\theta \theta}}{\partial \theta} = \; & \frac{\partial \kappa_{\perp \theta}}{\partial \theta},  \\
\frac{\partial \tens{\kappa}_{\phi r}}{\partial \phi} = \; & \left( \frac{\partial \kappa_{\perp r}}{\partial \phi} - \frac{\partial \kappa_{\parallel}}{\partial \phi} \right) \sin \psi \cos \psi , \\
\frac{\partial \tens{\kappa}_{\phi \phi}}{\partial \phi} = \; & \frac{\partial \kappa_{\parallel}}{\partial \phi} \, \sin^2 \psi + \frac{\partial \kappa_{\perp r}}{\partial \phi} \, \cos^2 \psi .
\end{align}
\end{subequations}
Similarly, for the perpendicular diffusion tensor (Eq.~\ref{eq:DperpTensor}),

\begin{subequations}
\begin{align}
\frac{\partial \tens{D}_{rr}^{\perp}}{\partial r} = \; & \frac{\partial D_{\perp r}(\mu)}{\partial r} \, \sin^2 \psi \; + \nonumber \\
 & 2 \tens{D}_{rr}^{\perp} \left( \frac{1}{r - r_{\odot}} - \frac{1}{v_{\rm sw}} \, \frac{\partial v_{\rm sw}}{\partial r} \right) \cos^2 \psi , \\
\frac{\partial \tens{D}_{r\phi}^{\perp}}{\partial r} = \; & \frac{\partial D_{\perp r}(\mu)}{\partial r} \, \sin \psi \cos \psi \; + \nonumber \\
 & \tens{D}_{r\phi}^{\perp} \left( \frac{1}{r - r_{\odot}} - \frac{1}{v_{\rm sw}} \, \frac{\partial v_{\rm sw}}{\partial r} \right) (\cos^2 \psi - \sin^2 \psi) , \\
\frac{\partial \tens{D}_{\theta \theta}^{\perp}}{\partial \theta} = \; & \frac{\partial D_{\perp \theta}(\mu)}{\partial \theta} , \\
\frac{\partial \tens{D}_{\phi r}^{\perp}}{\partial \phi} = \; & \frac{\partial D_{\perp r}(\mu)}{\partial \phi} \, \sin \psi \cos \psi , \\
\frac{\partial \tens{D}_{\phi \phi}^{\perp}}{\partial \phi} = \; & \frac{\partial D_{\perp r}(\mu)}{\partial \phi} \, \cos^2 \psi .
\end{align}
\end{subequations}
The derivatives with respect to $\phi$ are all zero for the DCs in Sect.~\ref{subsec:HMFSWDCs}.


\section{Model testing}
\label{apndx:TestModel}

\begin{figure*}[t!]
\includegraphics[trim=30mm 12mm 40mm 22mm, clip, scale=0.44]{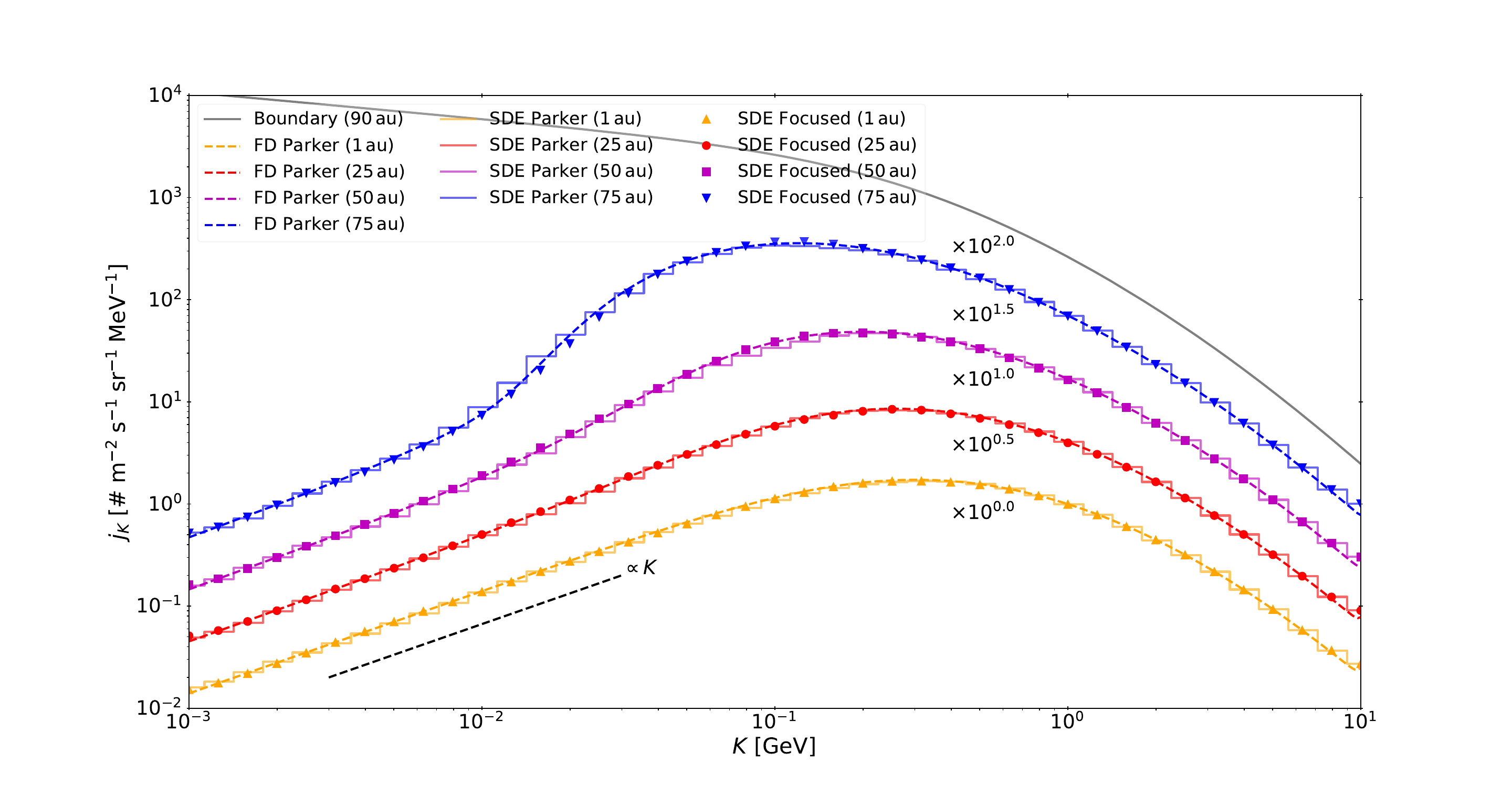}
\caption{\label{fig:TestSDEwithFD}Comparison of the spectra at different radial distances (different colours) as a function of kinetic energy between the SDE scheme of the 3D FTE (symbols), the SDE scheme of the 3D Parker TPE (steps), and an explicit FD scheme of the 1D Parker TPE (dashed lines). The spectra at various radial distances are multiplied by the indicated factor to clearly distinguish them.}
\end{figure*}

The SDE model presented in Sect.~\ref{sec:Model} was tested by comparing its results with those of an explicit finite-difference (FD) model that solves a 1D Parker TPE. A 1D model is used for simplicity, as incorporating more numerical dimensions into FD schemes is notoriously difficult. The model setup is the same as that presented in \citet{CaballeroLopezMoraal2004}, which is based on a previous model by \citet{GleesonUrch1971}. Therefore, consider a spherically symmetric heliosphere in which spatial variations depend solely on the radial coordinate. Spherical symmetry would be fully achieved if the advection velocity and diffusion tensor are independent of latitude or longitude. This requires $\vec{V}_{\rm d} = \vec{0}$ and $\vec{v}_{\rm d}(\mu) = \vec{0}$, as already assumed, and $\kappa_{\perp} = \kappa_{\parallel}$, so that the diffusion tensor is diagonal in spherical coordinates. If there are no temporal dependencies, the system will relax to a stationary state. Expanding the spatial and momentum derivatives of the stationary ($\partial F_0 / \partial t = 0$) version of Eq.~\ref{eq:ParkerTPE} and keeping only the radial components, yield
\begin{align}
\label{eq:ParkerTPE1D}
& - \frac{1}{3} \left( \frac{2 u_r}{r} + \frac{\partial u_r}{\partial r} \right) \frac{\partial F_0}{\partial s} \nonumber \\
& = \left( \frac{2 \kappa_{rr}}{r} + \frac{\partial \kappa_{rr}}{\partial r} - u_r \right) \frac{\partial F_0}{\partial r} + \kappa_{rr} \frac{\partial^2 F_0}{\partial r^2} ,
\end{align}
where $s = \ln p$. The SW speed and DC utilised for this comparison are those of \citet{GleesonUrch1971}, i.e.
\begin{equation}
\label{eq:TestSW}
u_r = V_{\rm sw} \left[ 1 - \eulernumb^{\eta (1 - r/r_{\odot})} \right]
\end{equation}
and
\begin{equation}
\label{eq:TestKappa}
\kappa_{rr} = \kappa_0 \, \frac{v}{c} \left( \frac{r}{r_{\oplus}} \right)^{\alpha} \left( \frac{P}{P_{\rm r}} \right)^{\rho} ,
\end{equation}
respectively, where $\kappa_0 = 4.38 \times 10^{22}~{\rm cm}^2~{\rm s}^{-1} = 0.7~{\rm au}^2~{\rm h}^{-1}$, $\alpha = 0$ governs the radial dependence of the DC, and $\rho = 1$. The spectrum at the outer boundary used for this comparison is that of \citet{CaballeroLopezMoraal2004},
\begin{equation}
\label{eq:TestVLIS}
j_{\rm o}(K) = \frac{21.1~\#~{\rm m}^{-2}~{\rm s}^{-1}~{\rm sr}^{-1}~{\rm MeV}^{-1}}{(K / K_{\rm r})^{2.8} \left[ 1 + 5.85 (K_{\rm r} / K)^{1.22} + 1.18 (K_{\rm r} / K)^{2.54} \right]} .
\end{equation}

The details of various FD schemes are discussed by \citet{GleesonUrch1971}, \citet{PressEA1992}, and \citet{Steenkamp1995}. In the FD scheme, Eq.~\ref{eq:ParkerTPE1D} is solved for $F_0(r,s)$ on an evenly spaced numerical grid, from the outer to the inner boundary (i.e. from $R_{\rm o} = 90~{\rm au}$ to $r_{\odot} = 0.005~{\rm au}$, spaced by $0.2~{\rm au}$) and from the highest to the lowest energy \citep[i.e. from $10~{\rm GeV}$ to $1~{\rm MeV}$, spaced according to the CFL condition; see][]{PressEA1992}. This is done because $F_0(R_{\rm o},s)$ is known (Eq.~\ref{eq:TestVLIS}), and it is assumed that the spectrum remains unmodulated at sufficiently high energies, such that $F_0(r,10~{\rm GeV}) = F_0(R_{\rm o},10~{\rm GeV})$ for all $r$. It was first verified that the fully explicit, fully implicit, and Crank-Nicholson schemes all yielded the same result (within $0.1\%$) for the CFL condition (not shown here). It was then verified that the fully explicit scheme produced the same results as those reported by \citet{CaballeroLopezMoraal2004} for different values of $\kappa_0$ and $\alpha$ \citep[also not shown;][show how a similar spectrum can be obtained at Earth with different values of $\alpha$ if $\kappa_0$ is normalised to the same modulation parameter]{CaballeroLopezMoraal2004}. The 3D SDE models were adapted to be compatible with the 1D FD model, as described above (i.e. by removing the latitudinal dependence of the SW and setting $\kappa_{\perp} = \kappa_{\parallel}$). Additionally, the solution was calculated only in the equator, and isotropic pitch-angle scattering (Eq.~\ref{eq:IsoDmm}) with pitch-angle-independent perpendicular diffusion (Eq.~\ref{eq:Dperp} with $h = 1$) was used in the FTE. For the solutions of the Parker and focused TPEs, $150$ and $400$ pseudo-particles were used per bin, respectively, with $\ell = 0.1$ for the adaptive time step (Eq.~\ref{eq:AdaptiveTimeStep}). The comparison between the fully explicit FD and SDE schemes is shown in Fig.~\ref{fig:TestSDEwithFD}. Note, however, that more pseudo-particles are necessary, as a three-point average in energy and radial position was applied to the SDE results to smooth them. The agreement between the two SDE schemes and the FD scheme is excellent. Although this is expected for the FD and Parker-SDE schemes, both of which solve the Parker TPE, it is reassuring that the FTE also produces the same result in this simplified setup. Fig.~\ref{fig:TestSDEwithFD} also indicates the adiabatic limit of $j_K \propto K$ by a dashed black line at low energies. This demonstrates that the physics of adiabatic energy losses, which dominate other processes at low energies, is correctly captured by the schemes \citep[see e.g.][]{Moraal2013}.


\section{Pitch-angle distribution of anisotropic pitch-angle scattering}
\label{apndx:PADanisoDmm}

To calculate Eq.~\ref{eq:DefineGmu} for the PADC of Eq.~\ref{eq:AnisoDmm},
\begin{align}
\label{eq:AnisoGint}
G(\mu) & = \frac{v}{2 L^{'}} \int_{-1}^{\mu} \frac{1}{D_{\rm A} \left[ |\eta| / (1 + |\eta|) + \epsilon \right]} \, \dinf \eta \nonumber \\
& = K_0 \, \xi \int_{-1}^{\mu} \frac{1 + |\eta|}{\epsilon + (1 + \epsilon) \, |\eta|} \, \dinf \eta ,
\end{align}
where $\xi = \lambda_{\parallel} / L^{'}$ and
\begin{align}
K_n = \; & \frac{2 (1 + \epsilon)^{4 - n}}{3} \left[ \frac{1}{6} + 2 \epsilon + \frac{5}{2} \epsilon^2 + \frac{2}{3} \epsilon^3 \right. + \nonumber \\
 & \left. (1 + 2 \epsilon) \ln \left( \frac{1 + 2 \epsilon}{\epsilon} \right) \right]^{-1} ,
\end{align}
the cases of negative and positive pitch cosine should be considered separately. When $\mu < 0$, the integral can be written as
\begin{align}
& \int_{-1}^{\mu} \frac{1 - \eta}{\epsilon - (1 + \epsilon) \, \eta} \, \dinf \eta \nonumber \\
= \; & \frac{1}{1 + \epsilon} \left[ \eta - \frac{1}{1 + \epsilon} \, \ln \left( \frac{\epsilon}{1 + \epsilon} - \eta \right) - \frac{\ln (1 + \epsilon)}{1 + \epsilon} \right]_{-1}^{\mu} \nonumber \\
= \; & \frac{1}{1 + \epsilon} \left\lbrace 1 + \mu + \frac{1}{1 + \epsilon} \, \ln \left[ \frac{1 + 2 \epsilon}{\epsilon - (1 + \epsilon) \, \mu} \right] \right\rbrace ,
\end{align}
while the integral can be written as
\begin{align}
& \int_{-1}^0 \frac{1 - \eta}{\epsilon - (1 + \epsilon) \, \eta} \, \dinf \eta + \int_0^{\mu} \frac{1 + \eta}{\epsilon + (1 + \epsilon) \, \eta} \, \dinf \eta \nonumber \\
= \; & \frac{1}{1 + \epsilon} \left[ 1 + \frac{1}{1 + \epsilon} \, \ln \left( \frac{1 + 2 \epsilon}{\epsilon} \right) \right] \, + \nonumber \\
& \frac{1}{1 + \epsilon} \left[ \eta + \frac{1}{1 + \epsilon} \, \ln \left( \frac{\epsilon}{1 + \epsilon} + \eta \right) + \frac{\ln (1 + \epsilon)}{1 + \epsilon} \right]_0^{\mu} \nonumber \\
= \; & \frac{1}{1 + \epsilon} \left\lbrace 1 + \mu + \frac{1}{1 + \epsilon} \, \ln \left[ \frac{\epsilon + (1 + \epsilon) \, \mu}{\epsilon^2 / (1 + 2 \epsilon)} \right] \right\rbrace
\end{align}
when $\mu > 0$, where the previous expression was used to evaluate the first integral \citep[see Eqs~2.111.1 and 2.111.5 of][for the integrals]{GradshteynRyzhik2007}. Hence, Eq.~\ref{eq:AnisoGint} is a piecewise continuous function, i.e.
\begin{equation}
G(\mu) = K_1 \, \xi \left[ 1 + \mu + \frac{1}{1 + \epsilon} \left\lbrace \begin{array}{lcl}
\ln \left[ \frac{1 + 2 \epsilon}{\epsilon - (1 + \epsilon) \, \mu} \right] & \mbox{if} & \mu \le 0 \\
\ln \left[ \frac{\epsilon + (1 + \epsilon) \, \mu}{\epsilon^2 / (1 + 2 \epsilon)} \right] & \mbox{if} & \mu > 0
\end{array} \right. \right] ,
\end{equation}
where $K_n = K_0 \, (1 + \epsilon)^{-n}$ was used.

Substituting this into Eq.~\ref{eq:StationaryPAD}, yields also a piecewise continuous function for the PAD, i.e.
\begin{equation}
\label{eq:AnisoPAD}
\tilde{f}_{\rm aniso}(\mu) = C \, \eulernumb^{K_1 \, \xi \, \mu} \left\lbrace \!\! \begin{array}{lcl}
\left[ (1 + \epsilon) \left( \frac{\epsilon}{1 + \epsilon} - \mu \right) \right]^{- K_2 \, \xi} & \mbox{if} & \mu \le 0 \\
\left[ \frac{1 + \epsilon}{\epsilon^2} \left( \frac{\epsilon}{1 + \epsilon} + \mu \right) \right]^{K_2 \, \xi} & \mbox{if} & \mu > 0
\end{array} \right. \! ,
\end{equation}
where
\begin{equation}
C = (1 + 2 \epsilon)^{K_2 \, \xi} \, \eulernumb^{K_1 \, \xi} \left[ \int_{-1}^1 \!\! \eulernumb^{G(\mu)} \, \dinf \mu \right]^{-1} .
\end{equation}
Because the function is piecewise continuous, the normalisation integral can be split into a negative and a positive range, so that
\begin{align}
& \int_{-1}^1 \!\! \eulernumb^{G(\mu)} \, \dinf \mu \nonumber \\
= \; & (1 + 2 \epsilon)^{K_2 \, \xi} \, \eulernumb^{K_1 \, \xi} \left[ (1 + \epsilon)^{- K_2 \, \xi} \int_{-1}^0 \!\! \eulernumb^{K_1 \, \xi \, \mu} \left( \frac{\epsilon}{1 + \epsilon} - \mu \right)^{- K_2 \, \xi} \! \dinf \mu \right. + \nonumber \\
& \left. \left( \frac{1 + \epsilon}{\epsilon^2} \right)^{K_2 \, \xi} \int_0^1 \!\! \eulernumb^{K_1 \, \xi \, \mu} \left( \frac{\epsilon}{1 + \epsilon} + \mu \right)^{K_2 \, \xi} \! \dinf \mu \right] \nonumber \\
= \; & \frac{(1 + 2 \epsilon)^{K_2 \, \xi} \, \eulernumb^{K_1 \, \xi}}{K_1 \, \xi} \nonumber \\
& \Re \left\lbrace (K_2 \, \xi)^{K_2 \, \xi} \, \eulernumb^{K_2 \, \epsilon \, \xi} \left[ \Gamma \left( 1 - K_2 \, \xi , K_1 \, \xi \left( \frac{\epsilon}{1 + \epsilon} - \mu \right) \right) \right]_{-1}^0 \right. + \nonumber \\
& \left. \frac{\eulernumb^{- K_2 \, \epsilon \, \xi}}{(- K_2 \, \epsilon^2 \, \xi )^{K_2 \, \xi}} \left[ \Gamma \left( 1 + K_2 \, \xi , - K_1 \, \xi \left( \frac{\epsilon}{1 + \epsilon} + \mu \right) \right) \right]_0^1 \right\rbrace \nonumber
\end{align}
\begin{align}
= \; & \Re \, \frac{(1 + 2 \epsilon)^{K_2 \, \xi} \, (K_2 \, \xi)^{K_2 \, \xi} \, \eulernumb^{K_2 \, (1 + 2 \epsilon) \, \xi}}{K_1 \, \xi} \nonumber \\
& \left[ \Gamma (1 - K_2 \, \xi , K_2 \, \epsilon \, \xi) - \Gamma (1 - K_2 \, \xi , K_2 \, (1 + 2 \epsilon) \, \xi) \right] \; + \nonumber \\
& \Re \, \frac{(1 + 2 \epsilon)^{K_2 \, \xi} \, \eulernumb^{K_2 \, \xi}}{K_1 \, \xi \, (- K_2 \, \epsilon^2 \, \xi )^{K_2 \, \xi}} \nonumber \\
& \left[ \Gamma (1 + K_2 \, \xi , - K_2 \, (1 + 2 \epsilon) \, \xi) - \Gamma (1 + K_2 \, \xi , - K_2 \, \epsilon \, \xi) \right]
\end{align}
where
\begin{equation}
\int \! \eulernumb^{ax} \, (c \pm x)^{\pm \, b} \, \dinf x = \Re \,  \frac{(\mp \, 1)^b \, \eulernumb^{\mp \, a c}}{a^{1 \pm \, b}} \, \Gamma \left( 1 \pm b , \mp \, a(c \pm x) \right)
\end{equation}
was used for the integrals \citep[calculated by][with the prompts {\tt integrate exp(a*x)*(c + x)\char94 b} or {\tt integrate exp(a*x)/(c - x)\char94 b}]{WolframAlpha2026}, with $\Gamma(\alpha,z)$ the upper incomplete Gamma function.

Similarly, the first-order anisotropy (Eq.~\ref{eq:A1}) can be calculated to be
\begin{align}
\label{eq:AnisoA1}
& A_1^{\rm aniso} = 3 \int_{-1}^1 \!\! \mu \, \tilde{f}_{\rm aniso}(\mu) \, \dinf \mu \nonumber \\
= \; & 3 \, C \left[ (1 + \epsilon)^{- K_2 \, \xi} \int_{-1}^0 \!\! \mu \, \eulernumb^{K_1 \, \xi \, \mu} \left( \frac{\epsilon}{1 + \epsilon} - \mu \right)^{- K_2 \, \xi} \! \dinf \mu \right. + \nonumber \\
& \left. \left( \frac{1 + \epsilon}{\epsilon^2} \right)^{K_2 \, \xi} \int_0^1 \!\! \mu \, \eulernumb^{K_1 \, \xi \, \mu} \left( \frac{\epsilon}{1 + \epsilon} + \mu \right)^{K_2 \, \xi} \! \dinf \mu \right] \nonumber \\
= \; & \left. \frac{3 \, C}{(K_1 \, \xi)^2} \, \Re \, \right\lbrace \left[ K_2 \, \epsilon \, \xi \, \Gamma \left( 1 - K_2 \, \xi , K_1 \, \xi \left( \frac{\epsilon}{1 + \epsilon} - \mu \right) \right) \right. - \nonumber \\
& \left. \Gamma \left( 2 - K_2 \, \xi , K_1 \, \xi \left( \frac{\epsilon}{1 + \epsilon} - \mu \right) \right) \right]_{-1}^0 (K_2 \, \xi)^{K_2 \, \xi} \, \eulernumb^{K_2 \, \epsilon \, \xi} \; - \nonumber \\
& \left[ K_2 \, \epsilon \, \xi \, \Gamma \left( 1 + K_2 \, \xi , - K_1 \, \xi \left( \frac{\epsilon}{1 + \epsilon} + \mu \right) \right) \right. + \nonumber \\
& \left. \left. \Gamma \left( 2 + K_2 \, \xi , - K_1 \, \xi \left( \frac{\epsilon}{1 + \epsilon} + \mu \right) \right) \right]_0^1 (- K_2 \, \epsilon^2 \, \xi)^{- K_2 \, \xi} \, \eulernumb^{- K_2 \, \epsilon \, \xi} \right\rbrace \nonumber \\
= \; & \Re \, \frac{3 \, C \, (K_2 \, \xi)^{K_2 \, \xi} \, \eulernumb^{K_2 \, \epsilon \, \xi}}{(K_1 \, \xi)^2} \nonumber \\
& \lbrace K_2 \, \epsilon \, \xi \left[ \Gamma (1 - K_2 \, \xi , K_2 \, \epsilon \, \xi) - \Gamma (1 - K_2 \, \xi , K_2 \, (1 + 2 \epsilon) \, \xi) \right] \, - \nonumber \\
& \left[ \Gamma (2 - K_2 \, \xi , K_2 \, \epsilon \, \xi) - \Gamma (2 - K_2 \, \xi , K_2 \, (1 + 2 \epsilon) \, \xi) \right] \rbrace \; - \nonumber \\
& \Re \, \frac{3 \, C}{(K_1 \, \xi)^2 (- K_2 \, \epsilon^2 \, \xi)^{K_2 \, \xi} \, \eulernumb^{K_2 \, \epsilon \, \xi}} \nonumber \\
& \lbrace K_2 \, \epsilon \, \xi \left[ \Gamma (1 + K_2 \, \xi , - K_2 \, (1 + 2 \epsilon) \, \xi) - \Gamma (1 + K_2 \, \xi , - K_2 \, \epsilon \, \xi) \right] \, + \nonumber \\
& \Gamma (2 + K_2 \, \xi , - K_2 \, (1 + 2 \epsilon) \, \xi) - \Gamma (2 + K_2 \, \xi , - K_2 \, \epsilon \, \xi) \rbrace
\end{align}
where
\begin{align}
& \int \! x \, \eulernumb^{ax} \, (c \pm x)^{\pm \, b} \, \dinf x \\
= \; & \Re \frac{(\mp \, 1)^{1 - b} \, \eulernumb^{\mp \, a c}}{a^{2 \pm \, b}} \left[ a c \, \Gamma \left( 1 \pm b , \mp \, a(c \pm x) \right) \pm \Gamma \left( 2 \pm b , \mp \, a(c \pm x) \right) \right] \nonumber
\end{align}
was used for the integrals \citep[calculated by][with the prompts {\tt integrate x*exp(a*x)*(c + x)\char94 b} or {\tt integrate (x*exp(a*x))/(c - x)\char94 b}]{WolframAlpha2026}.

\end{appendix}

\end{document}